\documentclass{pasa}%

\usepackage{graphicx}
\usepackage{amsmath}	
\usepackage{amssymb}	
\usepackage{rotating}   
\usepackage{longtable}  

\newcommand{\myaffil}[1]{$^{\rm #1}$}
\newcounter{inst}
\newcommand{\inst}[1]{\noindent%
   \refstepcounter{inst}\myaffil{\arabic{inst}\label{#1}}     
   }

\def\polyflux{{\sc Polygon\_Flux}}
\def\polysummary{Polygons drawn over GLEAM images to measure source and background flux densities for }
\def\polysuffix{ }
\def\spectrasummary{Spectral fitting over the GLEAM band for }
\def\spectrasuffix{ }
\newcommand{\perbeam}{\,beam\ensuremath{^{-1}}}
\newcommand{\SBunit}{\,W\,m\ensuremath{^{-2}}Hz\ensuremath{^{-1}}sr\ensuremath{^{-1}}}

\newcommand{\survarea}{2,860 }

\newcommand{\Fig}{Fig.}
\newcommand{\Figs}{Figs.}

\newcommand{\Sect}{Section}

\newcommand{\Tab}{Table}

\newcommand{\eqn}{equation}

\newcommand{\Eqn}{Equation}

\newcommand{\farcm}{\mbox{\ensuremath{.\mkern-4mu^\prime}}}
\newcommand{\fdg}{\mbox{\ensuremath{.\!\!^\circ}}}

\title[Candidate SNRs observed by GLEAM]{Candidate radio supernova remnants observed by the GLEAM survey over $345^\circ < l < 60^\circ$, $180^\circ < l < 240^\circ$ }

\author[Hurley-Walker~et~al.]{N.~Hurley-Walker\myaffil{\ref{ICRAR}},
B.~M.~Gaensler\myaffil{\ref{ASTRO3D},\ref{Toronto}},
D.~A.~Leahy\myaffil{\ref{Calgary}},
M.~D.~Filipovi\'c\myaffil{\ref{WSU}},
P.~J.~Hancock\myaffil{\ref{ICRAR},\ref{CAASTRO}},
T.~M.~O.~Franzen\myaffil{\ref{ASTRON}},
A.~R.~Offringa\myaffil{\ref{ASTRON}}, 
J.~R.~Callingham\myaffil{\ref{ASTRON}},
L.~Hindson\myaffil{\ref{Hert}},
C.~Wu\myaffil{\ref{UWA}},
M.~E.~Bell\myaffil{\ref{UTS}},
B.-Q.~For\myaffil{\ref{ASTRO3D},\ref{UWA}},
M.~Johnston-Hollitt\myaffil{\ref{ICRAR}},
A.~D.~Kapi\'nska\myaffil{\ref{NRAO}},
J.~Morgan\myaffil{\ref{ICRAR}},
T.~Murphy\myaffil{\ref{USyd},\ref{CAASTRO}},
B.~McKinley\myaffil{\ref{ICRAR}},
P.~Procopio\myaffil{\ref{CAASTRO},\ref{UMelb}},
L.~Staveley-Smith\myaffil{\ref{ASTRO3D},\ref{UWA}}
R.~B.~Wayth\myaffil{\ref{ICRAR},\ref{CAASTRO}},
Q.~Zheng\myaffil{\ref{SHAO}} \\
{\small \myaffil{}\,Email: nhw@icrar.org}\\
{\small \inst{ICRAR}\,International Centre for Radio Astronomy Research, Curtin University, Bentley, WA 6102, Australia}\\
{\small \inst{ASTRO3D}\,ARC  Centre  of  Excellence  for  All  Sky  Astrophysics  in  3  Dimensions  (ASTRO  3D)}\\
{\small \inst{Toronto}\,Dunlap Institute for Astronomy and Astrophysics, 50 St. George St, University of Toronto, ON M5S 3H4, Canada}\\
{\small \inst{Calgary}\,Department of Physics and Astronomy, Science B 605, University of Calgary, 2500 University Dr NW, Calgary, AB T2N 1N4 Canada}\\
{\small \inst{WSU}\,Department of Physics, 5 Second Ave, Kingswood NSW 2747, Australia}\\
{\small \inst{CAASTRO}\,ARC Centre of Excellence for All-sky Astrophysics (CAASTRO)}\\
{\small \inst{ASTRON}\,Netherlands Institute for Radio Astronomy (ASTRON), PO Box 2, 7990 AA Dwingeloo, The Netherlands}\\
{\small \inst{Hert}\,Centre for Astrophysics Research, School of Physics, Astronomy and Mathematics, University of Hertfordshire, College Lane, Hatfield AL10 9AB, UK}\\
{\small \inst{UWA}\,International Centre for Radio Astronomy Research, University of Western Australia, Crawley 6009, Australia}\\
{\small \inst{UTS}\,University of Technology Sydney, 15 Broadway, Ultimo NSW 2007, Australia}\\
{\small \inst{RRI}\,Raman Research Institute, Bangalore 560080, India}\\
{\small \inst{NRAO}\,National Radio Astronomy Observatory, P.O. Box O, Socorro, NM 87801, USA}\\
{\small \inst{USyd}\,Sydney Institute for Astronomy, School of Physics, The University of Sydney, NSW 2006, Australia}\\
{\small \inst{UMelb}\,School of Physics, The University of Melbourne, Parkville, VIC 3010, Australia}\\
{\small \inst{SHAO}\,Shanghai Astronomical Observatory, 80 Nandan Rd, Xuhui Qu, Shanghai Shi, China, 200000}\\
\\
}

\jid{PASA}
\doi{10.1017/pas.\the\year.xxx}
\jyear{\the\year}

\usepackage{aas_macros}
\usepackage{hyperref} 
\hypersetup{colorlinks,citecolor=blue,linkcolor=blue,urlcolor=blue}

\begin{document}

\begin{frontmatter}
\maketitle

\begin{abstract}
We examined the latest data release from the GaLactic and Extragalactic All-sky Murchison Widefield Array (GLEAM) survey covering $345^\circ < l < 60^\circ$, $180^\circ < l < 240^\circ$, using these data and that of the \textit{Widefield Infrared Survey Explorer} to follow up proposed candidate Supernova Remnant (SNR) from other sources. Of the 101 candidates proposed in the region, we are able to definitively confirm ten as SNR, tentatively confirm two as SNR, and reclassify five as \textsc{Hii} regions. A further two are detectable in our images but difficult to classify; the remaining 82 are undetectable in these data.
We also investigated the 18~unclassified Multi-Array Galactic Plane Imaging Survey (MAGPIS) candidate SNRs, newly confirming three as SNRs, reclassifying two as \textsc{Hii} regions, and exploring the unusual spectra and morphology of two others.
\end{abstract}

\end{frontmatter}

\section{Introduction}

Supernovae inject $\approx10^{51}$\,erg of kinetic energy into the surrounding interstellar medium (ISM), and a powerful blast wave propagates outward, sweeping up matter and distributing $\approx10-1000\mu$G magnetic fields throughout a roughly spherical volume. The cosmic rays and magnetic fields associated with the forward shock induce synchrotron emission, which is detectable at radio frequencies \citep{1950PhRv...78..616A}. The radio brightness is expected to increase secularly for young (age $t<100$\,yr) SNRs and thereafter decreases as the magnetic field is distributed over a larger volume \citep{2017ApJS..230....2B}. SNRs remain radio-bright until they merge with the ISM, $\approx200,000$\,yr after formation \cite[see][for a thorough review of the radio properties of SNRs]{2015A&ARv..23....3D}. 

While SNRs are also detectable at other wavelengths, some producing filamentary optical emission \citep{1954ApJ...119..206B}, others possessing central thermal X-ray emission \citep[see ][ for a review]{2012A&ARv..20...49V}, around $95$\,\% of known and candidate Galactic SNRs have been detected via their radio emission \citep{2015A&ARv..23....3D}. Follow-up observations via other means are often critical to determine whether the candidate truly is a SNR, such as determining whether the radio emission is thermal or non-thermal by examining its infrared (IR) brightness; finding an associated pulsar via radio or X-ray observations; measuring kinematic movement of the shell by \textsc{Hi} absorption; or determining the presence of a pulsar wind nebulae by examining the X-ray emission.

\cite{2014BASI...42...47G} published a compilation of SNR and candidate SNR detected by astronomers via all methods, including optical, radio, X-ray, and $\gamma$-rays, and regularly updates each SNR entry with new published work \footnote{See \protect\cite{2017yCat.7278....0G} for the latest version, also available at \href{http://www.mrao.cam.ac.uk/surveys/snrs/}{mrao.cam.ac.uk/surveys/snrs/}}. The June~2017 version contains 295~confirmed SNRs and $\approx250$~candidate SNRs which have not yet been confirmed by \citeauthor{2017yCat.7278....0G}. The objects in the latter category have only been detected by a single survey or published work, and would be easily classified by the addition of sensitive radio and infrared data.

The Murchison Widefield Array (MWA) offers a new view of the radio sky, as a Square Kilometer Array LOW precursor operating in the Murchison Radioastronomy Observatory of Western Australia \citep{2013PASA...30....7T}. The GaLactic and Extragalactic All-sky MWA \citep[GLEAM; ][]{2015PASA...32...25W} survey used the MWA to survey the sky at $\approx2'$ resolution, across a bandwidth of 72 to 231\,MHz. The latest data release of this survey unveils \survarea\,deg$^2$ of the Galactic Plane covering $345^\circ < l < 60^\circ$, $180^\circ < l < 240^\circ$, $|b|\leq 10^\circ$ (Hurley-Walker et al. submitted).\footnote{$240^\circ < l < 345^\circ$ will be published by Johnston-Hollitt et al. (in prep); $60^\circ < l < 180^\circ$ is inaccessible to the MWA due to being above its declination limit.}

This paper is the first of two papers studying SNRs in the region published by Hurley-Walker et al. (submitted), and is referred to as Paper~\textsc{I}. As the papers share common methods, datasets, and analysis, common material is presented in this paper under a single Methodology section (\Sect~\ref{sec:method}) and the material omitted from Paper~\textsc{II}, the focus of which is new SNRs detected in the region. Paper~\textsc{I} examines candidate SNRs detected by other methods, and attempts to confirm or invalidate them using the GLEAM data.

Within the published region, \cite{2017yCat.7278....0G} list 136~candidate SNRs (\Tab~\ref{tab:allsources})\footnote{http://www.mrao.cam.ac.uk/surveys/snrs/snrs.info.html\#S23}. Of particular note are the 35~unconfirmed candidates (of the original 49~suggested) by \cite{2006AJ....131.2525H}, the largest sample of unconfirmed candidates in this longitude range of \cite{2017yCat.7278....0G}. Examining the literature, 17 have already been reclassified by other work, leaving 18 to be investigated. This work uses the GLEAM Galactic Plane data release and other available data to follow up the 101 non-MAGPIS and 18~MAGPIS candidate SNRs to determine their nature. \Sect~\ref{sec:method} explains the methodology, including the datasets used in this work; \Sect~\ref{sec:results} details the findings for each candidate SNR detectable in GLEAM; \Sect~\ref{sec:discussion} examines the overall patterns in these data; and \Sect~\ref{sec:conclusions} summarises our conclusions.

Throughout the paper, we use the original names of SNR candidates as listed in their discovery publications, rather than renaming them to a consistent format. Positions are in J2000 unless otherwise specified. Single-frequency images use the `cubehelix' colour scale \citep{2011BASI...39..289G}. 

\section{Methodology}\label{sec:method}

\subsection{Data}\label{sec:data}

The primary resource for Papers~\textsc{I} and {II} is the GLEAM Galactic Plane data published by Hurley-Walker et al. (submitted), which consists of \survarea\,deg$^2$ covering $345^\circ < l < 60^\circ$, $180^\circ < l < 240^\circ$, $|b|\leq 10^\circ$\footnote{$240^\circ < l < 345^\circ$ will be published by Johnston-Hollitt et al. (in prep); $60^\circ < l < 180^\circ$ is inaccessible to the MWA due to being above its declination limit.} There are 24~frequencies available: $20\times7.68$-MHz ``narrowband'' images, and four ``wideband'' images at frequency ranges 72--103\,MHz, 103--134\,MHz, 139--170\,MHz and 170--231\,MHz. Thermal noise is significant in the narrowband images, varying from $\approx20$ to 500 mJy\perbeam~ over 72 to 213\,MHz (for $|b|>1^\circ$). The sensitivity of the wideband images, $\approx20$ to 50 mJy\perbeam, is largely limited by the quality of calibration and deconvolution, as well as confusion at low Galactic latitudes, due to the limited image resolution.

Given the wide bandwidth of this survey, for sources of integrated flux density $>3\times$ the local RMS in a narrowband image, it is possible to measure an in-band spectral index $\alpha$ assuming that the emission from the source follows a power-law spectrum of $S_\nu\propto\nu^\alpha$. Synchrotron sources such as shell-type SNR are expected to have ``falling'' spectra of $-1.1 < \alpha < 0$, depending on age and environment \citep[see][for a review]{2015A&ARv..23....3D}. \textsc{Hii} regions typically have flatter spectral indices of $-0.2 < \alpha < +2$ \citep{2016era..book.....C}, dominated by their thermal emission. At low frequencies, they become optically thick, and absorb the background diffuse synchrotron emission. In an RGB cube formed from the three lowest wideband GLEAM images (R $=72$--103\,MHz, G$=103$--134\,MHz, and B$=139$--170\,MHz), \textsc{Hii} regions are distinctively blue against the diffuse ``red'' synchrotron emission (see \Sect~\ref{sec:finding}).

The all-sky survey release of the \textit{Wide-field Infrared Survey Explorer} (\textit{WISE}) provides an extremely useful discriminant between thermal and non-thermal emission, specifically in discriminating \textsc{Hii} regions from SNR. \textsc{Hii} regions have a distinctive morphology in the lower two of the four \textit{WISE} bands: 22\,$\mu$m emission from stochastically heated small dust grains, surrounded by a 12\,$\mu$m halo, where polycyclic aromatic hydrocarbon (PAH) molecules fluoresce from the UV radiation \cite{2008ApJ...681.1341W, 2009ApJ...694..546W, 2010A&A...523A...6D}. The center is usually coincident with radio continuum emission from the ionised gas directly around the star. These data were used to catalog over 8000~Galactic \textsc{Hii} regions \citep{2014ApJS..212....1A}. Non-thermal emission of purely synchrotron origin, such as that from classic shell-type SNRs, should have no correlated emission in the mid-infrared, although there may be coincident emission from \textsc{Hii} regions in the same complex, or unrelated sources along the line-of-sight.

Two other radio surveys yield useful insights for our observations. The Bonn 11-cm (2.695\,GHz) Survey with the Effelsburg Telescope \citep[hereafter E11; ][]{1984A&AS...58..197R} is a single-dish survey covering $357\fdg4 \leq l \leq 76^\circ$, $b\leq |1\fdg5|$ with $4\farcm3$ resolution and 50\,mK (20\,mJy\perbeam) sensitivity, which is useful for total power measurements of larger SNR near the Galactic plane\footnote{E11 images were obtained here: \href{http://www3.mpifr-bonn.mpg.de/survey.html}{http://www3.mpifr-bonn.mpg.de/survey.html}}. Careful background subtraction is necessary to measure accurate flux densities (see \Sect~\ref{sec:fluxes}) but the large frequency lever arm between 200\,MHz and 2.695\,GHz yields excellent spectral indices even for measurements with large uncertainties. The Molonglo Synthesis Telescope (MOST) Galactic Plane Survey at 843\,MHz is an interferometric survey with $45"\times 45 \mathrm{cosec}|\delta|"$ resolution covering $245^\circ < l < 355^\circ$, $|b| < 1\fdg5$, with 1--2\,mJy\perbeam~ RMS noise. There have been two data releases, MGPS1 \citep{1999ApJS..122..207G} and MGPS2 \citep{2007MNRAS.382..382M,2014PASA...31...42G}; we find generally that our SNR candidates are more visible in the first data release than the second, and use MGPS1 throughout\footnote{\href{http://www.astrop.physics.usyd.edu.au/MGPS/}{http://www.astrop.physics.usyd.edu.au/MGPS/}}. The interferometric nature of the survey means that flux densities of larger objects may be underestimated but the resolution yields excellent morphological information.

Other ancillary datasets which frequently assist our search and classification are:
\begin{itemize}
\item{National Radio Astronomy Observatory Very Large Array (VLA) Sky Survey \citep[NVSS; ][]{1998AJ....115.1693C} at 1.4\,GHz;}
\item{The VLA Low-Frequency Sky Survey Redux \citep[VLSSr; ][]{2014MNRAS.440..327L};}
\item{The 2nd Digitized Sky Survey \citep[DSS2; ][]{2000ASPC..216..145M}\footnote{\href{http://www.cadc-ccda.hia-iha.nrc-cnrc.gc.ca/en/dss/}{cadc-ccda.hia-iha.nrc-cnrc.gc.ca/en/dss/}};}
\item{The Australia Telescope National Facility pulsar catalogue v1.59 \citep{2005AJ....129.1993M}\footnote{\href{http://www.atnf.csiro.au/research/pulsar/psrcat}{atnf.csiro.au/research/pulsar/psrcat/}};}
\item{Alternative Data Release 1 of the Tata Institute for Fundamental Research Giant Metrewave Radio Telescope Sky Survey \citep[TGSS-ADR1; ][]{2017A&A...598A..78I} at 150\,MHz.}
\end{itemize}


\subsection{SNR catalogues}\label{sec:snrcat}

The most comprehensive list of known SNRs is compiled by \cite{2014BASI...42...47G}, with the latest version compiled in 2017 \citep{2017yCat.7278....0G}\footnote{available at \href{http://www.mrao.cam.ac.uk/surveys/snrs/}{http://www.mrao.cam.ac.uk/surveys/snrs/}}, comprising 295~known SNRs and a similar number of candidates.
Within the Galactic longitude range of this data release, Green (priv. comm.) provided a machine-readable list of 136~candidate SNRs (\Tab~\ref{tab:allsources}). In this work, for each object, we used its position and diameter to generate a region file to overlay on Flexible Image Transport System (\textsc{FITS}) images using the viewing software DS9\footnote{\href{http://ds9.si.edu/site/Home.html}{ds9.si.edu}}. In Hurley-Walker et al. (submitted), we make use of the known and candidate SNR catalogues to exclude them from our search for new SNR, and for comparison with our detected SNRs. Hereafter, the known (non-candidate) SNRs in \citep{2017yCat.7278....0G} are denoted ``G17''.

We noted that 35~candidate SNR were derived from the Multi-Array Galactic Plane Imaging Survey \citep[MAGPIS; ][]{2006AJ....131.2525H}, the largest single contributor of as-yet un-confirmed SNRs. Given the homogeneity of this sample and the accessibility of its ancillary data\footnote{\href{https://third.ucllnl.org/gps/}{https://third.ucllnl.org/gps/}}, we obtained the full MAGPIS catalogue of 49 objects and included them all as potential objects to investigate. The results of this search are described in \Sect~\ref{sec:MAGPIS}.

\subsection{Finding SNRs}\label{sec:finding}

To find candidates which we are able to measure, we overlay the region files created in \Sect~\ref{sec:snrcat} on \textsc{FITS} images and search by-eye for objects with a shell-like morphology, both in the wide (170--231\,MHz) image, and an RGB cube formed from the 72--103\,MHz (R), 103--134\,MHz (G), and 139--170\,MHz (B) images. The resolution of the images is only $2'$--$4'$, so only SNR of extents larger then about $5'$ can be identified by eye. In this work, we search at the locations of candidate~SNRs, and in Paper~\textsc{II}, we search for objects that do not correspond to existing candidate or known SNR locations.

\begin{figure*}
    \centering
   \includegraphics[width=\textwidth]{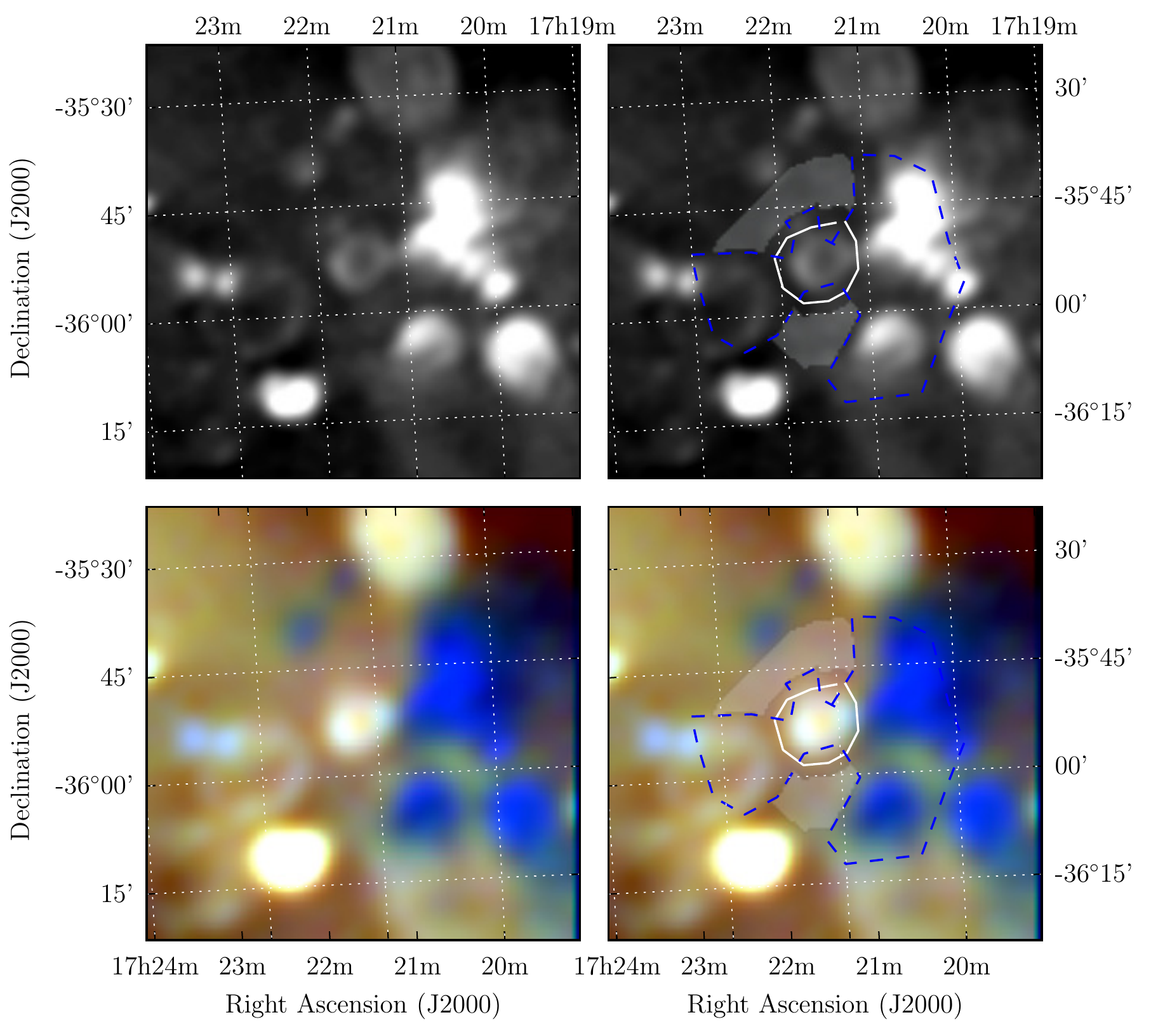}
    \caption{G\,$351.4+0.4$ as imaged, detected and measured in Paper~\textsc{II}, demonstrating the method used in that paper and this work. The top two panels show the GLEAM 170--231\,MHz images; the lower two panels show the RGB cube formed of the 72--103\,MHz (R), 103--134\,MHz (G), and 139--170\,MHz (B) images. G\,$351.4+0.4$ can be clearly discriminated as a white ellipse in the centre of the image, compared to the \textsc{Hii} regions surrounding it, which appear dark blue, due to their absorbing effect on the lowest frequency radio emission. The left two panels show the panels without annotation, while the right two panels show the use of the \polyflux~ software. The white lines indicate polygons drawn to encapsulate the SNR shell measured in this work; the blue dashed lines indicate polygons drawn to exclude regions from being used as background; the grey shading indicates areas used to calculate the background.}
    \label{fig:snr_rgb_example}
\end{figure*}

The wide spectral coverage makes it easy to discriminate between \textsc{Hii} regions and SNR, as the former appear blue in the RGB image, as they become optically thick below $\approx150$\,MHz and absorb the background Galactic synchrotron. \Fig~\ref{fig:snr_rgb_example} shows the SNR candidate  G\,$350.7+0.6$ (Paper~\textsc{II}, \Sect~3.18) as an example.

\subsection{Measuring SNRs}\label{sec:fluxes}

To effectively measure the flux densities of SNRs, we employed the \polyflux\footnote{\href{https://github.com/nhurleywalker/polygon-flux}{github.com/nhurleywalker/polygon-flux}} software (Hurley-Walker in prep). This presents an interactive view of the wide and RGB images, and allows the user to draw a polygon surrounding the object of interest. The user may also draw a polygon to encapsulate any regions which are thought to be contaminated with other objects which might interfere with a measurement of the background.

Once the polygons are selected, the flux densities of the SNR can be calculated from each of the 24~GLEAM mosaics, each convolved to the same resolution as the lowest-resolution image at 72--80\,MHz. A background is calculated from an annulus surrounding the first polygon, excluding any regions selected by the second polygon.
The total flux density inside the polygon is calculated and the average background level subtracted. This ensures that regardless of where the polygon is drawn, the total flux density of the SNR is calculated accurately. \Fig~\ref{fig:snr_rgb_example} shows an example of the polygon drawing method.

Some larger SNRs are seen to overlap with point sources (see e.g. \Sect~\ref{SNRG352.2-0.1}). A compact source near the centre of a SNR may be a pulsar or PWNe, but if it is off-centre, the origin is less clear. Extragalactic sources such as radio galaxies are fairly isotropically distributed and have varying brightnesses. At the resolution of the GLEAM survey, only one-third of radio galaxies are at all resolved, with a size more than 10\,\% greater than the local PSF, and the majority have non-thermal spectra with median $\alpha\approx-0.8\pm0.2$ \citep{2017MNRAS.464.1146H}). We therefore consider any non-central, unresolved source with a non-thermal spectrum to be a contaminating extragalactic radio source. We perform compact source-finding and fitting across the full GLEAM band as per Hurley-Walker et al. (submitted), forcing the background level to match that of the surrounding SNR. Sources thus measured are then subtracted from the flux density measurements of the SNR.

\begin{figure*}
    \centering
   \includegraphics[width=\textwidth]{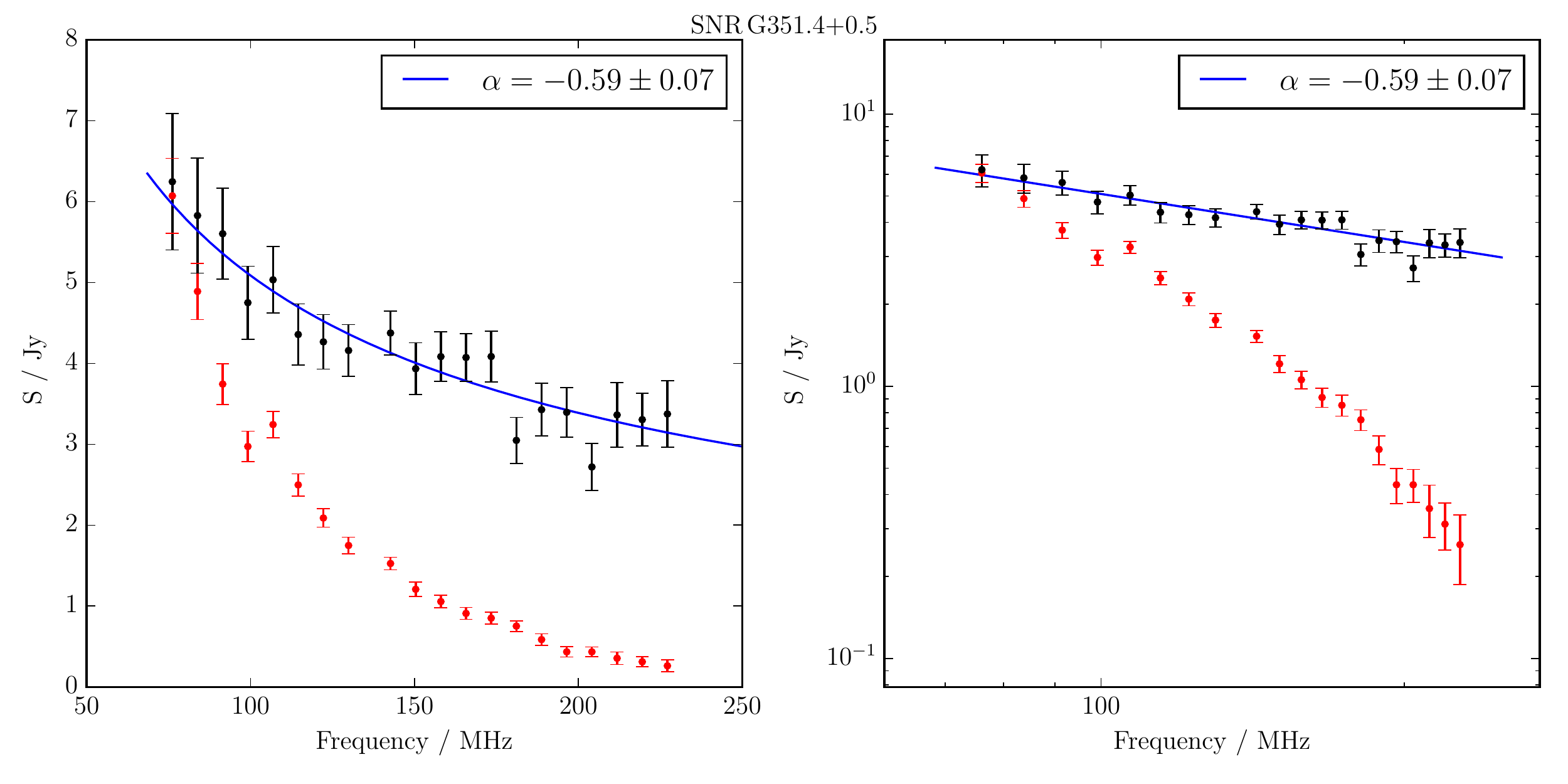}
    \caption{The spectrum of G\,$351.4+0.4$ as measured using the backgrounding and flux summing technique described in \Sect~\ref{sec:fluxes}. The left panel shows flux density against frequency with linear axes while the right panel shows the same data in log. (It is useful to include both when analysing the data as a log plot does not render negative data points, which occur for faint SNRs or negative background levels). The black points show the (background-subtracted) SNR flux density measurements, the red points show the measured background, and the blue curve shows a linear fit to the data above 150\,MHz (marked in black) in log-log space (i.e. $S_\nu \propto \nu^\alpha$). The fitted value of $\alpha$ is shown at the top right.}
    \label{fig:snr_spectrum_example}
\end{figure*}

Once these flux densities are calculated, a power-law SED is fit to the spectra, using only the 20~narrow-band measurements. If the reduced~$\chi^2>1.93$, indicating a poor fit, the calculation is repeated for the wide-band images, to improve signal-to-noise. If that also shows a poor fit, no spectral index can be reported for that SNR. All calculations are saved to a \textsc{FITS} table for future use. \Fig~\ref{fig:snr_spectrum_example} shows an example of the output from the spectral fitting routine.

For the new SNRs described in Paper~\textsc{II}, the software is used to perform measurements in the MGPS and E11 datasets by using exactly the same polygons as determined when examining the GLEAM images. This allows us to avoid contamination from thermal regions and point sources in the object and background flux density measurements, and for partial objects, means the same fraction of the objects is measured. For the candidate SNRs in this work, published flux density measurements are often used in conjunction with the GLEAM measurements to derive SEDs, although on occasion using the same \polyflux technique is useful.

\subsection{SNR evolution}

Multiple publications in the literature review the evolution of SNRs and suggest models of varying complexity to describe it. Given that from radio measurements alone, we generally know little about a given SNR aside from its radius, we restrict ourselves to modelling SNR evolution using the simplest descriptive models, following \cite{1972ARA&A..10..129W}, and making standard assumptions about their energetics. Over the lifetime of a SNR, it passes through several stages, where relatively simple equations can be used to predict the SNR radius as a function of time (or estimate the SNR age, given its radius). The initial stage is the ejecta-dominated stage of free expansion, in which the radius $R$ depends on velocity $v$ and time $t$ via:
\begin{equation}
R = v t
\label{eq:snr_r0}
\end{equation}
    
The velocity of the ejecta depends on the energy of the SNR $E$ and the ejecta mass $M_\mathrm{ejecta}$ via \citep{1972ARA&A..10..129W}:
\begin{equation}
v = \left(\frac{2E}{M_\mathrm{ejecta}}\right)
\approx 10^4\left(\frac{E}{10^{51}\mathrm{erg}}\right)^{1/2}\left(\frac{M_\mathrm{ejecta}}{M_\odot}\right)^{-1/2} \mathrm{km\,s}^{-1}
\label{eq:snr_r3}
\end{equation}
i.e. $10^4$\,km\,s$^{-1}$ for standard assumptions of $M_\mathrm{ejecta}=M_\odot$ and $E=10^{51}$\,erg.

As the ejecta collide with the ISM, a forward shock propagates outwards, and a reverse shock propagates toward the centre of the remnant. This adiabatic (or Sedov–Taylor) stage starts once mass of the swept-up ISM becomes comparable to the ejecta mass, at around 200\,years for typical SNR, and the SNR is fully in this phase once the ejecta are fully thermalised. The SNR radius $R$ is proportional to its age $t$ via \citep{sedov1959similarity}:
\begin{equation}
R=1.17\left(\frac{E}{\rho}\right)^{1/5}t^{2/5}
\label{eq:snr_r1}
\end{equation}
where $\rho$ is the density of the surrounding ISM. Expressed in units typical of spherical SNRs expanding into the Galactic ISM:
\begin{equation}
R\approx5\left(\frac{E}{10^{51}\,\mathrm{erg}}\right)^{1/5}\left(\frac{n_\mathrm{H}}{\mathrm{cm}^{-3}}\right)^{-1/5}\left(\frac{t}{1000\,\mathrm{years}}\right)^{2/5}\,\mathrm{pc}
\label{eq:snr_r2}
\end{equation}

The SNR enters its third phase of radiative expansion, and the shocked ISM begins to cool, starting when the SNR is $\approx$50,000 years old, and the shell velocity is around 200\,km\,s$^{-1}$. The SNR is driven only by internal pressure instead of the initial kinetic expansion, and its radius $R$ increases as $t^{2/7}$. Finally, the SNR cools and enters a stage where momentum drives the shell expansion, and eventually the shell velocity drops to the general ISM random values of $\approx10$\,km\,s$^{-1}$, and the SNR becomes indistinguishable from the surrounding ISM.

\section{Results}\label{sec:results}

We searched 101 non-MAGPIS candidate SNR positions and 18~MAGPIS candidate SNR positions (see \Sect~\ref{sec:snrcat}) to identify regions where the GLEAM data could serve as a useful discriminant. In the first category, we identified 19~regions, presented in \Sect~\ref{sec:candidates}; in the second, we identified seven, discussed in \Sect~\ref{sec:MAGPIS}.

\subsection{Candidate SNRs}\label{sec:candidates}

In this section we examine the 19~candidate SNR with useful GLEAM detections and compare to previous results in the literature, in order of Galactic longitude, first for the outer-Galactic ``oG'' region ($180^\circ < l < 240^\circ$) and then the inner-Galactic ``iG'' region ($345^\circ < l < 60^\circ$). \Tab~\ref{tab:candidates} summarises the vital properties of these objects.

\begin{sidewaystable*}
\caption{Summary of non-MAGPIS candidates, detailed in \Sect~\ref{sec:candidates}. Entries marked with a ``*'', and all entries in rows or columns so marked, were derived in this work. Italics indicate candidates that we believe should no longer be considered potential SNRs. The flux density and spectral index for G\,353.3-1.1 were calculated after the subtraction of contaminating radio sources, and extrapolation of the full shell morphology (see \Sect~\ref{SNRG353.3-1.1}). The two objects classed with ``SNR?'' are potentially SNRs but cannot be definitively proved so by this work. The object classed ``Both'' is a composite thermal and non-thermal source, the former a \textsc{Hii} region and the latter either a compact SNR, or a pulsar.\label{tab:candidates}}
\begin{tabular}{c|ccccccccccc}
Name & RA & Dec & l & b & major & minor & pa & $S_\mathrm{200\,MHz}$* & $\alpha_\mathrm{GLEAM}$* & Class* & Reference(s) \\
   & J2000 & J2000 & $^\circ$ & $^\circ$ & $'$ & $'$ & $^\circ$  & Jy  &                         &       &           \\
 \hline
\hyperref[SNRG189.6+3.3]{G189.6+3.3} & $06^\mathrm{h}19^\mathrm{m}22^\mathrm{s}$ & $ +22^\mathrm{d}13^\mathrm{m}08^\mathrm{s}$ & 189.6 & $3.3$ & $90$ & 90 & 0 & $170$ & $-1.5$ & SNR & {\cite{1994A+A...284..573A}} \\ 
\hyperref[SNRG345.1-0.2]{G345.1-0.2} & $17^\mathrm{h}05^\mathrm{m}21^\mathrm{s}$ & $ -41^\mathrm{d}26^\mathrm{m}04^\mathrm{s}$ & 345.1 & $-0.2$ & $6$ & 6 & 0 & $4.30\pm0.09$ & $-0.69\pm0.06$ & SNR & {\cite{1996A+AS..118..329W}} \\ 
\hyperref[SNRG345.1+0.2]{G345.1+0.2} & $17^\mathrm{h}03^\mathrm{m}40^\mathrm{s}$ & $ -41^\mathrm{d}05^\mathrm{m}11^\mathrm{s}$ & 345.1 & $0.2$ & $10$ & 10 & 0 & $1.6\pm0.2$ & $-0.57\pm0.10$ & SNR & {\cite{1996A+AS..118..329W}} \\ 
\hyperref[SNRG348.8+1.1]{G348.8+1.1} & $17^\mathrm{h}11^\mathrm{m}29^\mathrm{s}$ & $ -37^\mathrm{d}35^\mathrm{m}39^\mathrm{s}$ & 348.8 & $1.1$ & $10$ & 10 & 0 & $1.8\pm0.2$ & $-0.7\pm0.2$ & SNR & {\cite{1996A+AS..118..329W}} \\ 
\hyperref[SNRG352.2-0.1]{G352.2-0.1} & $17^\mathrm{h}26^\mathrm{m}05^\mathrm{s}$ & $ -35^\mathrm{d}33^\mathrm{m}29^\mathrm{s}$ & 352.2 & $-0.1$ & $27$ & 27 & 0 & $2.0\pm0.2$ & $-$ & SNR? & {\cite{2002ASPC..271...31M}} \\ 
\hyperref[SNRG353.3-1.1]{G353.3-1.1} & $17^\mathrm{h}33^\mathrm{m}09^\mathrm{s}$ & $ -35^\mathrm{d}11^\mathrm{m}34^\mathrm{s}$ & 353.3 & $-1.1$ & $60$ & 60 & 0 & $95\pm8$ & $-0.85\pm0.04$ & SNR & {\cite{1995MNRAS.277...36D,1997MNRAS.287..722D}} \\ 
\hyperref[SNRG354.5+0.1]{\textit{G354.46+0.07}} & $17^\mathrm{h}31^\mathrm{m}29^\mathrm{s}$ & $ -33^\mathrm{d}34^\mathrm{m}55^\mathrm{s}$ & 354.46 & $0.07$ & $1.6$ & 1.6 & 0 & $\approx2$ & $-$ & \textsc{Hii} region & {\cite{2013ApJ...774..150R}} \\
\hyperref[SNRG356.6+0.1]{G356.6+00.1} & $17^\mathrm{h}36^\mathrm{m}54^\mathrm{s}$ & $ -31^\mathrm{d}39^\mathrm{m}27^\mathrm{s}$ & 356.6 & $0.1$ & $7$ & 8 & ? & $1.9\pm0.3$ & $-1.1\pm0.3$ & SNR? & {\cite{1994MNRAS.270..847G}} \\ 
\hyperref[SNRG359.2-1.1]{G359.2-01.1} & $17^\mathrm{h}48^\mathrm{m}14^\mathrm{s}$ & $ -30^\mathrm{d}11^\mathrm{m}33^\mathrm{s}$ & 359.2 & $-1.1$ & $4$ & 5 & ? & $2.6\pm0.2$ & $-1.07\pm0.08$ & SNR & {\cite{1994MNRAS.270..847G}} \\ 
\hyperref[SNRG1.2-0.0]{\textit{G1.2-0.0}} & $17^\mathrm{h}48^\mathrm{m}26^\mathrm{s}$ & $ -27^\mathrm{d}54^\mathrm{m}36^\mathrm{s}$ & 1.2 & $0$ & $7.2$ & 4.2 & 10 & $\approx3$ & $-$ & \textsc{Hii} region & {\cite{2009PASJ...61S.209S}} \\ 
\hyperref[SNRG3.1-0.6]{G003.1-00.6} & $17^\mathrm{h}55^\mathrm{m}30^\mathrm{s}$ & $ -26^\mathrm{d}35^\mathrm{m}00^\mathrm{s}$ & 3.1 & $-0.6$ & $28$ & 52 & 10 & $20.0\pm0.9$ & $-0.87\pm0.12$ & SNR & {\cite{1994MNRAS.270..847G}} \\ 
\hyperref[SNRG5.3+0.1]{\textit{G005.3+0.1}} & $17^\mathrm{h}57^\mathrm{m}23^\mathrm{s}$ & $ -24^\mathrm{d}17^\mathrm{m}18^\mathrm{s}$ & 5.3 & $0.1$ & $2.5$ & 2 & ? & $\approx0.2$ & $-$ & \textsc{Hii} region & {\cite{2001ESASP.459..109T}} \\ 
\hyperref[SNRG7.5-1.7]{G7.5-1.7} & $18^\mathrm{h}09^\mathrm{m}58^\mathrm{s}$ & $ -23^\mathrm{d}11^\mathrm{m}49^\mathrm{s}$ & 7.5 & $-1.7$ & $98.4$ & 98.4 & 0 & $53\pm2$ & $-0.67\pm0.09$ & SNR & {\cite{2008ApJ...681..320R}} \\ 
\hyperref[SNRG12.75-0.15]{\textit{G12.75-0.15}} & $18^\mathrm{h}13^\mathrm{m}55^\mathrm{s}$ & $ -17^\mathrm{d}57^\mathrm{m}11^\mathrm{s}$ & 12.75 & $-0.15$ & $15$ & 15 & 0 & $\approx8$ & $-$ & \textsc{Hii} region & {\cite{1985SvA....29..128G}} \\ 
\hyperref[SNRG13.1-0.5]{G13.1-0.5} & $18^\mathrm{h}15^\mathrm{m}55^\mathrm{s}$ & $ -17^\mathrm{d}48^\mathrm{m}45^\mathrm{s}$ & 13.1 & $-0.5$ & $38$* & 28* & 15 & $28.6\pm2.3$ & $-0.57\pm0.03$ & SNR & {\cite{1988ApJS...68..715K,1990ApJ...364..187G}} \\ 
\hyperref[SNRG15.51-0.15]{G15.51-0.15} & $18^\mathrm{h}19^\mathrm{m}25^\mathrm{s}$ & $ -15^\mathrm{d}32^\mathrm{m}00^\mathrm{s}$ & 15.51 & $-0.15$ & $8$ & 9 & ? & $2.86\pm0.29$ & $-0.55\pm0.03$ & SNR & {\cite{2006ApJ...639L..25B}} \\ 
\hyperref[SNRG19.00-0.35]{\textit{G19.00-0.35}} & $18^\mathrm{h}26^\mathrm{m}53^\mathrm{s}$ & $ -12^\mathrm{d}32^\mathrm{m}11^\mathrm{s}$ & 19 & $-0.35$ & $30$ & 30 & 0 & $\approx17$ & $-$ & \textsc{Hii} region & {\cite{1985SvA....29..128G}} \\ 
\hyperref[SNRG35.40-1.80]{\textit{G35.40-1.80}} & $19^\mathrm{h}02^\mathrm{m}21^\mathrm{s}$ & $ +01^\mathrm{d}22^\mathrm{m}27^\mathrm{s}$ & 35.4 & $-1.8$ & $7$ & 7 & 0 & $1.00\pm0.09$ & $-0.85\pm0.02$ & Both & {\cite{1985SvA....29..128G}} \\ 
\hyperref[SNRG36.00+0.00]{G36.00+0.00} & $18^\mathrm{h}57^\mathrm{m}02^\mathrm{s}$ & $ +02^\mathrm{d}43^\mathrm{m}49^\mathrm{s}$ & 36 & $0$ & $20$ & 20 & 0 & $0.44\pm0.09$ & $-1.7\pm0.15$ & pulsar? & {\cite{2006IAUS..230..333U}} \\ 
\end{tabular}

\end{sidewaystable*}

\subsubsection{G$189.6+3.3$}\label{SNRG189.6+3.3}

G$189.6+3.3$ was detected by \cite{1994A+A...284..573A} as a faint X-ray excess overlapping G$189.1+3.0$ (IC443). \cite{2004AJ....127.2277L} performed a detailed radio continuum mapping of the region and noted the presence of a radio-bright arc coincident with the northern edge of the SNR candidate proposed by \cite{1994A+A...284..573A}, with a radio spectral index of $\alpha=-0.1$--$-0.6$. \cite{2008AJ....135..796L} used the VLA and Arecibo to investigate IC443 at 1420\,MHz and detected the arc of G$189.6+3.3$ where it intersects IC443. Our observations show about half of the shell that is visible in the X-ray observations. We were unable to extract a spectrum for the SNR candidate due to low signal-to-noise, and confusing effects from IC443. However, the peak flux density where the shell of G$189.6+3.3$ intersects with IC443, at RA\,$=6^\mathrm{h}18^\mathrm{m}30^\mathrm{s}$, Dec\,$=+22^\mathrm{d}50^\mathrm{m}$, can be measured both our own data, and that of \cite{2008AJ....135..796L}, who determine it to be $\approx6$\,mJy\perbeam~at 1420\,MHz \citep[\Fig~3 of ][]{2008AJ....135..796L} (c/f $\approx1$\,K or $\approx15$\,mJy\perbeam~by \cite{2004AJ....127.2277L}).

In our data, we first measure a background level of 60\,mJy\perbeam, from a nearby region not associated either with G$189.6+3.3$ or IC443. Our peak flux density measurement at the above RA and Dec is 230\,mJy\perbeam, and 170\,mJy\perbeam~after background subtraction. Combining $S_\mathrm{200MHz}=170$\,mJy\perbeam~and $S_\mathrm{1.4GHz}=10$\,mJy\perbeam, and assuming 30\,\% errors on each, yields a spectral index of $\alpha=-1.2\pm0.2$, consistent with an aged population of synchrotron-emitting electrons.

\begin{figure*}
    \centering
   \includegraphics[width=\textwidth]{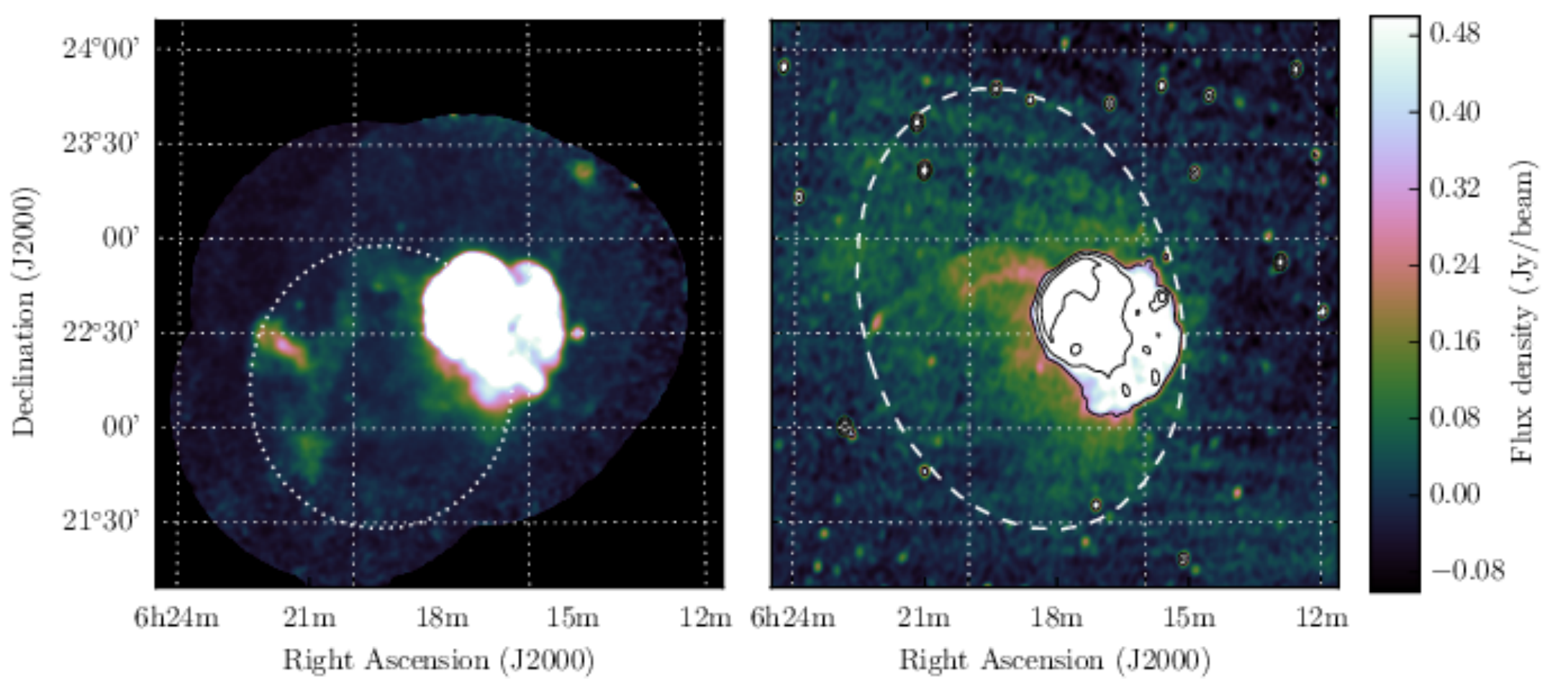}
    \caption{G$189.6+3.3$ as observed by ROSAT (left) and GLEAM at 200\,MHz (right, including color bar). The ROSAT X-ray image has been convolved with a Gaussian kernel 10~pixels wide in order to highlight the large-scale structure of SNR\,G$189.6+3.3$, which is marked with a dotted white ellipse. The dashed line in the right panel indicates the radio excess discussed in \Sect~\ref{SNRG189.6+3.3}. The right panel also contains five logarithmically-spaced thin black contours with levels between 0.3 and 10\,Jy\perbeam, inclusive, to highlight SNR~IC~433.}
    \label{fig:SNRG189.6+3.3}
\end{figure*}

We note that this 60\,mJy\perbeam~background level seems to be associated with an elliptical radio excess, with a centre at RA\,$=6^\mathrm{h}18^\mathrm{m}50^\mathrm{s}$, Dec$=+22^\mathrm{d}38^\mathrm{m}$, diameter $2.4^\circ \times 1.8^\circ$, and a position angle of $20^\circ$ (CCW from North), shown as a dashed ellipse in \Fig~\ref{fig:SNRG189.6+3.3}. The RMS noise at 200\,MHz in this region is about 20\,mJy\perbeam, but while the peak is low signal-to-noise, the total flux density is significant. In the lower-frequency wideband GLEAM images, the background flux density and RMS noise values in this area are 215, 90\,mJy\perbeam~(72--103\,MHz), 100, 40\,mJy\perbeam~(103--134\,MHz) and 68, 25\,mJy\perbeam~(139--170\,MHz). A spectral fit to these background levels (using the RMS values as errors) gives a $\alpha=-1.5\pm0.6$, making this object potentially another candidate SNR. Sidelobes from inadequate calibration and deconvolution of the nearby Crab~nebula increase the noise in the region; a more definitive statement could be made with more data processed to a higher quality.

\subsubsection{G$345.1-0.2$}\label{SNRG345.1-0.2}

\begin{figure*}
    \centering
   \includegraphics[width=\textwidth]{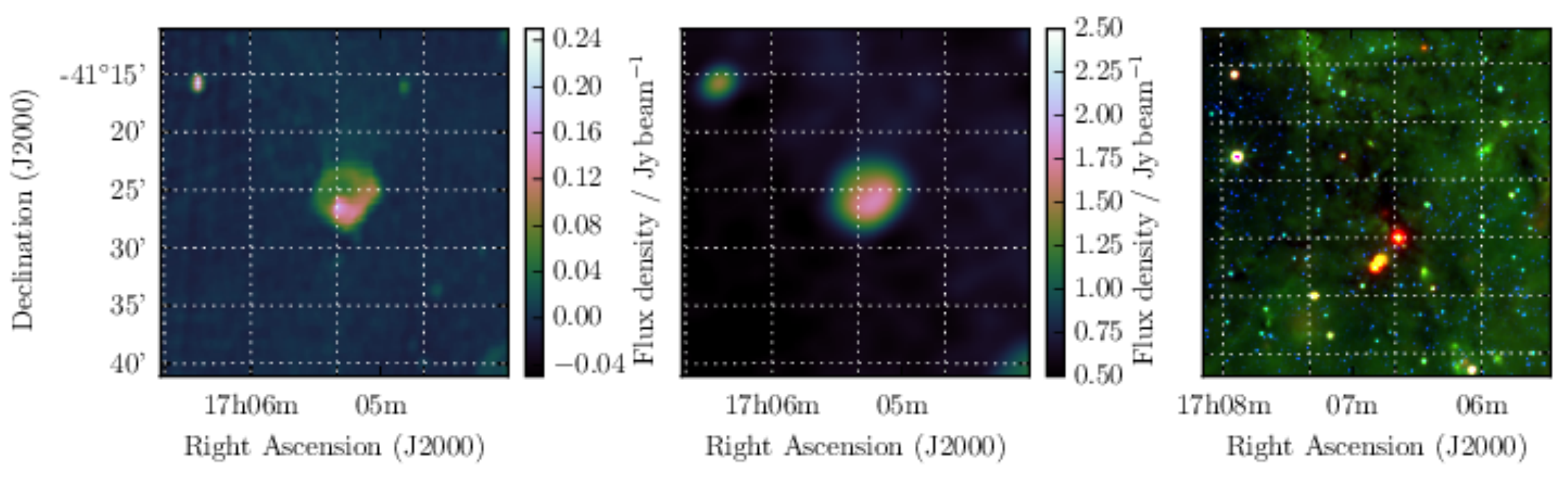}
    \caption{G$345.1-0.2$ as observed by MGPS at 843\,MHz (left), GLEAM at 200\,MHz (middle), and \textit{WISE} (right).}
    \label{fig:SNRG345.1-0.2}
\end{figure*}

\cite{1996A+AS..118..329W} presented 18~new SNRs and 16~candidate SNRs found in observations by the Molonglo Observatory Synthesis Telescope (MOST) at 843\,MHz. G$345.1-0.2$ falls into the candidate category and was described as a ``bright shell with point source'', a diameter of $6'\times6'$, a total flux density of 1.8\,Jy, and a mean surface brightness of $7.1\times10^{-21}$\SBunit. \Fig~\ref{fig:SNRG345.1-0.2} shows the MGPS 843\,MHz observation of G\,$345.1-0.2$, and the corresponding GLEAM image at 200\,MHz.

Despite the absence of any obvious emission in the \textit{WISE} 12- and 22-$\mu$m maps of this area, there is some evidence of a low-frequency turnover within the GLEAM band (\Fig~\ref{fig:SNR_G345.1-0.2_spectrum}). To avoid the effects of \textsc{Hii} region absorption on our spectral calculation, we use only those measurements with $\nu>150$\,MHz, and include the measurement of \cite{1996A+AS..118..329W} of $S_\mathrm{843MHz}=1.8$\,Jy, and find that the candidate has a slightly falling spectrum of $\alpha=-0.69\pm0.06$ and a fitted flux density of $S_\mathrm{200MHz} = 4.30\pm0.09$\,Jy. Its morphology and spectrum suggest that it truly is a SNR.

\subsubsection{G$345.1+0.2$}\label{SNRG345.1+0.2}

\begin{figure*}
    \centering
   \includegraphics[width=\textwidth]{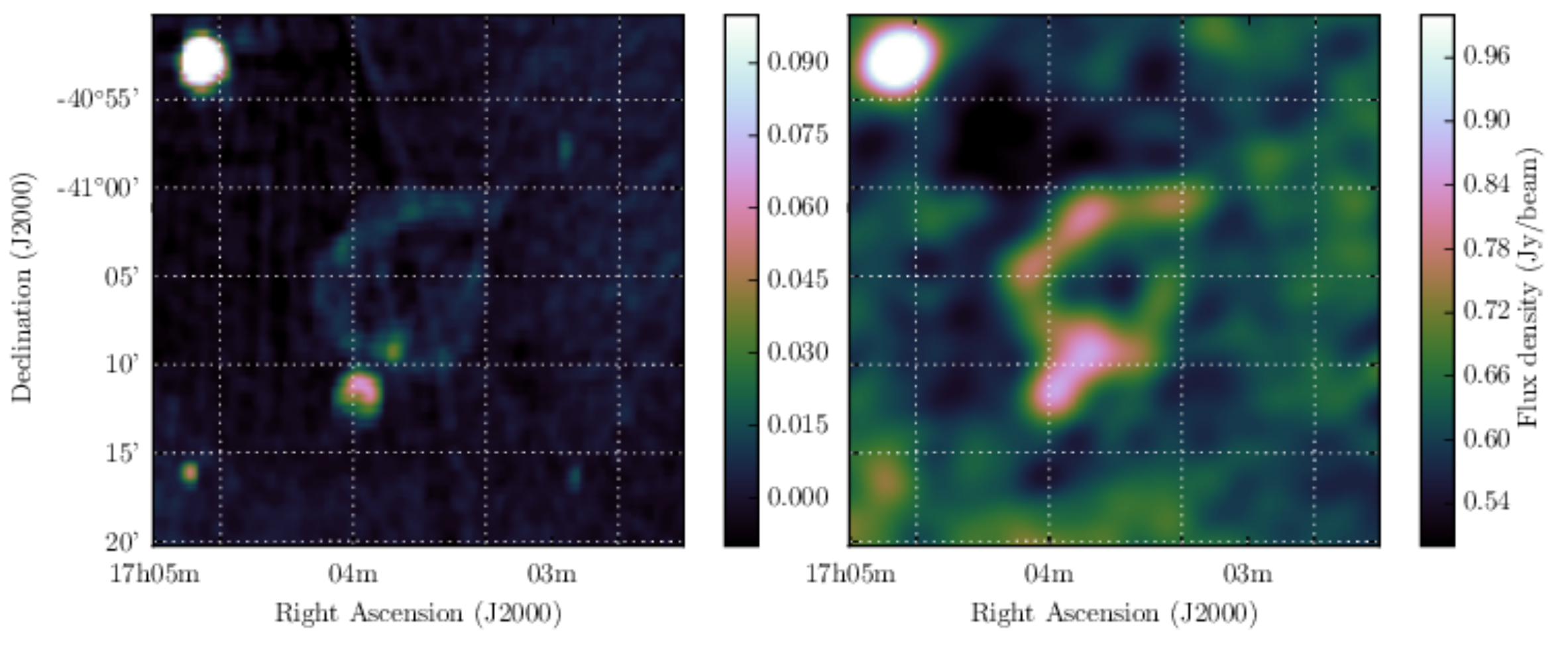}
    \caption{G$345.1+0.2$ as observed by MGPS at 843\,MHz (left) and GLEAM at 200\,MHz (right).}
    \label{fig:SNRG345.1+0.2}
\end{figure*}

Similarly to G\,$345.1-0.2$ (\Sect~\ref{SNRG345.1-0.2}), G$345.1+0.2$ was also classed as a potential SNR by \cite{1996A+AS..118..329W}, who gave its properties as total flux density of 0.7\,Jy, a $10'\times10'$ diameter, and a mean surface brightness of  $0.7\times10^{-21}$\SBunit, describing it as ``a faint shell''. \Fig~\ref{fig:SNRG345.1+0.2} shows the MGPS 843\,MHz observation of G\,$345.1+0.2$, and the corresponding GLEAM image at 200\,MHz.

We tentatively confirm this morphology, noting that the shell appears to have a gap in the north-west.
Background-subtracting our 200-MHz image and the 843-MHz MGPS image, and convolving the latter to the resolution of the former, we can form a spectral index map of the candidate. The shell has a spectral index of $-0.8$ to $-0.4$, while the source south-east of the shell has a flat spectrum of $-0.05$. This source does not appear to have a counterpart in \textit{WISE}, so is not a typical \textsc{Hii} region. It is difficult to separate the source from the shell at frequencies $<200$\,MHz in GLEAM due to the low resolution of the survey, and the background level in GLEAM is similar to the total flux density of the candidate, so the uncertainties on our narrow-band measurements are large. We therefore suggest using the combined GLEAM and MGS measurement as the most reliable estimator of its radio flux density and spectral index, calculating $S_\mathrm{200MHz}=1.6\pm0.2$\,Jy and $\alpha=-0.57\pm0.10$.

\subsubsection{G$348.8+1.1$}\label{SNRG348.8+1.1}

\begin{figure*}
    \centering
   \includegraphics[width=\textwidth]{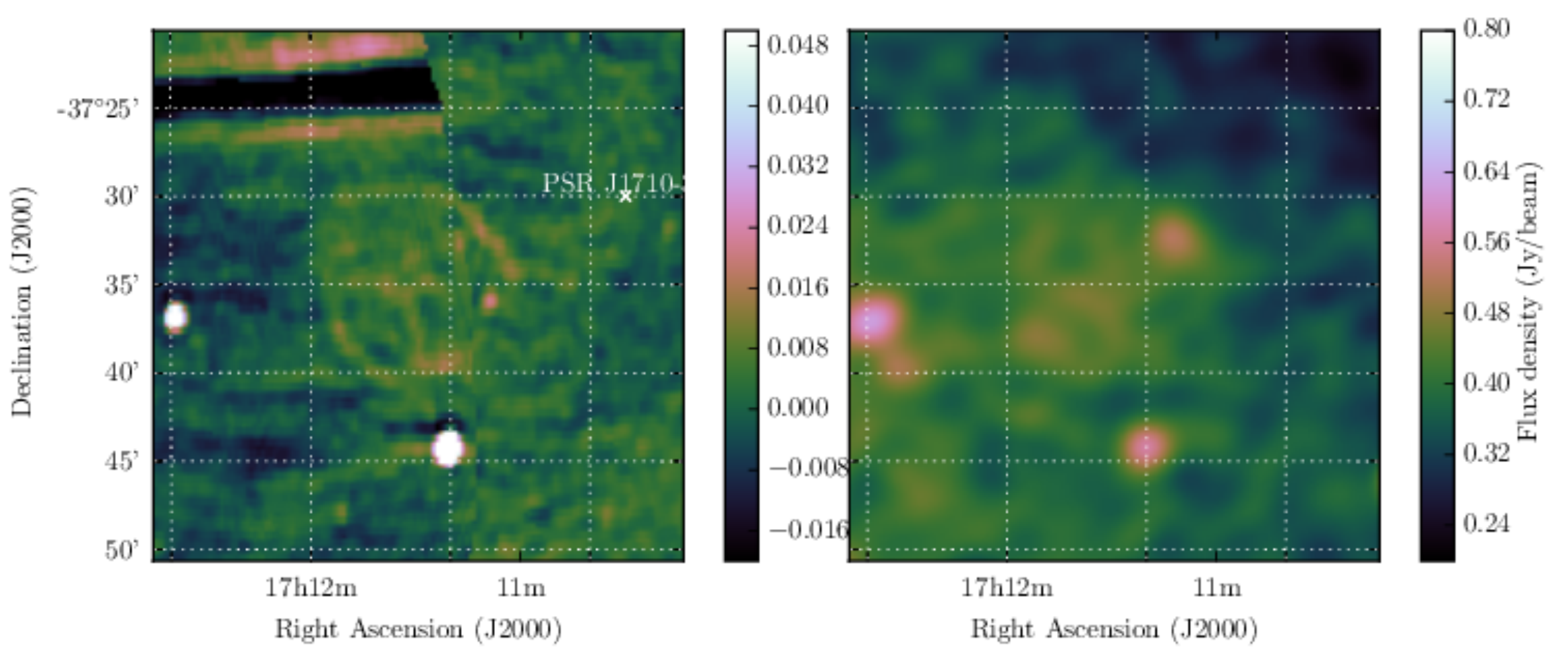}
    \caption{G$348.8+1.1$ as observed by MOST at 843\,MHz (left) and GLEAM at 200\,MHz (right).}
    \label{fig:SNRG348.8+1.1}
\end{figure*}

Concluding our selection from \cite{1996A+AS..118..329W} (see also \Sect~\ref{SNRG345.1-0.2} and \Sect~\ref{SNRG345.1+0.2}), G$348.8+1.1$ is a ``category C'' candidate, which described only as a ``faint, incomplete shell'' with $10'$ diameter, $S_\mathrm{843MHz}=0.1$\,Jy, and a mean surface brightness of $0.1\times10^{-21}$\SBunit. \Fig~\ref{fig:SNRG348.8+1.1} shows the MGPS 843\,MHz observation of G$348.8+1.1$, and the corresponding GLEAM image at 200\,MHz.

Our observations do not clearly show the same shell-like structure, but find regions of increased brightness coincident with the NW SE, S and central filaments seen in the MOST image. From our measurements we derive $\alpha=-0.7\pm0.2$ and $S_\mathrm{200MHz}=1.8\pm0.2$\,Jy. However, these values are inconsistent with those listed by \cite{1996A+AS..118..329W} in Table MSC.C; extrapolating to 843\,MHz, we would expect to see $\approx0.64$\,Jy of total flux density, instead of the listed 0.1\,Jy. However, if we use our own \polyflux~ software to measure the flux density in the MOST data, we find that the total flux density at 843\,MHz is 0.7\,Jy, consistent with our own observations, and thus suggest that \citeauthor{1996A+AS..118..329W} may have made a typographical error in their table. Given the existence in GLEAM, and spectrum of the candidate, we confirm it as a SNR.

PSR~J1710-37 lies outside of the shell of G$348.8+1.1$ by more than its radius, so is unlikely to be associated.

\subsubsection{G$352.2-0.1$}\label{SNRG352.2-0.1}

\begin{figure*}
    \centering
   \includegraphics[width=\textwidth]{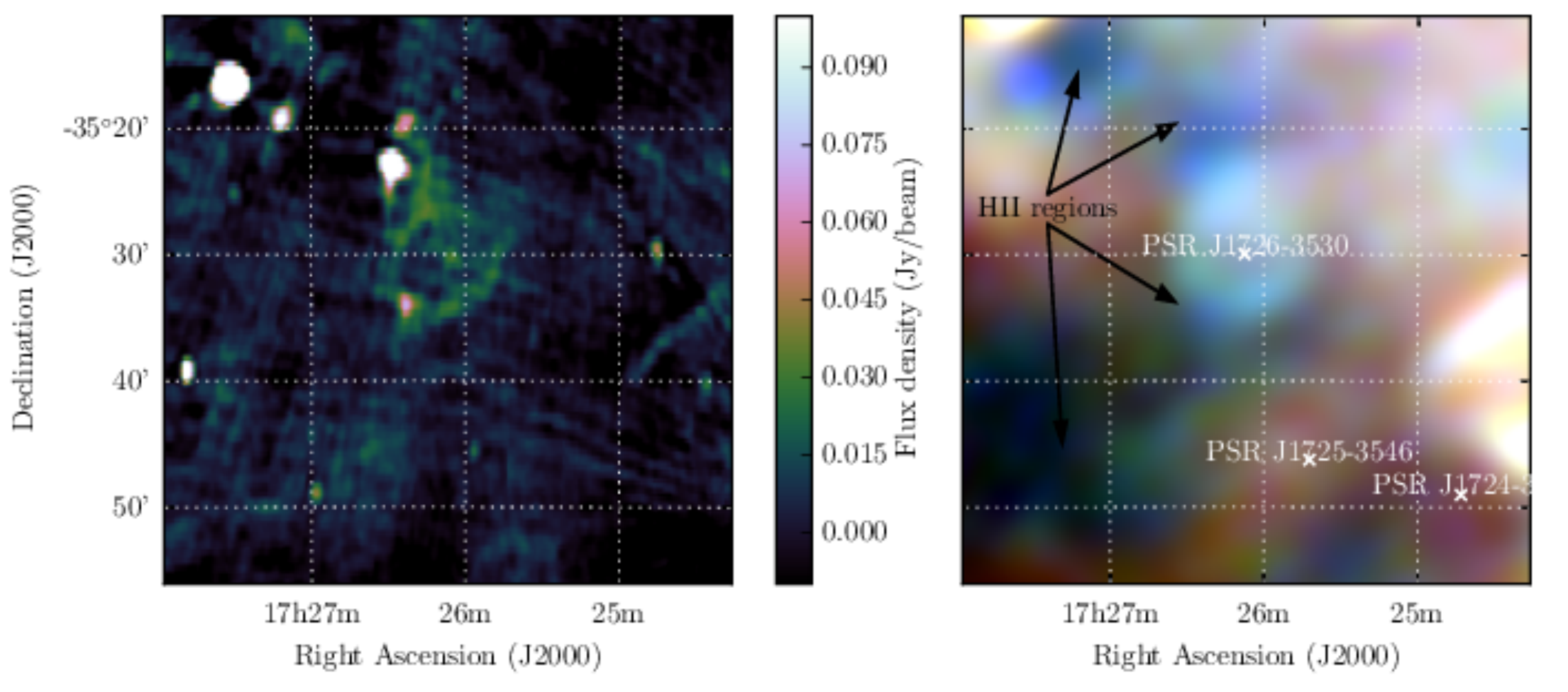}
    \caption{G$352.2-0.1$ as observed by MOST at 843\,MHz (left) and GLEAM at at 72--103\,MHz (R), 103--134\,MHz (G), and 139--170\,MHz (B) (right). The (linear) scales for the color ranges of these frequencies are 4.0--7.0, 2.3--3.6, and 1.0--1.7\,Jy\perbeam, respectively.}
    \label{fig:SNRG352.2-0.1}
\end{figure*}

\cite{2002ASPC..271...31M} used the Australia Telescope Compact Array and the MOST Galactic Plane Survey (MGPS) to search for SNR associated with pulsars detected in the Parkes Multibeam Survey, in this case PSR~J1726-3530. \Fig~\ref{fig:SNRG352.2-0.1} shows the MGPS 843\,MHz observation of G$352.2-0.1$, and the RGB GLEAM cube at 72--103\,MHz (R), 103--134\,MHz (G), and 139--170\,MHz (B). Our image shows strong absorption at 72--103\,MHz, indicating the presence of \textsc{Hii} regions. However, the candidate does appear to have a shell-like shape, indicating it could well be a SNR. 

\textit{WISE} shows increased 12- and 22-$\mu$m emission just northeast of the edge of this shell, corresponding to one of the most highly absorbing regions in our image. This region is sufficiently complex that we cannot derive a spectrum for the candidate, only the foreground \textsc{Hii} regions (see \Fig~\ref{fig:SNR_G352.2-0.1_spectrum}). We estimate $S_\mathrm{200MHz}=2.0\pm0.2$\,Jy without any correction for contaminating \textsc{Hii} regions. Using \polyflux, we measure $S_\mathrm{843MHz}=1.3\pm0.2$\,Jy from MGPS, deriving $\alpha=-0.30\pm0.05$ from these two measurements alone.

Using the pulsar dispersion measure of $718$\,cm$^{-3}$\,pc \citep{2013MNRAS.435.1610P} and the electron density model of \cite{2017ApJ...835...29Y}, a distance of 4.7\,kpc can be assigned to PSR~J1726-3530\footnote{Reduced from the original estimate of 10\,kpc of \cite{2002ASPC..271...31M}, who used the electron density model of \cite{1993ApJ...411..674T}.}.
If PSR~J1726-3530 and G$352.2-0.1$ are associated and there is no line-of-sight difference in distances, the SNR is 8\,pc in diameter. PSR~J1726-3530 has $P\approx1.11$\,s and $\dot{P}\approx1.2\times10^{-12}$\,s\,s$^{-1}$ \citep{2001MNRAS.328...17M}, giving it a characteristic age of 14,500 years, in which it is likely to be in the Sedov-Taylor phase, described by \Eqn~\ref{eq:snr_r2}.

After 14,500 years, a typical SNR in this phase would therefore be $\approx29$\,pc in diameter, more than three times larger than the observed diameter. There are multiple factors which could cause such an inconsistency: the distance estimate may yet be wrong (having already been revised by a factor of two in a decade); the SNR may have an unusually ($\approx530\times$) low energy (unlikely to be the sole factor given SNR energies range from $10^{50}$--$10^{52}$\,erg; see \citealt{2017ApJ...837...36L}); or the ISM may be overdense by a similar factor. Additionally, pulsar characteristic ages are usually an overestimate since they assume a spindown from $P=0$ \citep[see e.g. ][]{2001ApJ...560..371K}, but not usually by the necessary factor of $\approx20$, particularly for a slow pulsar such as PSR~J1726-3530. We cannot easily reconcile these values and suggest some combination of factors would be necessary to explain the discrepancy. Given the known pulsar spatial density in the region, there is also a $\approx40$\,\% chance that the pulsar lies within the shell of the SNR purely through chance geometrical alignment.




\subsubsection{G$353.3-1.1$}\label{SNRG353.3-1.1}

\begin{figure*}
    \centering
   \includegraphics[width=\textwidth]{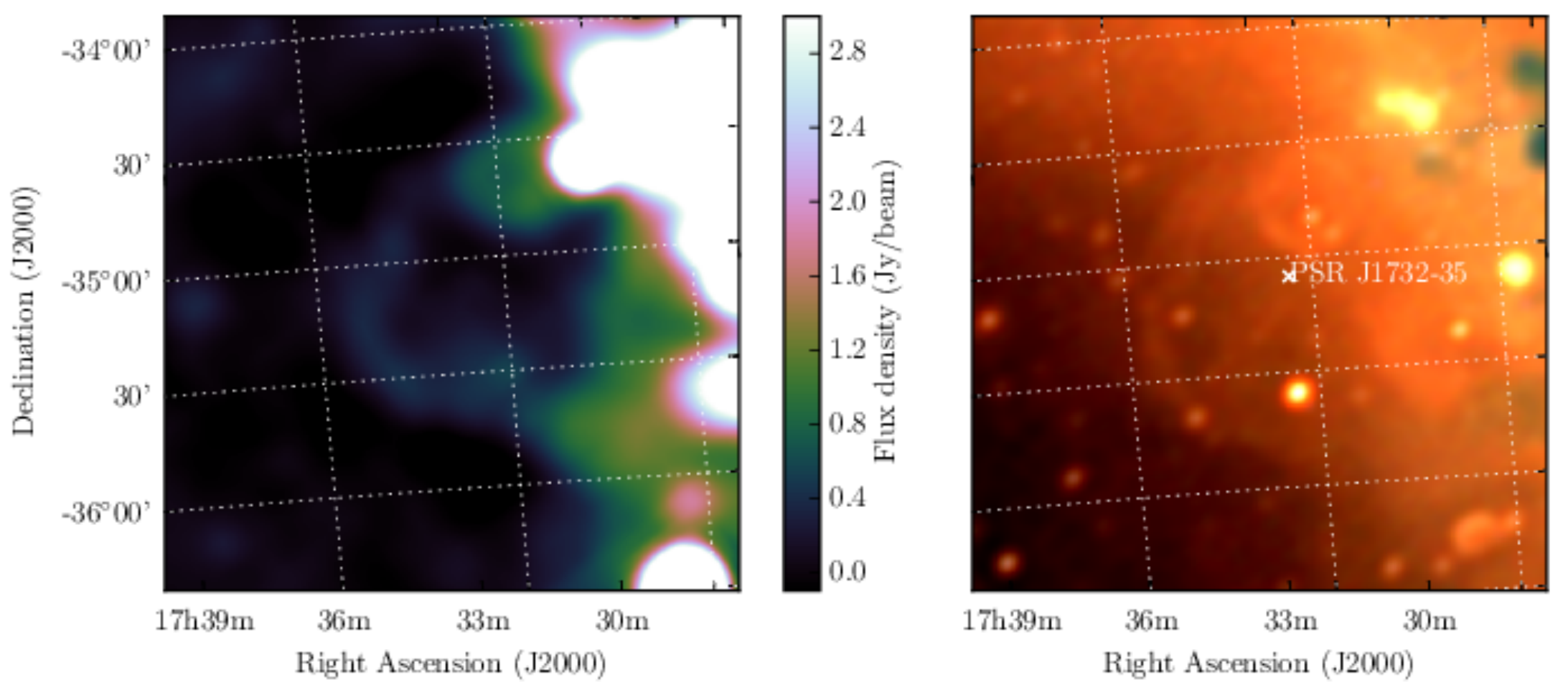}
    \caption{G$353.3-1.1$ as observed by Parkes at 2.4\,GHz (left) and GLEAM at at 72--103\,MHz (R), 103--134\,MHz (G), and 139--170\,MHz (B) (right). All frequencies are set to an identical linear color scale of 0.5 to 5.0\,Jy\perbeam.}
    \label{fig:SNRG353.3-1.1}
\end{figure*}

Using the Parkes radio telescope at 2.4\,GHz, \cite{1995MNRAS.277...36D} conducted a survey of the Galactic plane in continuum and polarisation, publishing 25~supernova remnant candidates \citep{1997MNRAS.287..722D}. Many of these have since been confirmed as SNRs, but the large angular diameter ($\approx1^\circ$) and partial, poorly-resolved morphology of G$353.3-1.1$ has made it difficult to confirm. \Fig~\ref{fig:SNRG353.3-1.1} shows the Parkes 2.4\,GHz continuum image of G$353.3-1.1$, and the GLEAM RGB cube of the same region.

The GLEAM observations confirm this object as a SNR, clearly resolving the shell structure against the diffuse Galactic background and despite the presence of other SNRs and contaminating extragalactic radio sources. Given that half of the shell is obscured by other SNRs, we use \polyflux~ to measure half of the shell (\Fig~\ref{fig:SNR_G353.3-1.1_poly}). We also use \polyflux~ to measure G$353.3-1.1$ in the Parkes 2.4\,GHz continuum image\footnote{http://www.atnf.csiro.au/research/surveys/2.4Gh\_Southern/block1.html}, across the same region fitted in GLEAM. To deal with the contaminating compact radio sources, we performed compact source-finding and fitting across the full GLEAM band as per Hurley-Walker et al. (submitted), finding six unresolved sources with $S_\mathrm{200MHz}>5\times$ the local RMS noise. In total, these had $S_\mathrm{200MHz}=6.3$\,Jy and a median $\alpha=-0.84$.

We extrapolate the radio source spectra across 72\,MHz to 2.4\,GHz and subtract them from all measurements. Two of the sources have flat or rising spectra with large error bars, so we do not extrapolate these to 2.4\,GHz. There is some evidence for contamination by \textsc{Hii} regions in the spectrum, as it begins to flatten at $\nu<150$\,MHz. We therefore fit to the source-subtracted data for $\nu>150$\,MHz, resulting in $\alpha = -0.85\pm0.04$ and 
$S_\mathrm{200MHz} = 43\pm4$\,Jy (\Fig~\ref{fig:SNRG353.3-1.1_spectrum}). Since we are measuring approximately half of the shell, we postulate that the total flux density is $95\pm8$\,Jy.

One pulsar lies 10\farcm5 from the centre of the shell: \cite{2015MNRAS.450.2922N} detected PSR~J1732-35 using the Parkes radio telescope at 1.4\,GHz as part of the High Time Resolution Universe (HTRU) survey. The pulsar has a spin period of $P\approx127$\,ms, a dispersion measure (DM) of $340\pm2$\,cm$^{-3}$pc, and therefore a distance of 4.1\,kpc \citep{2017ApJ...835...29Y}\footnote{Reduced from the original estimate of 5.1\,kpc of \cite{2015MNRAS.450.2922N}, who used the NE2001 electron density model \citep{2002astro.ph..7156C}}. Timing solutions are not yet precise enough to determine whether the pulsar is isolated or in a binary system, so it is not possible to tell if this spin rate is due to partial recycling or simply youth. The period derivative of the pulsar has not yet been measured, so an age cannot be estimated. However if $\dot{P}\gtrsim 10^{-13}$s\,s$^{-1}$, its $P$ and $\dot{P}$ would be typical of pulsars with SNR associations.

If the pulsar distance estimate is correct and the pulsar and SNR are associated, G$353.3-1.1$ is 71\,pc in diameter, and the pulsar has moved 12\,pc since birth.  Reversing \Eqn~\ref{eq:snr_r2}, we can use the SNR radius to predict the age, finding $t\approx130,000$\,yr. At this age the SNR is not likely to still be in the Sedov-Taylor phase, and is likely in the ``snow plough'' stage, where its radius will evolve more slowly with respect to time ($R\propto t^{2/7}$). This age could therefore be considered a lower limit. The pulsar line-of-sight velocity would therefore be $<90$\,km\,s$^{-1}$, reasonable considering that 32\,\% of young pulsars have velocities distributed in a Maxwellian with an average velocity of $\sigma\sqrt{8/\pi}=130$\,km\,s$^{-1}$ \citep{2017A&A...608A..57V}. Given the pulsar spatial density in this region and the large size of the shell, we would expect at least one pulsar to lie within the shell purely through geometric coincidence. A measurement of the pulsar's $\dot{P}$ or true velocity would help to confirm its association with the SNR.



\subsubsection{G$354.46+0.07$}\label{SNRG354.5+0.1}

\cite{2013ApJ...774..150R} used the GMRT to observe G$354.46+0.07$ at 330\,MHz and 1.4\,GHz. They made long-baseline and short-baseline images in order to separate extended ($>2'$) \textsc{Hii} region emission from what they identified as a young SNR shell. They attributed to the ``shell'' flux densities of $S_\mathrm{300MHz}=0.9\pm0.1$\,Jy and $S_\mathrm{1400MHz}=0.7\pm0.1$\,Jy, and thereby determined a spectral index of $\alpha=-0.2\pm0.1$, which they note is ``quite flat and unexpected for a shell-type SNR''. We also note that in Figure~3, they plot these flux densities multiplied by a factor of four, but have not concomitantly multiplied the error bars.

\begin{figure*}
    \centering
    \includegraphics[width=\textwidth]{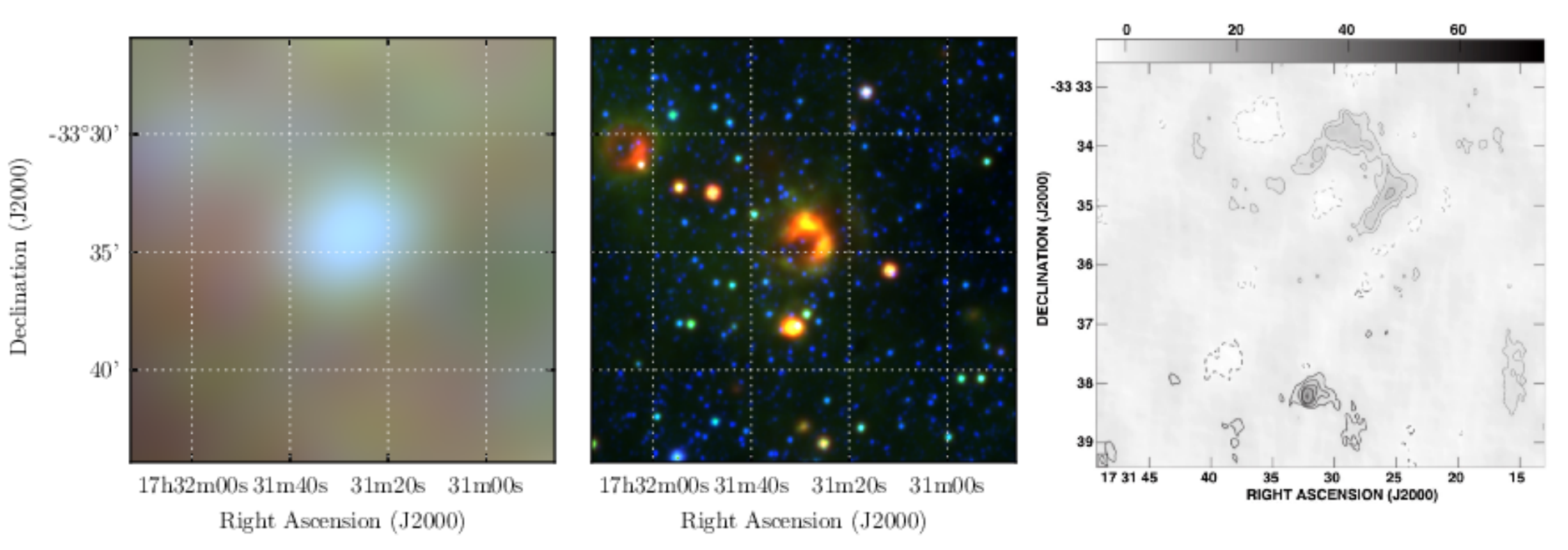}
    \caption{G$354.5+0.1$ as observed by GLEAM (left) at 72--103\,MHz (R), 103--134\,MHz (G), and 139--170\,MHz (B), by \textit{WISE} (middle) at 22\,$\mu$m (R), 12\,$\mu$m (G), and 4.6\,$\mu$m (B), and with the GMRT at 1.4\,GHz with $(u,v)>1000\lambda$ \citep[left panel of Figure~2 from ][]{2013ApJ...774..150R}. The colour scales for the GLEAM RGB cube are 4.3--8.6, 2.0--4.4, and 0.8--2.3\,Jy\perbeam, respectively.}
    \label{fig:SNRG354.5+0.1}
\end{figure*}

Searching the \textit{WISE} data, we find that this ``shell'' has the same morphology as a 12- and 22-$\mu$m emitting region, strongly indicating it is actually an \textsc{Hii} region. Our GLEAM observations would not be able to resolve the shell were it to be present, but they do show that there is no emission on a $4'$-scale, which is what \cite{2013ApJ...774..150R} claim as the extent of the \textsc{Hii} region. \Fig~\ref{fig:SNRG354.5+0.1} shows the \textit{WISE}, GLEAM, and GMRT high-resolution data for this region, while \Fig~\ref{fig:SNR_G354.5+0.1_spectrum} shows the GLEAM spectrum, with a distinct low-frequency turnover indicating absorption of Galactic synchrotron from an \textsc{Hii} region. We argue that G$354.5+0.1$ has been mistakenly identified as a SNR and is in fact an \textsc{Hii} region.

\subsubsection{G$356.6+0.1$}\label{SNRG356.6+0.1}

\begin{figure*}
    \centering
    \includegraphics[width=\textwidth]{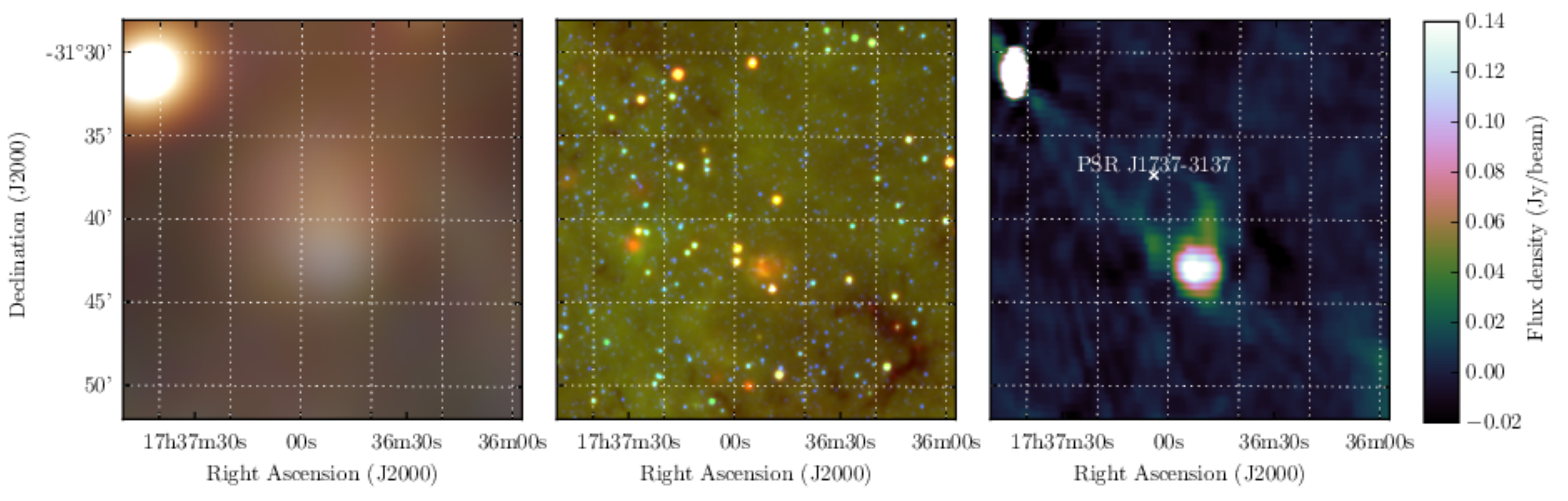}
    \caption{G$356.6+0.1$ as observed by GLEAM (left) at 72--103\,MHz (R), 103--134\,MHz (G), and 139--170\,MHz (B), by \textit{WISE} (middle) at 22\,$\mu$m (R), 12\,$\mu$m (G), and 4.6\,$\mu$m (B), and with MGPS at 843\,MHz (right). The colour scales for the GLEAM RGB cube are all 2--10\,Jy.}
    \label{fig:SNRG356.6+0.1}
\end{figure*}

\cite{1994MNRAS.270..847G} searched the Galactic Centre using the MOST and detected 24~candidate SNR, many of which are not visible in the GLEAM images, in many cases because the Galactic Centre is such a complex region that disentangling individual objects becomes difficult. \citeauthor{1994MNRAS.270..847G} identify G\,$356.6+0.1$ and suggest $S_\mathrm{843MHz}=3.7$\,Jy, although with ``much uncertainty''. \textit{WISE} shows a small circular region of increased 22-$\mu$m emission to the south-west of the object, which is likely an \textsc{Hii} region. This is also noted by \citeauthor{1994MNRAS.270..847G} as IRAS\,17335$-3136$. 
The GLEAM observations do not resolve the source into the two components but the GLEAM RGB image (\Fig~\ref{fig:SNRG356.6+0.1}) shows that it has a flatter spectrum toward the southwest, consistent with this part being a \textsc{Hii} region. Running \polyflux~ on MGPS data, we can separate the object into the likely \textsc{Hii} region component and the potential shell component, finding $S_\mathrm{843MHz,\textsc{Hii}}=0.72\pm0.7$\,Jy and $S_\mathrm{843MHz,shell}=0.37\pm0.4$\,Jy. Assuming the \textsc{Hii} region has a flat spectrum of $\alpha=0.0$, we can subtract the first value from all GLEAM flux density measurements. Restricting the fit to $\nu>150$\,MHz to avoid any absorbing effect from the \textsc{Hii} region, we find $S_\mathrm{200MHz}=1.9\pm0.3$\,Jy and $\alpha=-1.1\pm0.3$, distinctly non-thermal. However, this analysis assumes that the MGPS data and our \polyflux~ measurement is superior to that of \cite{1994MNRAS.270..847G}, and they overestimated the 843-MHz flux density by a factor of $\approx2$.

PSR~J1737-3137 lies just on the edge of the shell (\Fig~\ref{fig:SNRG356.6+0.1}) and has $P\approx$450\,ms and $\dot{P}\approx1.4\times10^{-13}$\,s\,s$^{-1}$\citep{2002MNRAS.335..275M}, giving it a characteristic age of $\approx51,000$\,years. If it is associated with the SNR, then its distance of 4.2\,kpc \citep{2017ApJ...835...29Y} would constrain the SNR to be 9\,pc in diameter. \Eqn~\ref{eq:snr_r2} predicts a radius of 24\,pc for a SNR of this age, i.e. about five times larger than expected if the pulsar association is correct. The factors in \Sect~\ref{SNRG352.2-0.1} would have to be even more extreme in order to explain this discrepancy, so we suspect the SNR and pulsar are not related, despite the low probability ($\approx3$\,\%) that the pulsar and SNR are in chance geometrical alignment.



\subsubsection{G$359.2-1.1$}\label{SNRG359.2-1.1}

\begin{figure*}
    \centering
    \includegraphics[width=\textwidth]{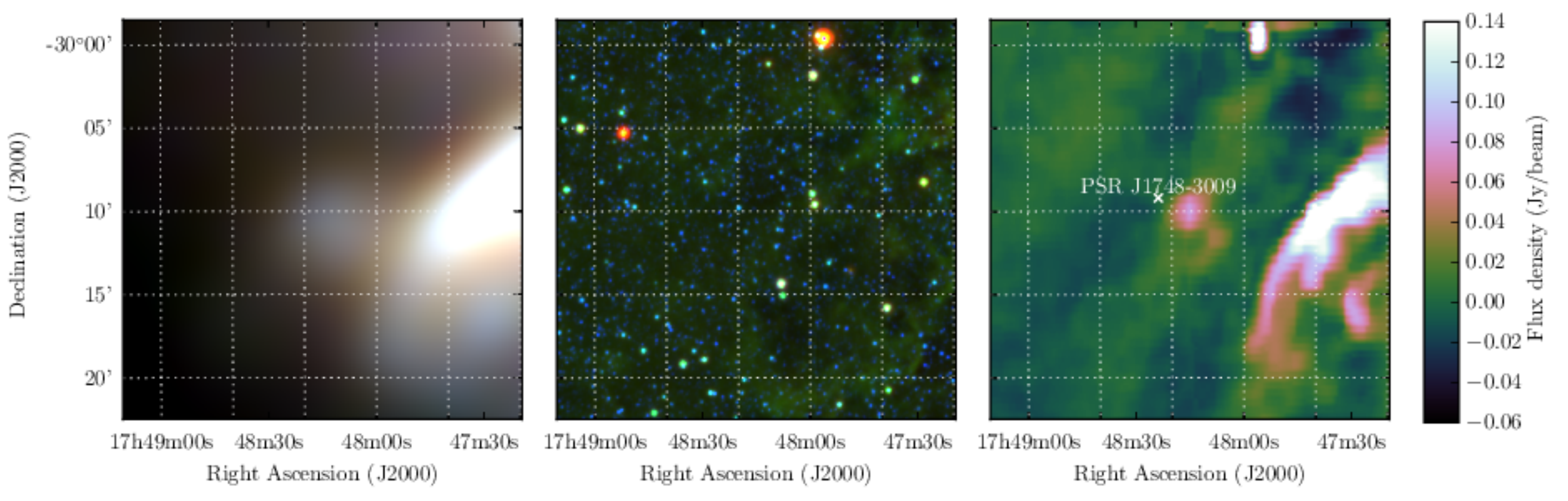}
    \caption{G$359.2-1.1$ as observed by GLEAM (left) at 72--103\,MHz (R), 103--134\,MHz (G), and 139--170\,MHz (B), by \textit{WISE} (middle) at 22\,$\mu$m (R), 12\,$\mu$m (G), and 4.6\,$\mu$m (B), and with MGPS at 843\,MHz. The colour scales for the GLEAM RGB cube are 7.1--19.7, 3.1--10.1, and 1.1--4.7\,Jy\perbeam for R, G, and B, respectively.}
    \label{fig:SNRG359.2-1.1}
\end{figure*}

Another candidate detected by \cite{1994MNRAS.270..847G}, its higher Galactic latitude makes SNR\,G$359.2-1.1$ easier to distinguish, although at $4'\times5'$ it is close to unresolved by GLEAM. \citeauthor{1994MNRAS.270..847G} describe it as asymmetric, and note that its lack of sharp edges potentially makes it less likely to be a SNR. ``With considerable uncertainty", they estimate $S_\mathrm{843MHz}=1.3$\,Jy. Using \polyflux~ on the MGPS data, we find $S_\mathrm{843MHz}=0.56$\,Jy. \citeauthor{1994MNRAS.270..847G} note the lack of IR emission from IRAS; in \textit{WISE} there is no obvious diffuse 12- or 22-$\mu$m emission. PSR~J1748-3009 lies nearby but with a period of $P\approx9$\,ms, it is likely an unrelated recycled pulsar ($\dot{P}$ has not yet been measured). \Fig~\ref{fig:SNRG359.2-1.1} displays the GLEAM, \textit{WISE}, and MGPS images for this region.

The GLEAM observations yield $S_\mathrm{200MHz}=2.6\pm0.2$\,Jy and $\alpha=-0.83\pm0.2$, which would predict $S_\mathrm{843MHz}=0.8$\,Jy, slightly more consistent with our own measurement on MGPS than the measurement of \cite{1994MNRAS.270..847G} (similar to G$356.6+0.1$, \Sect~\ref{SNRG356.6+0.1}). Given the low S/N of this source in both our data and the MGPS, we fit a single power law SED to the GLEAM flux densities and our measurement from MGPS, and find $S_\mathrm{200MHz} = 2.6\pm0.2$
and $\alpha = -1.07\pm0.08$, clearly non-thermal. Higher-resolution observations will be necessary to confirm the morphology, so we only tentatively confirm this candidate (\Tab~\ref{tab:candidates}).

\subsubsection{G$1.2-0.0$}\label{SNRG1.2-0.0}

\cite{2009PASJ...61S.209S} investigated the Sagittarius~D \textsc{Hii} region with the \textit{Suzaku} X-ray telescope. They identified "Diffuse Source 1" (DS1) via its diffuse X-ray emission, and proposed it as a previously-unknown SNR. They identified this also as a region of emission at 18\,cm from observations made by \cite{1998ApJ...493..274M}. From 3.5\,cm and 6.0\,cm observations taken by \cite{2008ApJS..177..255L} with the GBT they calculated the spectral index of this emission to be $\alpha=-0.5$.

\begin{figure*}
    \centering
    \includegraphics[width=\textwidth]{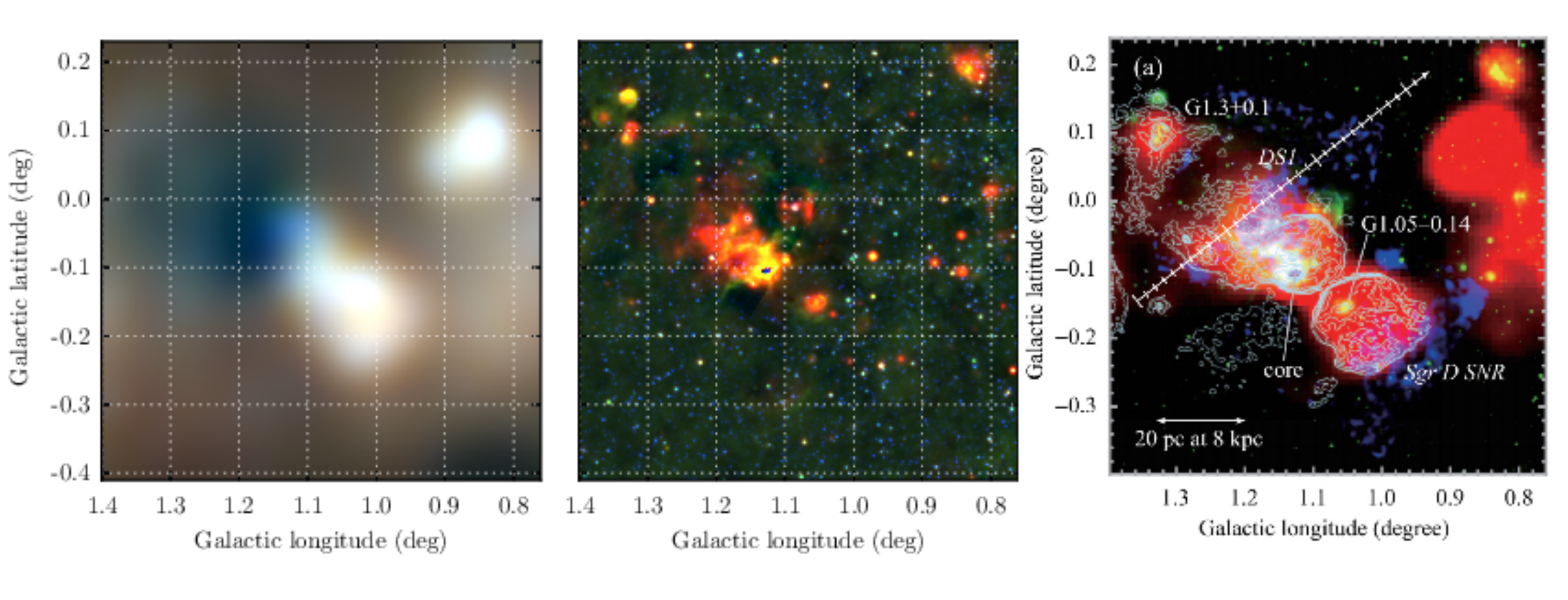}
    \caption{G$1.2-0.0$ as observed by GLEAM (left) at 72--103\,MHz (R), 103--134\,MHz (G), and 139--170\,MHz (B), by \textit{WISE} (middle) at 22\,$\mu$m (R), 12\,$\mu$m (G), and 4.6\,$\mu$m (B) and by \protect\cite{2009PASJ...61S.209S} with Suzaku X-ray (0.7--5.5\,keV) in blue, Spitzer MIR (24\,$\mu$m) in green, and GBT radio (6.0\,cm) in red. The slice for the calculated radio spectral index ($\alpha=-0.5$) is shown with a ticked vector. Objects are labeled in Italic for SNRs and in Roman for \textsc{Hii} regions. The colour scales for the GLEAM RGB cube are 11--38, 5--21, and 2--11\,Jy\,beam$^{-1}$ for R, G, and B, respectively.}
    \label{fig:SNRG1.2-0.0}
\end{figure*}

However, GLEAM observations reveal this region to have large amounts of low-frequency absorption, and a rising spectrum, indicative of \textsc{Hii} regions. \textit{WISE} images show 12- and 22-$\mu$m emission in the same location, also indicating a thermal origin to the radio emission. We argue that G$1.2-0.0$ has been mistakenly identified as a SNR and is in fact several superposed and complex \textsc{Hii} regions. \Fig~\ref{fig:SNRG1.2-0.0} shows the GLEAM and \textit{WISE} data for this region, and a reproduction of Figure~6(a) from \cite{2009PASJ...61S.209S}.

\subsubsection{G$3.1-0.6$}\label{SNRG3.1-0.6}

G\,$3.1-0.6$ is the second-brightest object in the candidate sample of \cite{1994MNRAS.270..847G}, with $S_\mathrm{843\,MHz}=6$\,Jy, and is described as being $28'\times52'$, with ``extensive filaments''. \cite{2002MNRAS.329..775R} used the GMRT at 330\,MHz to follow up G$3.1-0.6$ and found similar filamentary structures. They also found increased emission toward the south-west of the region, and a potential shell-like structure in the southernmost part. We argue however, that this is not a separate structure, and merely a chance coincidence of filaments.

\begin{figure*}
    \centering
    \includegraphics[width=\textwidth]{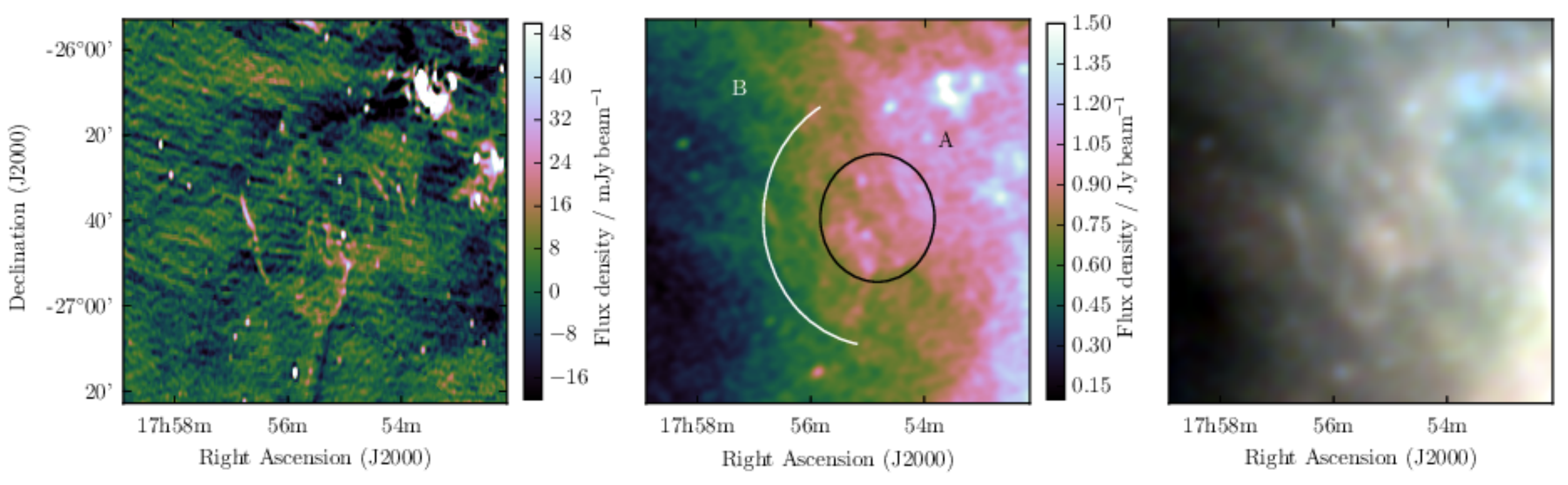}
    \caption{G$3.1-0.6$ as observed by MGPS at 843\,MHz (left), GLEAM at 200\,MHz (middle), and at 72--103\,MHz (R), 103--134\,MHz (G), and 139--170\,MHz (B) (right). The colourscales for the MGPS and GLEAM 200\,MHz data are shown on the figure, while the colour scales for the GLEAM RGB cube are 4.8--13.2, 1.8--6.3, and 0.5--2.8\,Jy\perbeam~for R, G, and B, respectively. On the middle panel, we have overlaid a black ellipse (``A'') and a white arc (``B'') indicating the two potential expanding shells of different radii propagating from the same stellar explosion (see \Sect~\ref{SNRG3.1-0.6}.}
    \label{fig:SNRG3.1-0.6}
\end{figure*}

\Fig~\ref{fig:SNRG3.1-0.6} shows the MGPS 843\,MHz observation of G$3.1-0.6$, the 200-MHz GLEAM image, and the RGB GLEAM cube at 72--103\,MHz (R), 103--134\,MHz (G), and 139--170\,MHz (B). The MGPS data resolves out the large-scale structure of the region, and indeed only filamentary edges are visible. However, the GLEAM data show the large-scale structure of this candidate, and it is clear the morphology is very complex, with two distinct regions:  a spherical object ``A'' centred at
$\mathrm{RA}=17^\mathrm{h}54^\mathrm{m}50^\mathrm{s}$, $\mathrm{Dec}=-26^\mathrm{d}39^\mathrm{m}30^\mathrm{s}$,
with radius $15'$, and a more elliptical object ``B'', which is consistent with a filled arc of equivalent radius $30'$ centred at the same position; these are marked in \Fig~\ref{fig:SNRG3.1-0.6}. The ratio of flux densities between A and B is almost unity and its uncertainty is dominated by the subjective decision of how to separate the components. The chance of two short-lived SNR of such similar brightnesses and spectral indices (identical within errors) just happening to overlap along the line-of-sight is very small, so we believe these two structures are part of the same object.

 
This SNR is reminiscent of the Cygnus Loop \citep[e.g. \Fig~1 of][]{1997AJ....114.2081L}, or, even more so, VRO42.05.01 \citep[see right panel of \Fig~2 of][]{2005A&A...440..929L}. \cite{1982ApJ...261L..41L} examine this latter remnant in detail, concluding that its asymmetric morphology is most likely due to a single SNR expanding into two slabs of very different densities: the star explodes into a dense slab of gas, forming a circular shell; on the west side, the shock front breaks out of the dense slab and expands rapidly into a lower-density medium, forming a larger ``wing''-like object, very similar to object ``B''. Further interactions with denser regions of the ISM increase the brightness of the edge of the wing \citep[see ][for further details]{1985A&A...151...52P, 1987ApJ...315..580P}. We postulate that G\,$3.1-0.6$ is undergoing a similar type of expansion. Like \cite{1982ApJ...261L..41L}, we may estimate the density discontinuity between the two slabs by assuming the remnant is in the adiabatic phase and the SNR energy has been divided equally between the two components. The ratio of radii of the two components is 2:1, and the Sedov-Taylor relation (\eqn~\ref{eq:snr_r1}) implies $R_\mathrm{A}^5\rho_\mathrm{A} = R_\mathrm{B}^5\rho_\mathrm{B}$. Therefore, the $\rho_\mathrm{A}/\rho_\mathrm{B}\approx32$.

\citep{2019A&A...622A...6A} postulate an alternative explanation for the shape of VRO42.05.01, where the progenitor star was moving at supersonic velocities from a high to low-density environment, creating a bow-shock in the latter. After forming a supernova, the ``wing'' expands outward into the bow-shock cavity, while the circular shell expands in the cavity left behind in the higher-density medium. In the case of G\,$3.1-0.6$, the edge of the ``wing'' appears more circular than triangular, and the MGPS image (left panel of \Fig~\ref{fig:SNRG3.1-0.6}) shows filaments reconnecting the wing toward the centre of the explosion, which may be less compatible with a bow-shock interpretation. In any case, we cannot explain the morphology and non-thermal spectrum of the candidate by any mechanism other than a SNR and therefore confirm G\,$3.1-0.6$ as a SNR.

\subsubsection{G$5.3+0.1$}\label{SNRG5.3+0.1}

\cite{2001ESASP.459..109T} searched NVSS data for polarised features which were part of shell-like objects in the continuum images, and followed them up with the RATAN-600 telescope at 0.96, 2.3, and 3.9\,GHz. From these data they selected 15~candidate SNR, of which SNR\,G$5.3+0.1$ is one. We attempted to extract a spectrum for this SNR from the listed website \footnote{\href{http://www.sao.ru/cats/snr\_spectra.html}{sao.ru/cats/snr\_spectra.html}} but found it was not available. \cite{2001ESASP.459..109T} indicate that has size $2\farcm5\times2\farcm0$, and therefore nearly unresolved in GLEAM.

\begin{figure*}
    \centering
    \includegraphics[width=\textwidth]{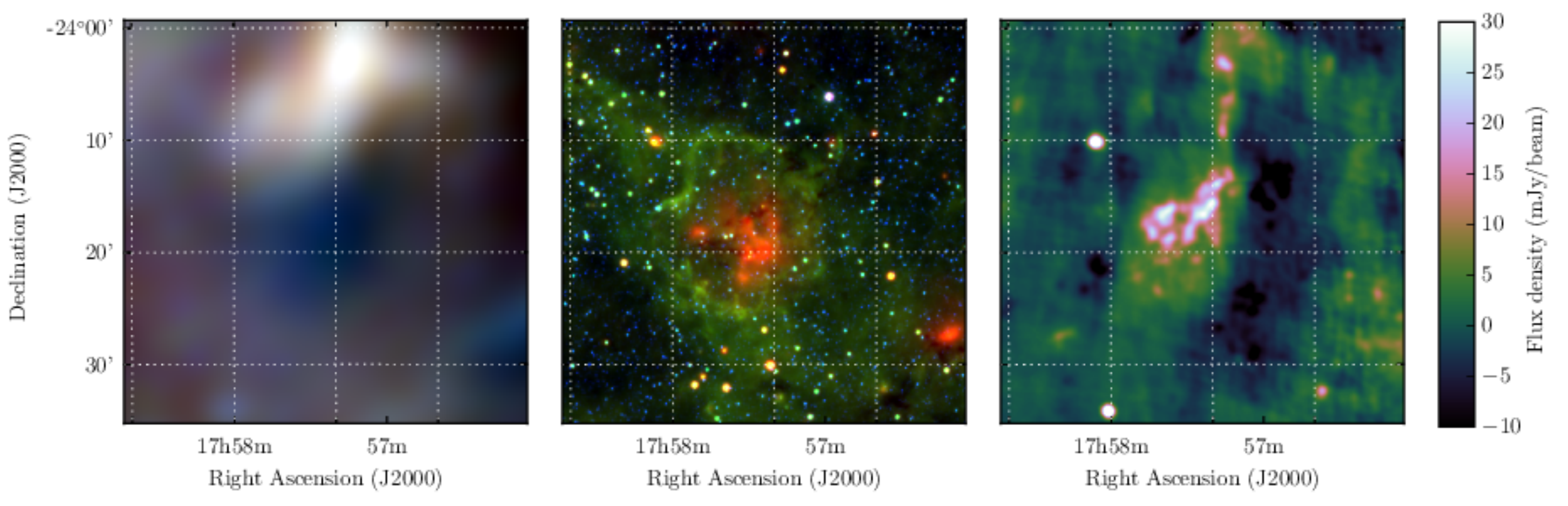}
    \caption{G$5.3+0.1$ as observed by GLEAM (left) at 72--103\,MHz (R), 103--134\,MHz (G), and 139--170\,MHz (B), by \textit{WISE} (middle) at 22\,$\mu$m (R), 12\,$\mu$m (G), and 4.6\,$\mu$m (B) and NVSS at 1.4\,GHz (right). The colour scales for the GLEAM RGB cube are 6.6--15.1, 3.4--7.8, and 1.3--3.5\,Jy\perbeam~for R, G, and B, respectively.}
    \label{fig:SNRG5.3+0.1}
\end{figure*}

\textit{WISE} shows a classic \textsc{Hii} region in this location, with strong central 22-$\mu$m emission, and an encircling ring of 12-$\mu$m emission. The GLEAM data show very strong absorption at low frequencies at this location. However, these features are $\approx15'$ in size, and there is no sign of compact emission on a $2'$ scale. The NVSS image does have some compact features, but the quality is poor due to inadequate sampling and deconvolution of features on larger scales.  We conclude that G$5.3+0.1$ is not a SNR, but part of a \textsc{Hii} region, only partly resolved by the VLA and RATAN-600. \Fig~\ref{fig:SNRG5.3+0.1} shows the GLEAM, \textit{WISE}, and NVSS images for this region.

\subsubsection{G\,$7.5-1.7$}\label{SNRG7.5-1.7}

\begin{figure*}
    \centering
    \includegraphics[width=\textwidth]{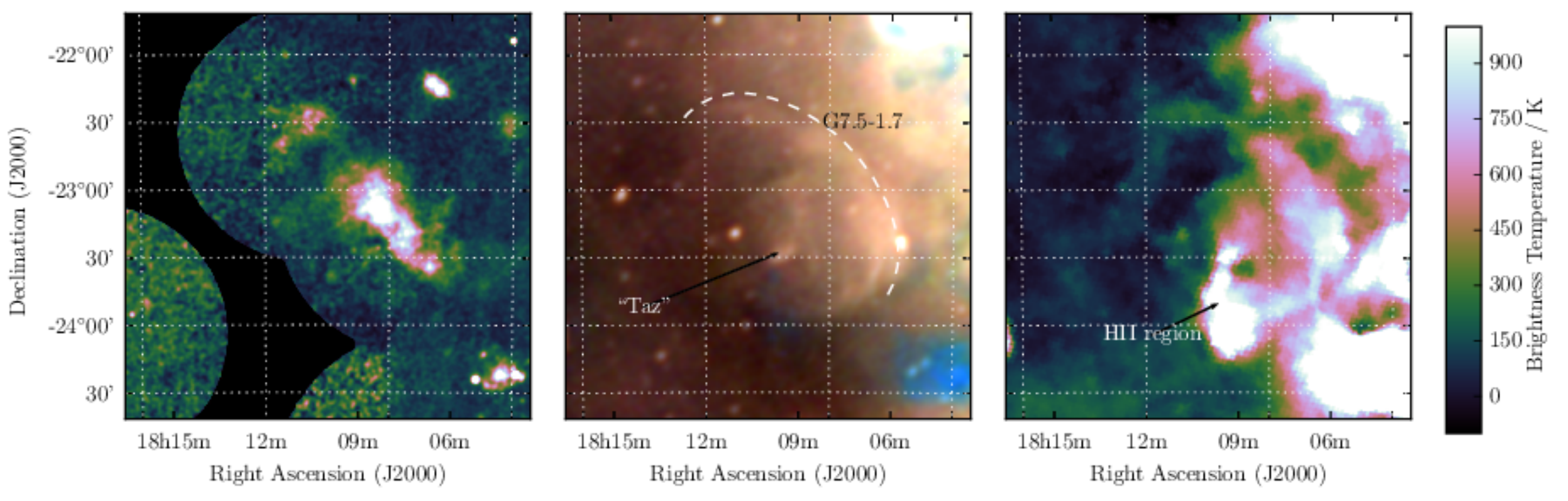}
    \caption{G$7.5-1.7$ as observed by ROST PSPC at 0.1--2.4 keV (left), GLEAM at 72--103\,MHz (R), 103--134\,MHz (G), and 139--170\,MHz (B) (middle), and by Effelsberg at 2695\,MHz (right). The colour scales for the GLEAM RGB cube are 0.5--6.0, 0.1--3.5, and $-0.1$--2.0\,Jy\perbeam~for R, G, and B, respectively.}
    \label{fig:SNRG7.5-1.7}
\end{figure*}

\cite{2008ApJ...681..320R} observed G$7.5-1.7$ as an irregular shell surrounding the PWNe ``Taz'', which itself is centred on the variable $\gamma$-ray source 3EG\,J1809-2328. \citeauthor{2008ApJ...681..320R} followed up the X-ray excess they observed in archival ROSAT and ASCA data with a VLA observation at 324.84\,MHz, and also used the Effelsberg Bonn 11-cm (2695\,MHz) Survey \citep{1984A&AS...58..197R} to examine the emission in this region. They found evidence for a shell of radio emission with a steep ($\alpha\approx-0.8$) spectral index. \Fig~\ref{fig:SNRG7.5-1.7} shows the ROSAT Position-Sensitive Proportionate Counter (PSPC) 0.1--2.4 keV image, with 2695\,MHz contours, and the GLEAM RGB cube of the same region.

GLEAM reveals the larger angular scales of this object quite clearly, showing a large irregular ellipse with a NW edge coincident with the steep-spectrum shell identified in \Fig~3 of \cite{2008ApJ...681..320R} (dashed arc on \Fig~\ref{fig:SNRG7.5-1.7}). There is a faint inner circular ring in the GLEAM image, centred about $10'$ to the East of ``Taz''. This is reminiscent of the shape of extended PWNe such as the Crab Nebula \citep[\Fig~10 of ][]{2017ApJ...840...82D}, although more circular. We confirm G\,$7.5-1.7$ as a Crab-like SNR with a shell.

\subsubsection{G$12.75-0.15$}\label{SNRG12.75-0.15}

\cite{1985SvA....29..128G} used  publicly-available radio surveys at 408\,MHz \citep{1970AuJPA..14...77S}, 1.42\,GHz \citep{1983AISAO..17...67A}, 4.875\,GHz \citep{1979A&AS...35...23A}, 5\,GHz \citep{1970AuJPA..14....1G}, and 10.7\,GHz \citep{1968ApJ...154..833M} to determine spectral indices of 14~extended Galactic sources, and suggested that the eight with non-thermal spectra may be supernova remnants. Of these eight, G$12.75-0.15$ was listed as having $\alpha=-0.61$, and $S_\mathrm{408MHz}=53\pm8$\,Jy, which would lead to $S_\mathrm{200MHz}=82\pm12$\,Jy. With an angular extent of $15'$, this should be overwhelmingly obvious in the GLEAM images.

\begin{figure*}
    \centering
    \includegraphics[width=\textwidth]{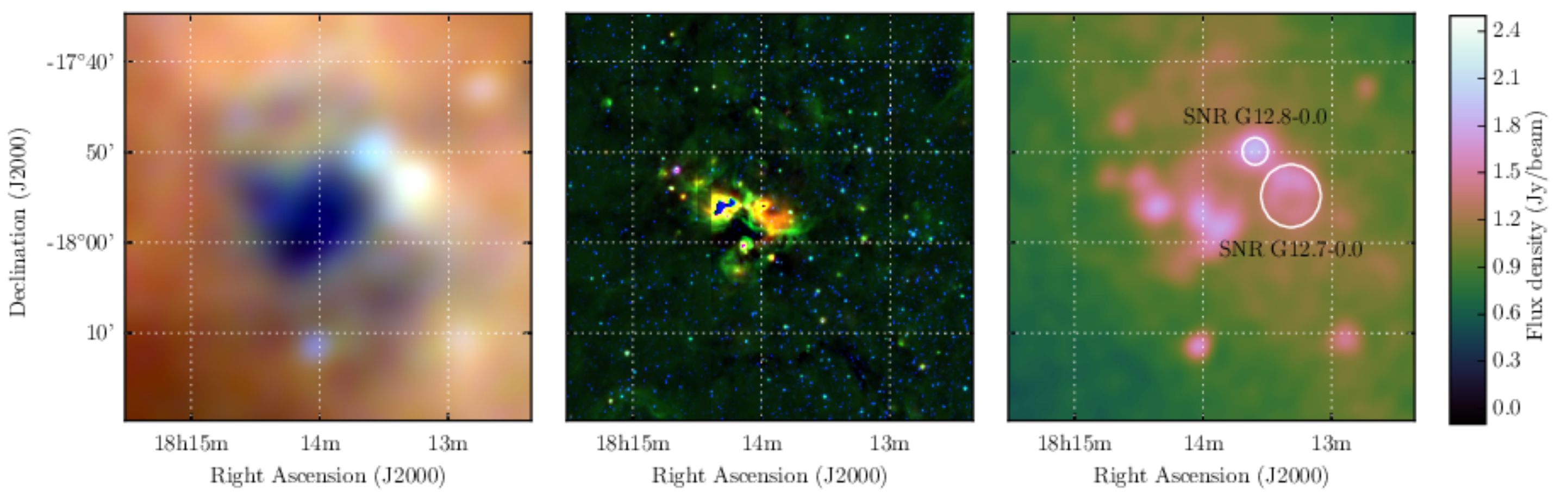}
    \caption{G$12.75-0.15$ as observed by GLEAM (left) at 72--103\,MHz (R), 103--134\,MHz (G), and 139--170\,MHz (B), by \textit{WISE} (middle) at 22\,$\mu$m (R), 12\,$\mu$m (G), and 4.6\,$\mu$m (B) and GLEAM at 200\,MHz (right). The colour scales for the GLEAM RGB cube and wideband 200-MHz image are $1.9$--8.0, $1.9$--4.8, $1.2$--2.7, and $-0.1$--2.5Jy\perbeam, respectively.}
    \label{fig:SNRG12.75-0.15}
\end{figure*}

\Fig~\ref{fig:SNRG12.75-0.15} shows an image of this region; the only feature on a $15'$ scale is an irregular \textsc{Hii} region of dimensions $15'\times12'$, clearly shown in absorption in the low-frequencies of GLEAM (left panel), and in 12- and 22-$\mu$m emission in \textit{WISE} (middle panel). Two known SNRs of diameters $3'$ and $6'$ are also labelled in the 200-MHz image (right panel) and appear unrelated to the \textsc{Hii} region. \cite{1985SvA....29..128G} fitted Gaussian templates to try to separate the Galactic background level from the observed features, and perhaps were unable to correctly separate thermal from non-thermal emission in this case. We suggest that G$12.75-0.15$ is not a SNR, but a \textsc{Hii} region.

\subsubsection{G$13.1-0.5$}\label{SNRG13.1-0.5}

The Clark Lake (CL) survey of the Galactic plane at 30.9\,MHz \citep{1988ApJS...68..715K} was analysed by \cite{1990ApJ...364..187G}, who compared it with higher-frequency radio data in order to distinguish candidate SNR from \textsc{Hii} regions, which are seen in absorption at 30.9\,MHz. The resolution of the CL survey was $11'\times13'$ with a sensitivity of 2\,Jy, and many of these candidates have since been confirmed as true SNRs. G$13.1-0.5$ was measured as $49'\times46'$ in size, with a flux density of $S_\mathrm{31MHz}=79$\,Jy, and having ``clear association with distinct radio flux and polarization features and no obvious \textsc{Hii} region confusion'', no obvious IRAS 60\,$\mu$m counterpart, and in the Palomar sky survey, having ``distinct nebulosity with possible SNR morphology''.

\begin{figure*}
    \centering
    \includegraphics[width=\textwidth]{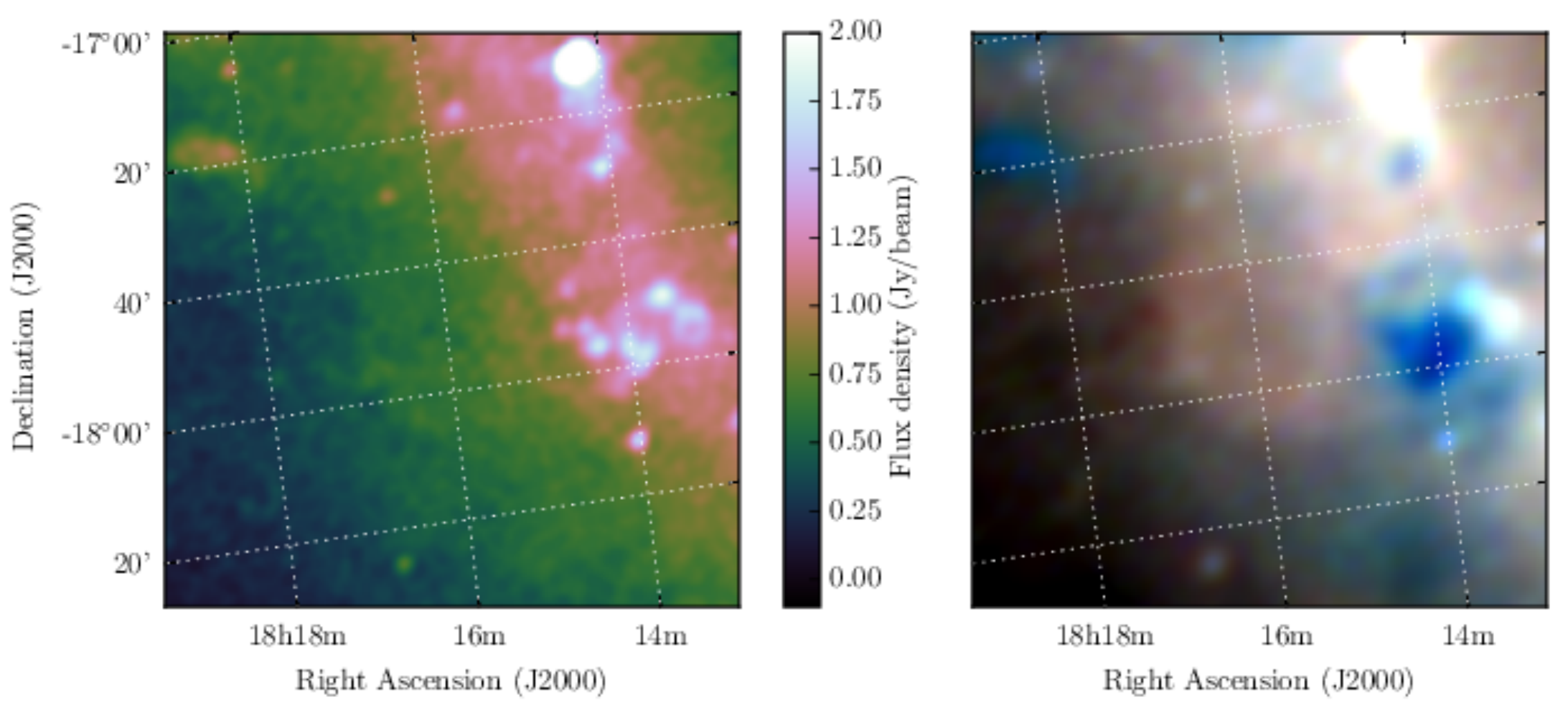}
    \caption{G$13.1-0.5$ as observed by GLEAM at 170--231\,MHz (left) and 72--103\,MHz (R), 103--134\,MHz (G), and 139--170\,MHz (B) (right). The colour scales for the GLEAM wideband image and elements of the RGB cube are $-0.1$--2.0, $3.1$--8.3, $1.3$--4.6, and $0.5$--2.3Jy\perbeam, respectively}
    \label{fig:SNRG13.1-0.5}
\end{figure*}

G$13.1-0.5$ has low surface brightness in the GLEAM observations, but being so large, high enough integrated flux density in each narrow-band image to make it possible to fit a reliable spectrum. We find $S_\mathrm{200MHz}=28.6\pm2.3$\,Jy and $\alpha=-0.57\pm0.03$, entirely consistent with the CL measurement. \Fig~\ref{fig:SNRG13.1-0.5} shows the GLEAM images for this SNR, which are the first to show the full extent of the SNR. Based on the morphology in our images, we suggest that the true extent of the SNR is $38'\times28'$, with a position angle of $15^\circ$ CCW from North.

\subsubsection{G$15.51-0.15$}\label{SNRG15.51-0.15}

\cite{2006ApJ...639L..25B} discovered 35~SNRs between $4.5^\circ < l < 22.0^\circ$ and $|b|<1.25^\circ$, using the VLA at 330\,MHz. These remnants were ranked by the likelihood of being true SNRs, with Class~\textsc{III} being the least likely. G$15.51-0.15$ fell into this class, despite meeting their three criteria for admission into the sample: a shell or partial-shell morphology, a negative spectral index, and no correlation with bright mid-IR $\mu$m emission. It was not further discussed, except to clarify that as Class~\textsc{III}, it must be very faint, very confused, or does not exhibit a typical SNR morphology. It is just visible in the 11\,cm Bonn survey of the Galactic Plane but at low significance and slightly confused with the other emission to the East. As one of a sample of 75~SNRs and six candidates, \cite{2009ApJ...694L..16H} searched G$15.51-0.15$ for maser emission, but did not make a detection, to a limit of $\approx25$\,mJy at $0.53$\,km\,s$^{-1}$ velocity resolution. 
No associated X-ray emission is seen in ROSAT, nor is there any significant 12 or 22\,$\mu$m emission in \textit{WISE}.

\begin{figure*}
    \centering
    \includegraphics[width=\textwidth]{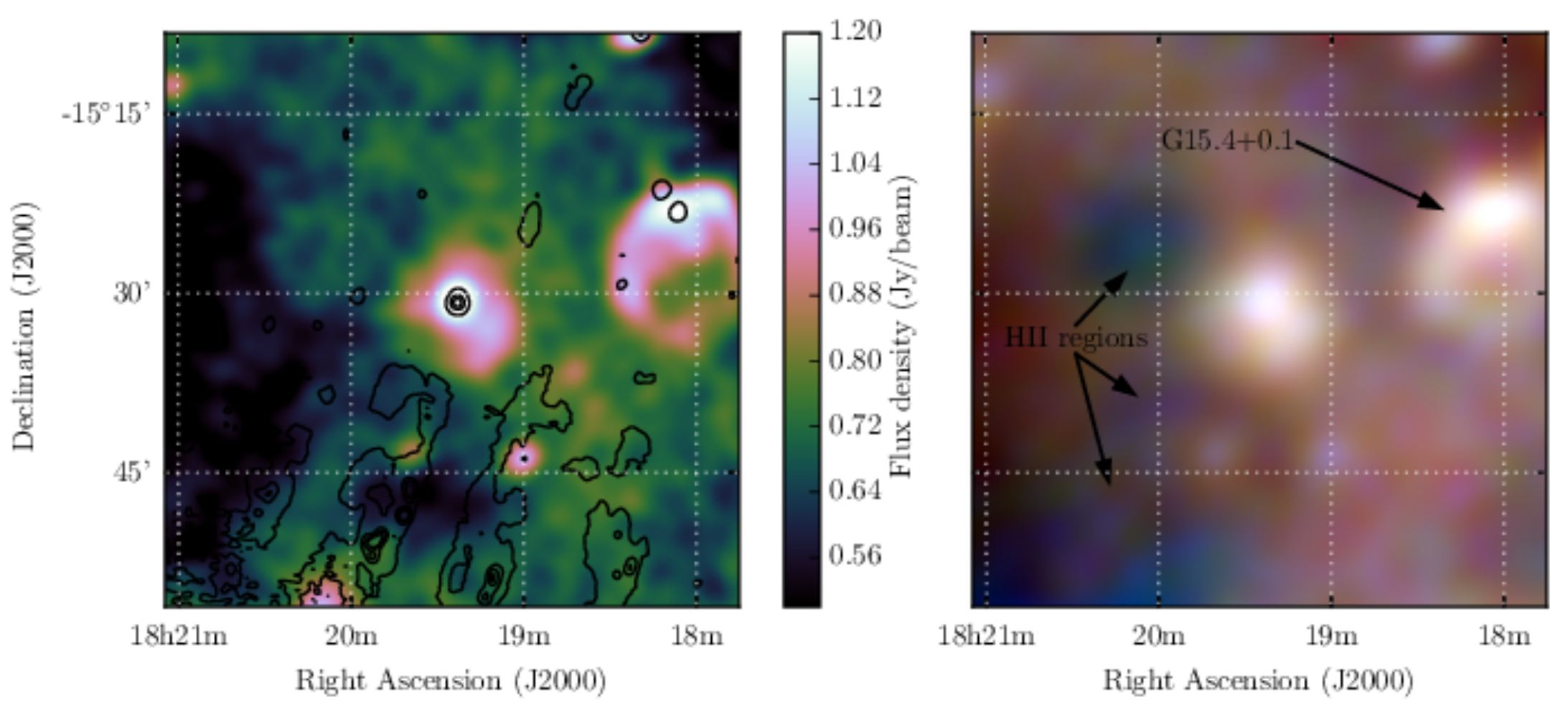}
    \caption{G$15.51-0.15$ as observed by GLEAM at 200\,MHz (left) and at at 72--103\,MHz (R), 103--134\,MHz (G), and 139--170\,MHz (B) (right). The colour scales for the RGB cube are 3--7, 1.7--4, and 0.8--2\,Jy\,beam$^{-1}$, respectively. Black contours on the left panel show the NVSS data for the region, highlighting the compact source in the centre of the remnant. Levels are at 5, 35, 65, and 95\,mJy\,beam$^{-1}$. The local NVSS RMS noise level is 2\,mJy\,beam$^{-1}$.}
    \label{fig:SNRG15.51-0.15}
\end{figure*}

Our observations show that this source lies in a complex and confusing region, with two areas of low-frequency absorption (likely \textsc{Hii} regions) to the East and South, and the known SNR~G$15.4+0.1$  to the West. Morphologically, it appears to be a complete and filled shell. \Fig~\ref{fig:SNRG15.51-0.15} shows the 200-MHz GLEAM image, and the RGB GLEAM cube at 72--103\,MHz (R), 103--134\,MHz (G), and 139--170\,MHz (B), for G\,$15.51-0.15$. 

There is also a potentially unrelated point source within the shell; running source-finding on the NVSS postage stamp for this region, we can find it has position 
RA\,$=18^\mathrm{h}19^\mathrm{m}23^\mathrm{s}$
Dec\,$=+15^\mathrm{d}30^\mathrm{m}48^\mathrm{s}$, size $1'\times1'$ with 1.4-GHz peak and integrated flux densities of $93\pm3$\,mJy\perbeam~and $163\pm6$\,mJy, respectively. It is therefore unresolved by GLEAM; in the 170--231\,MHz image it has a peak flux density of 1.5\,Jy\perbeam~and the local background (consisting of the shell of G$15.5-0.2$) is 1\,Jy\perbeam, so its flux density at 200\,MHz is 0.5\,Jy. Using the NVSS 1.4\,GHz and GLEAM 200\,MHz integrated flux densities, we calculate a spectral index of $\alpha=-0.58$. The extrapolated flux density of this source is subtracted from each of the GLEAM measurements.

Excluding the nearby \textsc{Hii} regions and SNR~G$15.4+0.1$ from the background calculation of G$15.5-0.2$, and subtracting the contaminating point source, yields a good spectral fit to the integrated flux density measurements, with $S_\mathrm{200MHz}=2.86\pm0.29$\,Jy and $\alpha=-0.55\pm0.03$, typical for a SNR. This is very similar to the spectral index of the central radio source, perhaps indicating a common origin of the emission. Based on the spectral index, morphology, and lack of IR emission, we confirm G$15.51-0.15$ as a SNR.

\subsubsection{G$19.00-0.35$}\label{SNRG19.00-0.35}

\begin{figure*}
    \centering
    \includegraphics[width=\textwidth]{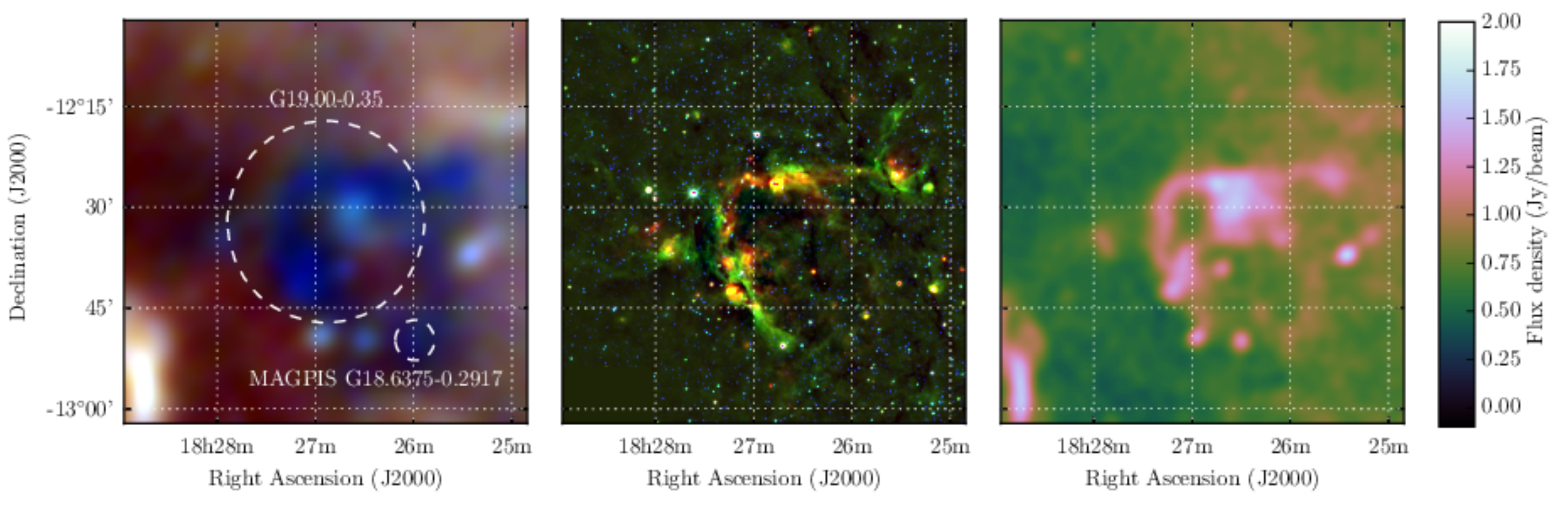}
    \caption{G$19.00-0.35$ as observed by GLEAM (left) at 72--103\,MHz (R), 103--134\,MHz (G), and 139--170\,MHz (B), by \textit{WISE} (middle) at 22\,$\mu$m (R), 12\,$\mu$m (G), and 4.6\,$\mu$m (B) and GLEAM at 200\,MHz (right). The colour scales for the GLEAM RGB cube and wideband 200-MHz image are $2.8$--9.0, $2.0$--4.6, $1.0$--2.4, and $-0.1$--2.0Jy\perbeam, respectively. The candidate MAGPIS\,$18.6375-0.2917$ is discussed in \Sect~\ref{sec:MAGPIS}.}
    \label{fig:SNRG19.00-0.35}
\end{figure*}

Another candidate proposed by \cite{1985SvA....29..128G} (see \Sect~\ref{SNRG12.75-0.15}), G$19.00-0.35$ is described as having a $30'$ diameter, with $S_\mathrm{1.42GHz}=56$\,Jy and $\alpha=-0.48$, therefore predicting $S_\mathrm{200MHz}=143$\,Jy. It is revealed as a complicated region of thermal and non-thermal emission by the GLEAM and \textit{WISE} data, with $S_\mathrm{200MHz}=28\pm2$\,Jy (see \Fig~\ref{fig:SNRG19.00-0.35}). The dominant \textsc{Hii} region is visible in the \textit{WISE} data (middle panel) as a bright loop of 12- and 22-$\mu$m emission, much brighter on the northwest side than the southeast; this is mirrored by the 200-MHz GLEAM data (right panel). In the GLEAM RGB cube (left panel), the free-free absorption of the low-frequency radio background over the whole loop becomes clear. Four or more non-thermal sources appear to be embedded in the complex. It is understandable that the Gaussian template fitting of \cite{1985SvA....29..128G} was unable to disentangle the complexity of this region. Now, however, we can state categorically that G$19.00-0.35$ is not a SNR.

\subsubsection{G$35.40-1.80$}\label{SNRG35.40-1.80}

Also proposed by \cite{1985SvA....29..128G} as a SNR candidate, G$35.40-1.80$ was described as having $7'$ extent, $S_\mathrm{408MHz}=7.9\pm1.2$\,Jy, and $\alpha=-0.42$, which would predict $S_\mathrm{200MHz}=10.7\pm1.5$\,Jy. At the time of their observations, they did not consider this source to be part of the W48~\textsc{Hii} region complex, and specified W48 as a reference source to assist their flux calibration. \cite{1994ApJ...426..249O} classed all of the objects in this area as part of W48, with G$35.40-1.80$ labeled as W48C and W48D\footnote{W48A and W48B are often grouped together in the literature as ``W48'', while W48E is another distinct \textsc{Hii} region lying to the West.}. Using the VLA, they attempted to detect RRLs toward W48A--E, and were successful in all cases except for W48C. Despite this, the classification of W48C as an \textsc{Hii} region appears to have persisted in the literature.

\begin{figure*}
    \centering
    \includegraphics[width=\textwidth]{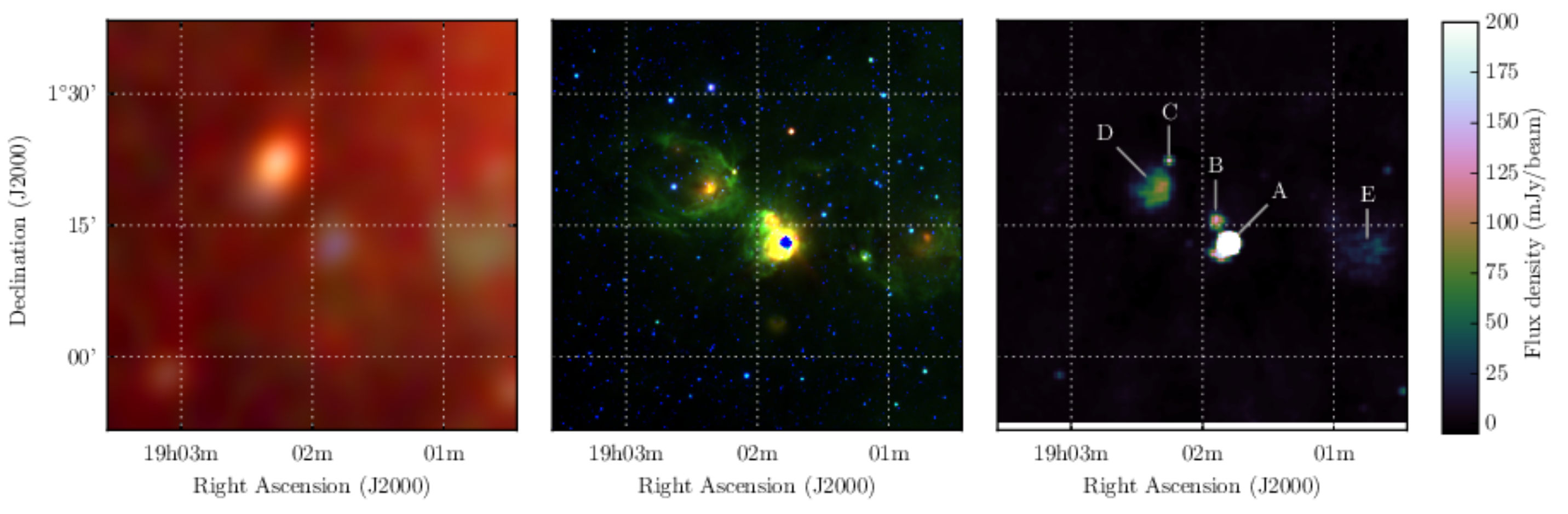}
    \caption{G$35.40-1.80$ and the W48 region, as observed by GLEAM at 72--103\,MHz (R), 103--134\,MHz (G), and 139--170\,MHz (B) (left), \textit{WISE} (middle), and NVSS (right). The colour scales for the GLEAM RGB cube are $-0.1$--2.5\,Jy\perbeam. The five components "A"--"E" of W48 identified by \protect\cite{1994ApJ...426..249O} are labeled on the right panel; C and D together make up the object G$35.40-1.80$ identified by \protect\cite{1985SvA....29..128G} as a SNR candidate.}
    \label{fig:G35.40-1.80}
\end{figure*}

The compact source-finding of Hurley-Walker et al. (submitted) detected G$35.40-1.80$ as two distinct components, GLEAM\,J190215+012219 ($S_\mathrm{200MHz}=0.97\pm0.08$, $\alpha=-0.82\pm0.06$) and GLEAM\,J190222+011904, for which $S_\mathrm{200MHz}=1.4\pm0.1$\,Jy, and a distinct low-frequency turnover is observed. These objects are, respectively, the same as W48C and W48D, as classified by \cite{1994ApJ...426..249O}. 
The non-thermal source, GLEAM\,J190215+012219, is also detected in VLSSr
, TGSS-ADR1
, and NVSS
; \Fig~\ref{fig:G35.40-1.80} shows a plot of the spectrum containing these and the GLEAM flux density measurements: combining them yields $\alpha = -0.85\pm0.02$ and $S_\mathrm{200MHz} = 1.00\pm0.09$\,Jy.

\begin{figure*}
    \centering
    \includegraphics[width=\textwidth]{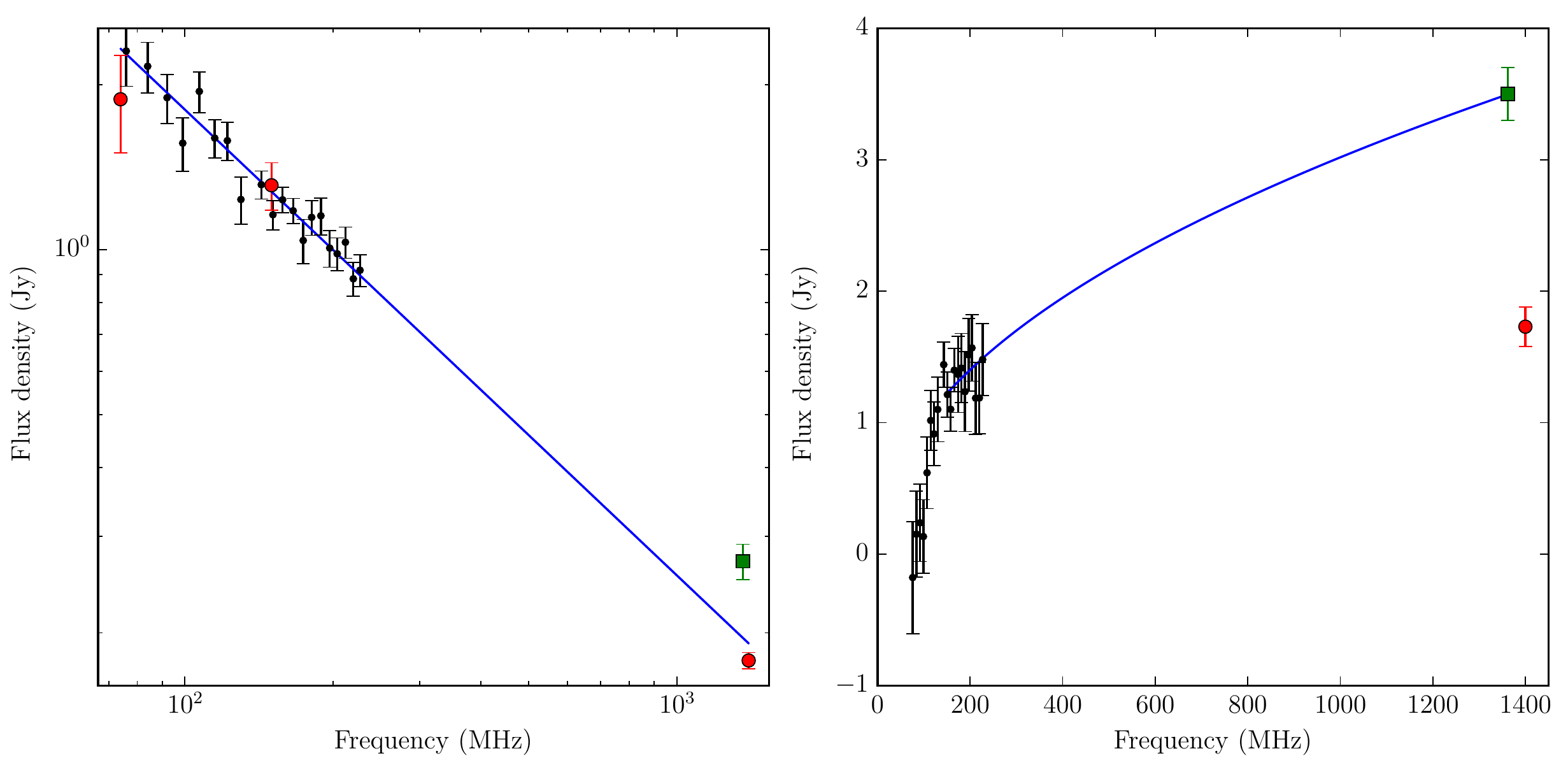}
    \caption{The spectra of the two components of G$35.40-1.80$: the left panel shows the non-thermal source GLEAM\,J190215+012219 (W48C) and the right panel shows the \textsc{Hii} region GLEAM\,J190222+011904 (W48D). Black points indicate GLEAM measurements; red points indicate VLSSr (74\,MHz), TGSS-ADR1 (150\,MHz), and NVSS (1.4\,GHz), while green squares show 1.362\,GHz measurements made by \protect\cite{1994ApJ...426..249O}. In the left panel, the NVSS point is taken from the NVSS catalogue, while in the right panel, it has been measured using \polyflux. Blue lines indicate least-squares power-law fits to the data; the left fit uses all plotted data points, while the right fit excludes the NVSS point and uses only the GLEAM data with $\nu>150$\,MHz.}
    \label{fig:G35.40-1.80_spectrum}
\end{figure*}

The thermal source, GLEAM\,J190222+011904, is resolved out by VLSSr and TGSS, and while it appears in NVSS images, is clearly not catalogued correctly by the automated source-finding. We use \polyflux~to measure $S_\mathrm{1.4GHz}=1.73\pm0.15$\,Jy from the NVSS image, but suspect that this is an underestimate of the true flux density due to the large angular extent ($6'$) of the source. \Fig~\ref{fig:G35.40-1.80_spectrum} shows a plot of this measurement, that of \cite{1994ApJ...426..249O}, and the GLEAM flux density measurements.

Based on the low-frequency free-free absorption, thermal radio spectrum and the strong 12- and 22-$\mu$m emission seen in \textit{WISE}, we confirm that W48D / GLEAM\,J190222+011904 is a \textsc{Hii} region. W48C / GLEAM\,J190215+012219 has no optical or IR counterpart; the nearest discrete source is an apparently unrelated galaxy at $\mathrm{RA}=19^\mathrm{h}02^\mathrm{m}14.5^\mathrm{s}$, $\mathrm{Dec}=+01^\mathrm{d}22^\mathrm{m}38^\mathrm{s}$, 14" away. This unresolved source with no RRLs, and no optical or IR counterpart,  $\alpha=-0.85$, is a good candidate pulsar, or even a high-redshift radio galaxy. However, given $\alpha>-1.3$, it is still potentially a SNR, perhaps very young. Higher-resolution and/or pulsar timing observations will be necessary to reveal the nature of this source.






\subsubsection{G$36.00+0.00$}\label{SNRG36.00+0.00}


\cite{2006IAUS..230..333U} suggested G$36.00+0.00$ as a candidate SNR based on a diffuse X-ray excess of size $\approx12'\times10'$ in the \textit{Advanced Satellite for Cosmology and Astrophysics (ASCA)} Galactic Plane Survey.

\begin{figure*}
    \centering
    \includegraphics[width=\textwidth]{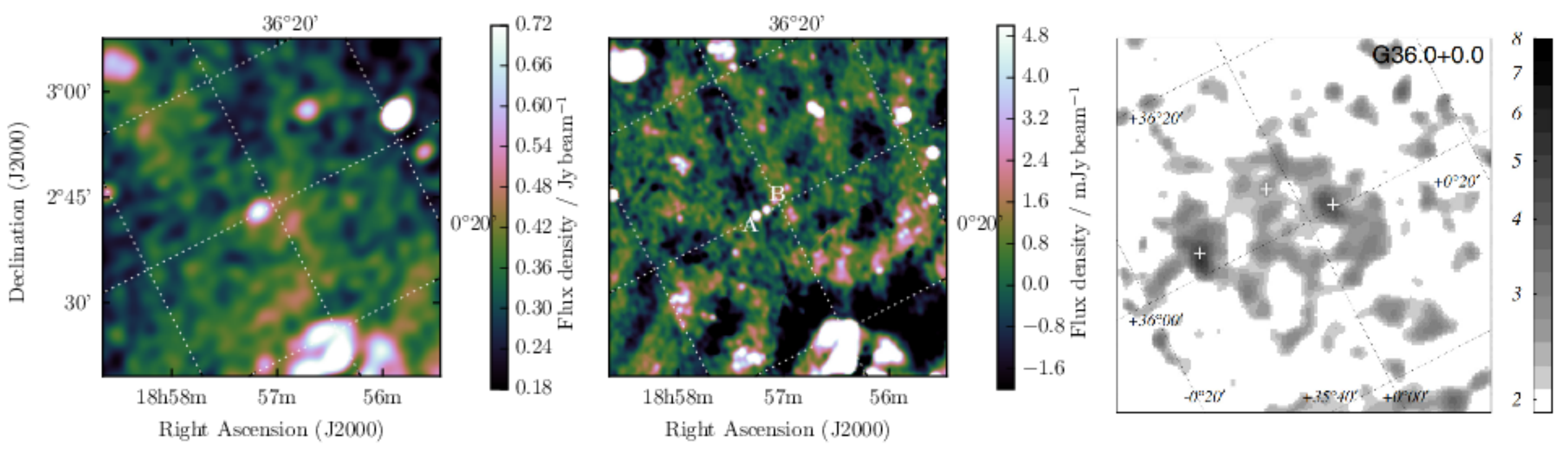}
    \caption{G$36.00+0.00$ as observed by GLEAM (left) at 200\,MHz, by NVSS (middle) at 1.4\,GHz, and by \protect\cite{2006IAUS..230..333U} with \textit{ASCA} X-ray (2.0--7.0\,keV) (right). Linear colourscales for GLEAM and NVSS are shown in the figure, while for \textit{ASCA}, the scale is logarithmic and the numbers next to the scale bars correspond to the surface brightness in $\times10^{-6}$counts\,cm$^{-2}$s$^{-1}$arcmin$^{-2}$. Dashed lines, white in the left and middle panel, and black in the right panel, indicate Galactic co-ordinates. In the middle panel, the two compact radio sources discussed in the text are labelled A and B. In the right panel, the X-ray sources detected by \protect\cite{2001ApJS..134...77S} are designated with white crosses.}
    \label{fig:SNRG36.00+0.00}
\end{figure*}

In the GLEAM images, there is no sign of diffuse emission on these scales, only an isolated point source. However, in TGSS-ADR1 and NVSS, this source is resolved into two individual sources, appearing at $\mathrm{RA}=18^\mathrm{h}57^\mathrm{m}15.77^\mathrm{s}$, $\mathrm{Dec}=+02^\mathrm{d}42^\mathrm{m}21.39^\mathrm{s}$, and $\mathrm{RA}=18^\mathrm{h}57^\mathrm{m}09.70^\mathrm{s}$, $\mathrm{Dec}=+02^\mathrm{d}43^\mathrm{m}10.21^\mathrm{s}$, and labelled A and B respectively in \Fig~\ref{fig:SNRG36.00+0.00}. Neither appear in the NVSS catalogue so we use \textsc{Aegean} to make measurements in the NVSS images, while we obtain TGSS flux densities directly from the ADR1 catalogue. Source~A has $S_\mathrm{150MHz}=55\pm10$\,mJy and $S_\mathrm{1.4GHz}=24.8\pm0.1$\,mJy, giving the source a spectral index of $\alpha=-0.36\pm0.08$. Source~B has $S_\mathrm{150MHz}=540\pm60$\,mJy and $S_\mathrm{1.4GHz}=13.5\pm0.1$\,mJy, giving the source a spectral index of $\alpha=-1.65\pm0.04$. For both sources combined, we measure $S_\mathrm{200MHz}=490\pm90$\,mJy and $\alpha=-1.62\pm0.15$ across the GLEAM band, consistent with the TGSS-ADR1 and NVSS measurements.

There is no obvious counterpart in \textit{WISE} or DSS2 to either the diffuse X-ray source or the compact radio sources, and there are no known pulsars within the excess X-ray emission identified by \cite{2006IAUS..230..333U}. We cannot confirm the status of G$36.00+0.00$ as a SNR, as we do not detect any diffuse component, but based on its steep spectrum and compact nature, we suggest Source~B is a candidate pulsar.


\subsection{MAGPIS candidates}\label{sec:MAGPIS}

\cite{2006AJ....131.2525H} imaged $5^\circ < l < 32^\circ$; $|b| < 0.8^\circ$ with the VLA in multiple configurations at 1.4\,GHz, to a noise level of $\approx2$\,mJy\perbeam, to create the Multi-Array Galactic Plane Imaging Survey (MAGPIS). They compared this data to 330\,MHz VLA data and the 21\,$\mu$m MSX Point Source Catalog version 2.3 ``MSX6C'' data set \citep{2003AAS...203.5708E}, in order to detect new compact Galactic objects and distinguish \textsc{Hii} regions from SNR. They presented 49~``high-probability'' candidate SNR, requiring the object to be undetected in MSX6C, brighter at 330\,MHz than at 1.4\,GHz, and have a shell-like or PWNe-like morphology. Due to the high resolution of the VLA survey, these are all relatively compact objects, with diameters $\leq14'$ and typically $\approx3'$, and are therefore usually unresolved by GLEAM.

The MSX6C data set and 330\,MHz VLA data appear not to have been sensitive enough to completely discriminate SNR from \textsc{Hii} regions. The \textit{WISE} data is $\approx300\times$ more sensitive than the MSX data
and reveals that 31 of these candidate SNRs are associated with strong 12- and sometimes 22-$\mu$m emission, which is morphologically similar to the radio emission. Three are coincident with known \textsc{Hii} regions observed in the canonical survey of \cite{1989ApJS...71..469L}. \cite{2009AJ....138.1615J} measured the \textsc{HI} absorption spectra toward the 41~candidates to attempt to measure their distances, and reclassified a further nine as \textsc{Hii} regions based on strong RRL emission, 8-$\mu$m GLIMPSE, or 24-$\mu$m MIPSGAL emission. \cite{2017A+A...605A..58A} reclassified a further eight MAGPIS SNR candidates as known \textsc{Hii} regions, based on their IR emission in \textsc{WISE}, as well as another MAGPIS candidate as a known planetary nebula (PN). 10~SNRs were accepted to the catalogue of \cite{2014BASI...42...47G}. \Tab~\ref{tab:helfand} summarises the MAGPIS candidates, extending \Tab~4 of \cite{2006AJ....131.2525H} to include the GLEAM measurements and morphological classifications, where possible. The total number of candidates previously confirmed as SNRs is 10, re-classified as \textsc{Hii}~regions is 20, with one further object reclassified as a PN.

Of the 10~SNR added to the catalogue of \cite{2014BASI...42...47G}, we detect and measure eight, including spectral indices for three, which are uniformly non-thermal. MAGPIS\,$29.366700+0.100000$ is noted as a potential PWNe by \cite{2006AJ....131.2525H}; the GLEAM spectrum of $\alpha=-0.09\pm,0.14$ is consistent with a PWNe interpretation. Morphologically in the MAGPIS data, it resembles a wide-angle tail radio galaxy, but there is no obvious host visible in \textit{WISE} or DSS2. There is also a somewhat filled shell visible in both MAGPIS and GLEAM. We therefore tentatively confirm this SNR as a composite SNR with central emission and a surrounding shell.

Of the 20~candidates already reclassified as \textsc{Hii} regions, we observe eight as regions absorbing low frequencies in GLEAM, further increasing the likelihood that they are \textsc{Hii} regions. We have noted their rising spectra in \Tab~\ref{tab:helfand} ($\alpha>0$), but do not attempt to fit power-law spectral indices as these do not sufficiently describe the spectra. The upcoming publication Su et al. (in prep) will publish the GLEAM spectra for these objects, as well as all other detected \textsc{Hii} regions in the data release of Hurley-Walker et al. (submitted).

\begin{figure*}
    \centering
   \includegraphics[width=\textwidth]{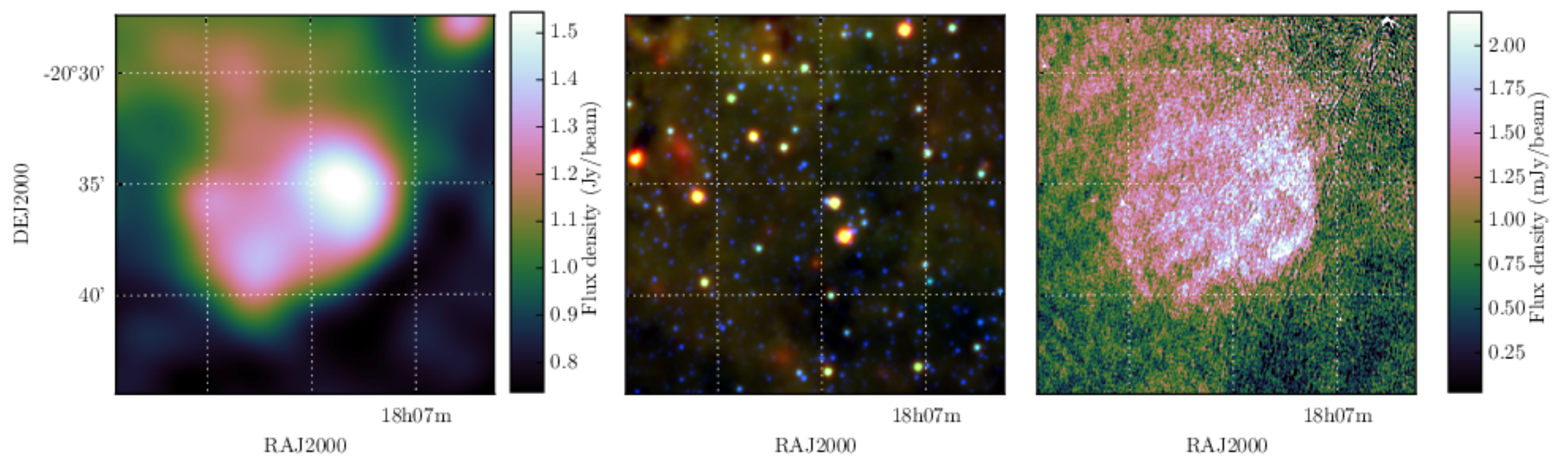}
    \caption{MAGPIS\,$9.683300-0.066700$ as observed by GLEAM at 200\,MHz (left), \textit{WISE} at 22\,$\mu$m (R), 12\,$\mu$m (G), and 4.6\,$\mu$m (B), and by MAGPIS at 1.4\,GHz (right).}
    \label{fig:SNR_G9.7-0.1}
\end{figure*}

\begin{figure*}
    \centering
   \includegraphics[width=\textwidth]{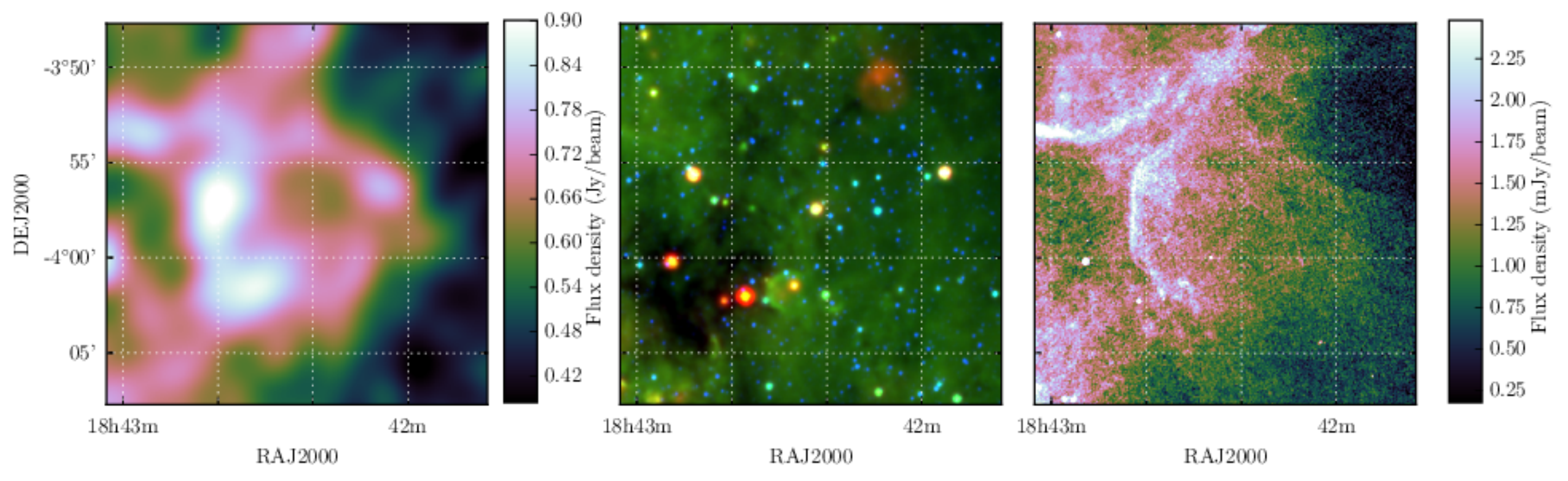}
    \caption{MAGPIS\,$28.375000+0.202800$ as observed by GLEAM at 200\,MHz (left), \textit{WISE} at 22\,$\mu$m (R), 12\,$\mu$m (G), and 4.6\,$\mu$m (B), and by MAGPIS at 1.4\,GHz (right).}
    \label{fig:SNR_G28.4+0.2}
\end{figure*}

\begin{figure*}
    \centering
   \includegraphics[width=\textwidth]{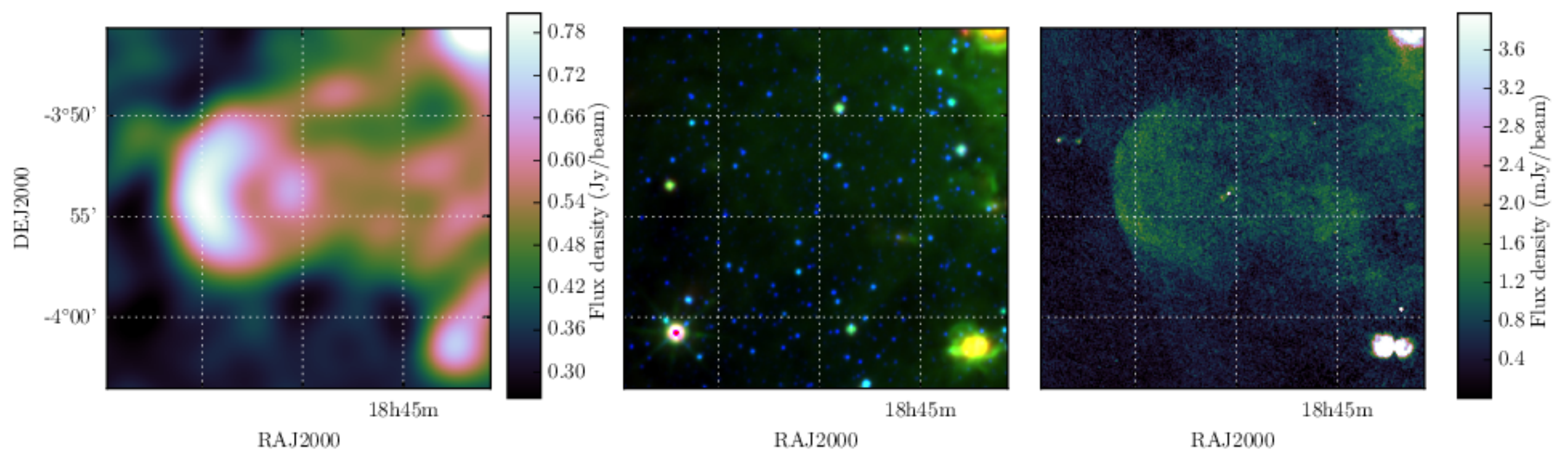}
    \caption{MAGPIS\,$28.7667-0.4250$ as observed by GLEAM at 200\,MHz (left), \textit{WISE} at 22\,$\mu$m (R), 12\,$\mu$m (G), and 4.6\,$\mu$m (B), and by MAGPIS at 1.4\,GHz (right).}
    \label{fig:SNR_G28.8-0.4}
\end{figure*}

Of the 18~unconfirmed and not-previously-reclassified objects, we detect seven in GLEAM. MAGPIS\,$9.683300-0.066700$, MAGPIS\,$28.375000+0.202800$, and MAGPIS\,$28.7667-0.4250$ have non-thermal spectra and shell-like morphology, and are easily confirmed as SNRs (\Figs~\ref{fig:SNR_G9.7-0.1}, \ref{fig:SNR_G28.4+0.2}, and \ref{fig:SNR_G28.8-0.4}, respectively). MAGPIS\,$18.6375-0.2917$ appears in absorption and overlaps with the candidate G\,$19.00-0.35$ (\Sect~\ref{SNRG19.00-0.35}), which we reclassify as a \textsc{Hii} region; its location is marked on \Fig~\ref{fig:SNRG19.00-0.35}.
Similarly, MAGPIS\,$20.4667+0.1500$ also has a thermal spectrum with absorption at low frequencies (\Fig~\ref{fig:SNR_G20.5+0.2_spectrum}), so we class it as a \textsc{Hii} region. The spectrum of these two candidates is indicated with $\alpha>0$ in \Tab~\ref{tab:helfand}.

\begin{figure*}
    \centering
   \includegraphics[width=\textwidth]{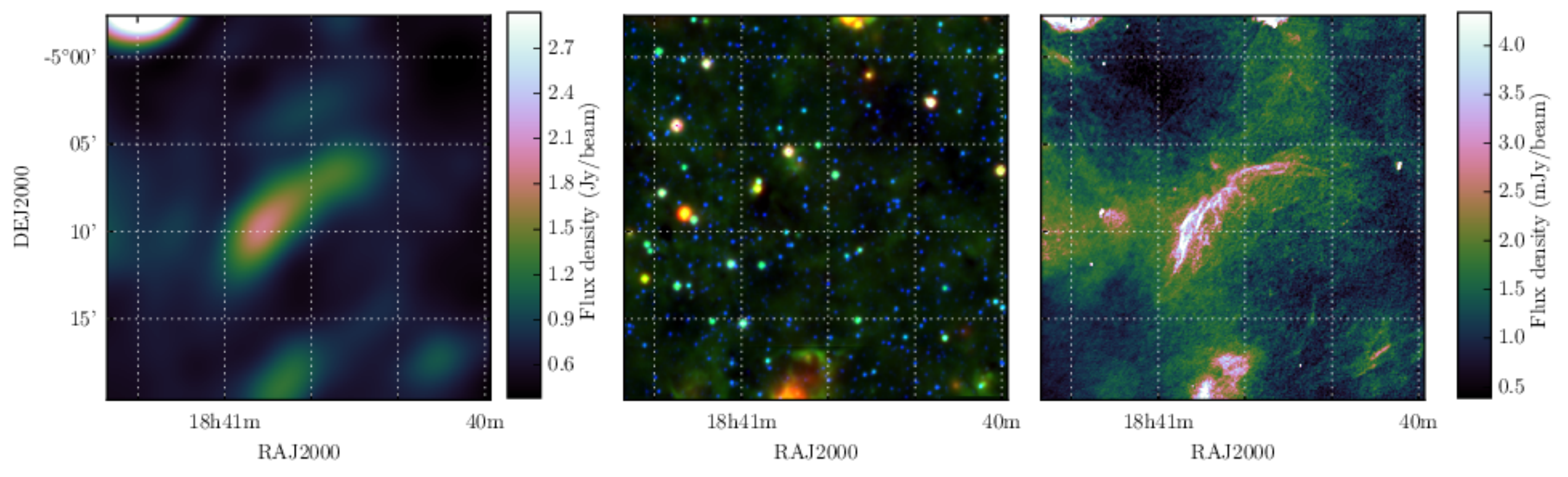}
    \caption{MAGPIS\,$27.133300+0.033300$ as observed by GLEAM at 200\,MHz (left), \textit{WISE} at 22\,$\mu$m (R), 12\,$\mu$m (G), and 4.6\,$\mu$m (B), and by MAGPIS at 1.4\,GHz (right).}
    \label{fig:SNR_G27.1+0.0}
\end{figure*}

MAGPIS\,$27.133300+0.033300$ appears to be a single lone arc of emission, reminiscent of part of a SNR shell (\Fig~\ref{fig:SNR_G27.1+0.0}). There is a \textsc{Hii} region to the south which could be confused as a counterpart arc were it not for the \textit{WISE} data, which shows strong 12- and 22-$\mu$m emission. There is another non-thermal region of emission to the south-west which has a similar filamentary morphology in MAGPIS, and is likely the other half of the shell. Unfortunately the region is strongly confused in GLEAM, with many nearby \textsc{Hii} regions, making it difficult to extract reliable flux density measurements. \Tab~\ref{tab:helfand} therefore has only a 200-MHz measurement for this source, and only for the single arc as seen in MAGPIS.

MAGPIS\,$18.758300-0.073600$ is small and located in a confused region, so while we may make a flux density measurement, it is unresolved by GLEAM. It is noted as a potential PWNe by \cite{2006AJ....131.2525H}. Due to its small size ($1\farcm6$), we can use NVSS to measure its 1.4-GHz flux density as $22.8\pm1.5$\,mJy, implying a spectral index between 200 and 1400 MHz of $-1.8$, which is highly incompatible with a PWNe interpretation, which we have indicated in \Tab~\ref{tab:helfand}.


It would be ideal to use the MAGPIS 1.4-GHz integrated flux density measurement to constrain the spectra of all 13~SNR candidates measurable in GLEAM, but \cite{2006AJ....131.2525H} note that their flux density scale has a large amount of uncertainty, and that integrated flux densities are unreliable and often overestimated by large and varying factors. Our measurements confirm this, finding that for the five sources where we are able to derive full spectra, the resulting 1.4-GHz flux density prediction is of order 6--25\,\% of the values obtained by MAGPIS.




\begin{table*}
\small
\caption{MAGPIS SNR candidates; the first four columns are taken from \Tab~4 of \protect\cite{2006AJ....131.2525H}; the next four columns are calculated in \Sect~\ref{sec:MAGPIS}. ``Class'' is determined from the literature where possible, or this work if the GLEAM spectrum and/or morphology are clear: a ``--'' indicates that the GLEAM data does not improve our understanding of this candidate.
\label{tab:helfand}}
\begin{tabular}{cccccccc}
MAGPIS & Diam & $S_p$ & $S_i$ & $S_\mathrm{200\,MHz}$ & $\alpha_\mathrm{GLEAM}$ & Class & References \\
       & $'$    & mJy\perbeam~& Jy    & Jy &               & & \\
 \hline
06.4500-0.5583 & 3.3 & 4.7 & 6.64 & -- & -- & -- & -- \\
06.5375-0.6028 & 5.0 & 16.9 & 9.42 & -- & -- & -- & -- \\
07.2167+0.1833 & 6.5 & 17.2 & 9.1 & $0.59\pm0.36$ & -- & SNR & \cite{2006ApJ...639L..25B} \\
08.3083-0.0861 & 3.0 & 263.5 & 7.0 & -- & -- & \textsc{Hii} region & \cite{1989ApJS...71..469L} \\
08.8583-0.2583 & 4.0 & 3.4 & 4.79 & -- & -- & -- & --  \\
09.6833-0.0667 & 8.5 & 7.2 & 12.5 & $6.03\pm0.19$ & $-0.48\pm0.07$ & SNR  & This work \\
10.8750+0.0875 & 2.8 & 4.9 & 4.05 & -- & -- & \textsc{Hii} region & \cite{1989ApJS...71..469L} \\
11.1639-0.7167 & 7.0 & 1.9 & 3.11 & $1.19\pm0.51$ & -- & SNR & \cite{2004AJ....127..355B,2006ApJ...639L..25B} \\
11.2000+0.1167 & 7.5 & 8.3 & 10.8 & $2.78\pm0.27$ & -- & SNR & \cite{2004AJ....127..355B,2006ApJ...639L..25B} \\
11.5500+0.3333 & 4.5 & 2.5 & 4.34 & -- & -- & -- & -- \\
11.8903-0.2250 & 3.5 & 4.2 & 3.03 & -- & -- & -- & -- \\
12.2694+0.2972 & 4.0 & 2.8 & 2.62 & $1.39\pm0.11$ & $-0.39\pm0.18$ & SNR & \cite{2006ApJ...639L..25B} \\
12.7167+0.0000 & 4.5 & 7.0 & 8.56 & $3.45\pm0.15$ & -- & SNR & \cite{2006ApJ...639L..25B} \\
12.8208-0.0208 & 2.0 & 5.1 & 3.4 & $2.49\pm0.11$ & -- & SNR & \cite{2006ApJ...639L..25B} \\
12.9139-0.2806 & 1.5 & 16.4 & 1.39 & -- & -- & \textsc{Hii} region & \cite{2004ApJ...605..285S} \\
13.1875+0.0389 & 2.5 & 27.0 & 7.58 & -- & $>0$ & \textsc{Hii} region & \cite{1989ApJS...71..469L} \\
16.3583-0.1833 & 2.8 & 7.9 & 2.48 & -- & -- & -- & -- \\
17.0167-0.0333 & 4.0 & 4.5 & 2.63 & -- & -- & SNR & \cite{2006ApJ...639L..25B} \\ 
17.3361-0.1389 & 1.8 & 2.9 & 0.282 & -- & -- & SNR & \cite{2006ApJ...639L..25B} \\ 
18.1500-0.1722 & 7.0 & 7.8 & 9.69 & $9.03\pm0.35$ & -- & SNR & \cite{1986AJ.....92.1372O} \\
18.2536-0.3083 & 3.5 & 15.8 & 8.51 & -- & $>0$ & \textsc{Hii} region & \cite{2009AJ....138.1615J,2017A+A...605A..58A} \\
18.6375-0.2917 & 4.0 & 4.1 & 4.31 & -- & $>0$ & \textsc{Hii} region & This work \\ 
18.7583-0.0736 & 1.6 & 10.2 & 1.34 & $0.76\pm0.09$ & $-1.8$ & not PWNe & This work \\
19.4611+0.1444 & 6.0 & 105.9 & 8.09 & -- & -- & \textsc{Hii} region & \cite{2017A+A...605A..58A} \\
19.5800-0.2400 & 3.2 & 306.6 & 6.61 & -- & $>0$ & \textsc{Hii} region & \cite{2009AJ....138.1615J,2017A+A...605A..58A} \\
19.5917+0.0250 & 0.8 & 2.9 & 0.245 & -- & -- & \textsc{Hii} region & \cite{2017A+A...605A..58A} \\
19.6100-0.1200 & 4.5 & 8.5 & 5.87 & -- & -- & \textsc{Hii} region & \cite{2009AJ....138.1615J,2017A+A...605A..58A} \\
19.6600-0.2200 & 4.5 & 10.2 & 4.8 & -- & $>0$ & \textsc{Hii} region & \cite{2009AJ....138.1615J,2017A+A...605A..58A} \\
20.4667+0.1500 & 5.5 & 4.6 & 6.56 & -- & $>0$ & \textsc{Hii} region & This work \\
21.5569-0.1028 & 4.0 & 1.5 & 1.9 & -- & -- & -- & -- \\
21.6417+0.0000 & 2.8 & 5.4 & 1.54 & -- & -- & \textsc{Hii} region & \cite{2017A+A...605A..58A} \\
22.3833+0.1000 & 7.0 & 5.5 & 4.66 & -- & -- & -- &  -- \\
22.7583-0.4917 & 3.8 & 20.8 & 6.01 & -- & -- & \textsc{Hii} region & \cite{2009AJ....138.1615J,2017A+A...605A..58A} \\
22.9917-0.3583 & 3.8 & 15.2 & 5.1 & -- & -- & \textsc{Hii} region & \cite{2009AJ....138.1615J,2017A+A...605A..58A} \\
23.5667-0.0333 & 9.0 & 9.0 & 23.3 & -- & -- &  \textsc{Hii} region & \cite{2009AJ....138.1615J,2017A+A...605A..58A} \\
24.1803+0.2167 & 5.2 & 44.5 & 5.39 & -- & -- & \textsc{Hii} region & \cite{2009AJ....138.1615J,2017A+A...605A..58A} \\
25.2222+0.2917 & 2.0 & 3.2 & 1.42 & -- & -- &  \textsc{Hii} region & \cite{2017A+A...605A..58A} \\
27.1333+0.0333 & 11.0 & 5.7 & 17.8 & $4.93\pm0.14$ & -- & SNR arc? & This work \\ 
28.3750+0.2028 & 10.0 & 11.4 & 14.9 & $4.18\pm0.26$ & $-0.72\pm0.10$ & SNR  & This work \\
28.5167+0.1333 & 14.0 & 26.6 & 13.0 & -- & -- & -- & -- \\
28.5583-0.0083 & 3.0 & 17.5 & 5.37 & -- & -- &  --  & -- \\ 
28.7667-0.4250 & 9.5 & 5.4 & 10.9 & $3.17\pm0.23$ & $-0.79\pm0.12$ & SNR & This work \\
29.0667-0.6750 & 8.0 & 46.1 & 7.56 & -- & $>0$ & \textsc{Hii} region & \cite{2009AJ....138.1615J,2017A+A...605A..58A} \\
29.0778+0.4542 & 0.7 & 8.1 & 0.657 & -- & -- & PN & \cite{2017A+A...605A..58A} \\
29.3667+0.1000 & 9.0 & 6.5 & 16.6 & $2.96\pm0.16$ & $0.09\pm0.14$ & PWNe & \cite{2006AJ....131.2525H} \\
30.8486+0.1333 & 2.2 & 90.4 & 3.81 & -- & $>0$ & \textsc{Hii} region & \cite{2017A+A...605A..58A} \\
31.0583+0.4833 & 4.5 & 11.6 & 4.86 & -- & $>0$ & \textsc{Hii} region & \cite{2017A+A...605A..58A} \\
31.6097+0.3347 & 3.1 & 3.7 & 1.74 & -- & $>0$ & \textsc{Hii} region & \cite{2017A+A...605A..58A} \\
31.8208-0.1222 & 1.8 & 3.2 & 0.896 & -- & -- & \textsc{Hii} region & \cite{2017A+A...605A..58A} \\
\end{tabular}
\end{table*}

\section{Discussion}\label{sec:discussion}

Of the 101 non-MAGPIS candidates proposed in the region, 82 are undetectable in these data. Given the range of different origins of these candidates, it is difficult to draw any particular conclusions about how one might improve the detection rate. Certainly, higher sensitivity by increasing the integration time may be useful for the larger and fainter objects. 51 have diameters $\leq5'$, so arcminute or better resolution may help reveal the nature of these sources. The upcoming GLEAM-eXtended (GLEAM-X; Hurley-Walker et al. in prep) survey using the upgraded MWA will have up to $10\times$ the sensitivity of GLEAM, and double the resolution. This should be a valuable resource for determining the nature of many of the remaining candidates.


Low frequencies appear to offer a decided advantage in finding new SNRs over high frequencies, where both thermal and non-thermal emission have similar contributions to the Galactic brightness. 30 of the 49 MAGPIS candidates are in fact \textsc{Hii} regions, while Hurley-Walker et al. (submitted) show 27 new SNRs detected at low frequencies with no IR counterparts, many with pulsar associations, and Johnston-Hollitt et al. (in prep) finds similar results over the region $240^\circ < l < 345^\circ$.

The wide bandwidth of the MWA has been particularly useful in discriminating between types of emission, due to the distinct absorption signature of the \textsc{Hii} regions. However, wide bandwidths at higher frequencies, such as those available to the Australian Square Kilometer Array Pathfinder \citep[ASKAP; ][]{2014PASA...31...41H}, should also allow useful discrimination between thermal and non-thermal emission. Indeed, the combination of upcoming radio surveys in the Southern Hemisphere from both the MWA and ASKAP will offer powerful insight into Galactic astrophysics and should result in many more SNR detections.  Flux density calibration across multiple epochs and instruments is key: the poor flux calibration of MAGPIS appears to have hindered their ability to produce reliable spectra, contaminating their candidate sample.

As the population of known Galactic SNRs grows, more SNR will be found in line-of-sight overlap, and require careful analysis to separate the different objects. There are likely a further $\approx700$ SNRs yet to be discovered, and many will naturally be overlapping. High resolution and high surface brightness sensitivity will become exponentially more important as the counts of SNRs increase.

\section{Conclusions}\label{sec:conclusions}

We examined the latest data release from the GaLactic and Extragalactic All-sky Murchison Widefield Array (GLEAM) survey covering $345^\circ < l < 60^\circ$, $180^\circ < l < 240^\circ$, using these data and that of the \textit{Widefield Infrared Survey Explorer} to follow up proposed candidate SNR from other sources. Of the 101 candidates proposed in the region, we are able to definitively confirm ten as SNR, tentatively confirm two as SNR, and reclassify five as \textsc{Hii} regions. A further two are detectable in our images but difficult to classify; the remaining 82 are undetectable in these data.
We also investigated the 18~unclassified Multi-Array Galactic Plane Imaging Survey (MAGPIS) candidate SNRs, newly confirming three as SNRs, reclassifying two as \textsc{Hii} regions, and exploring the unusual spectra and morphology of two others.

\begin{acknowledgements}
This scientific work makes use of the Murchison Radio-astronomy Observatory, operated by CSIRO. We acknowledge the Wajarri Yamatji people as the traditional owners of the Observatory site. Support for the operation of the MWA is provided by the Australian Government (NCRIS), under a contract to Curtin University administered by Astronomy Australia Limited. We acknowledge the Pawsey Supercomputing Centre which is supported by the Western Australian and Australian Governments. We acknowledge the work and support of the developers of the following following python packages: Astropy \citet{TheAstropyCollaboration2013}, Numpy \citep{vaderwalt_numpy_2011}, and Scipy \citep{Jones_scipy_2001}. We also made extensive use of the visualisation and analysis packages DS9\footnote{\href{http://ds9.si.edu/site/Home.html}{ds9.si.edu/site/Home.html}} and Topcat \citep{Taylor_topcat_2005}.
This publication makes use of data products from the Wide-field Infrared Survey Explorer, which is a joint project of the University of California, Los Angeles, and the Jet Propulsion Laboratory/California Institute of Technology, funded by the National Aeronautics and Space Administration.
\end{acknowledgements}

\begin{appendix}

\section{List of candidate SNRs searched}

\onecolumn
\begin{longtable}{ccccc}
\caption{SNR candidates searched for in this work, ordered first by detection method, second by date detected, and third by $l$. \label{tab:allsources}}\\
\hline
l & b & diameter & detection method & Reference(s) \\
$^\circ$ & $^\circ$ & arcmin & & \\
\hline
57.10 & +1.70 & 40.0 & Radio & \cite{1983A+A...123L...5G} \\
12.75 & -0.15 & 15.0 & Radio & \cite{1985SvA....29..128G}  \\
19.00 & -0.35 & 30.0 & Radio &     "    \\
35.40 & -1.80 & 7.0 & Radio &     " \\
38.05 & -0.05 & 8.0 & Radio &     " \\
359.90 & -0.10 & 10.0 & Radio & \cite{1985ApJ...288..575H} \\
7.60 & -0.60 & 16.0 & Radio & \cite{1986AJ.....92.1372O} \\
57.20 & +1.10 & 100.0 & Radio & \cite{1988ApJ...326..751R} \\
13.10 & -0.50 & 47.0 & Radio & \cite{1990ApJ...364..187G} \\
14.20 & -0.90 & 57.0 & Radio &     " \\
14.50 & +1.20 & 41.0 & Radio &     " \\
25.80 & +0.90 & 25.0 & Radio &     " \\
27.90 & -1.20 & 30.0 & Radio &     " \\
29.40 & +1.40 & 33.0 & Radio &     " \\
34.80 & +1.00 & 43.0 & Radio &     " \\
36.00 & -0.20 & 51.0 & Radio &     " \\
41.30 & -1.30 & 30.0 & Radio &     " \\
44.20 & +0.50 & 22.0 & Radio &     " \\
44.60 & +0.10 & 37.0 & Radio &     " \\
51.70 & -0.80 & 34.0 & Radio &     " \\
356.60 & +0.10 & 7.0 & Radio & \cite{1994MNRAS.270..847G} \\
357.10 & -0.20 & 8.0 & Radio &   " \\
358.70 & +0.70 & 18.0 & Radio &   " \\
359.20 & -1.10 & 5.0 & Radio &   " \\
3.10 & -0.60 & 40.0 & Radio & " \\
4.20 & +0.00 & 4.0 & Radio &     "   \\
352.60 & +2.20 & 30.0 & Radio & \cite{1995MNRAS.277...36D} \\
353.30 & -1.00 & 60.0 & Radio & \cite{1995MNRAS.277...36D,1997MNRAS.287..722D} \\
345.10 & -0.20 & 6.0 & Radio & \cite{1996A+AS..118..329W} \\
345.10 & +0.20 & 10.0 & Radio &   " \\
348.80 & +1.10 & 10.0 & Radio &   " \\
359.00 & -0.00 & 13.0 & Radio & \cite{2000AJ....119..207L} \\
346.50 & -0.10 & 12.0 & Radio & \cite{2001ApJ...559..963G} \\
4.20 & -0.30 & 15.0 & Radio & \cite{2001ESASP.459..109T} \\
5.30 & +0.10 & 2.5 & Radio &     " \\
5.70 & -0.20 & 2.5 & Radio &     " \\
29.80 & +2.10 & 10.0 & Radio &     " \\
39.70 & +0.50 & 12.0 & Radio &     " \\
40.40 & +0.70 & 10.0 & Radio &     " \\
44.00 & -0.10 & 26.0 & Radio &     " \\
44.50 & -1.30 & 35.0 & Radio &     " \\
54.50 & +1.20 & 25.0 & Radio &     " \\
55.10 & +0.10 & 5.0 & Radio &     " \\
55.60 & +0.60 & 14.0 & Radio &     " \\
353.00 & +2.20 & 6.0 & Radio & " \\
203.20 & +7.90 & 95.0 & Radio &   \cite{2002nsps.conf....1R,2005ASPC..343..286S} \\
206.80 & +6.20 & 180.0 & Radio &     " \\
352.20 & -0.10 & 6.0 & Radio & \cite{2002ASPC..271...31M} \\
359.56 & -0.08 & 2.5 & Radio & \cite{2007A+A...462.1065M} \\
354.46 & +0.07 & 1.6 & Radio & \cite{2013ApJ...774..150R} \\
346.20 & -1.00 & 7.0 & Radio & \cite{2014PASA...31...42G} \\
354.10 & +0.30 & 11.0 & Radio &  " \\
351.60 & +0.20 & 12.0 & Radio & \cite{2015MNRAS.453.2082D}  \\
41.40 & +1.20 & 25.0 & Radio & \cite{1993AJ....105..314G} \\
45.90 & -0.10 & 27.0 & Radio & \cite{1992AJ....103..931T} \\
8.50 & -6.70 & 60.0 & Radio & \cite{1998A+AS..128..423C} \\
10.20 & -3.50 & 90.0 & Radio &     " \\
10.70 & -5.40 & 120.0 & Radio &     " \\
11.90 & -3.60 & 45.0 & Radio &     " \\
12.70 & -3.90 & 60.0 & Radio &     " \\
32.60 & +7.30 & 240.0 & Radio & \cite{2000A+A...364..552P} \\
6.50 & -12.00 & 480.0 & Radio & \cite{2001A+A...366.1047C} \\
6.80 & -0.20 & 15.0 & Radio & \cite{2000ApJ...540..842Y} \\
43.50 & +0.60 & 20.0 & Radio &  \cite{2002ApJ...566..378K} \\
41.90 & +0.00 & 3.0 & Radio & \cite{2002ApJ...566..378K,2003AcASn..44S.183Z}  \\
47.80 & +2.00 & 6.0 & Radio &       "  \\
5.71 & -0.08 & 10.0 & Radio &  \cite{2006ApJ...639L..25B} \\
6.31 & +0.54 & 6.0 & Radio &       " \\
15.51 & -0.15 & 9.0 & Radio &       " \\
19.13 & +0.90 & 21.0 & Radio &       " \\
6.45 & -0.55 & 3.3 & Radio &  \cite{2006AJ....131.2525H} \\
6.53 & -0.60 & 5.0 & Radio &     " \\
8.85 & -0.25 & 4.0 & Radio &     " \\
10.87 & +0.08 & 2.8 & Radio &     " \\
11.55 & +0.33 & 4.5 & Radio &     " \\
12.91 & -0.28 & 1.5 & Radio &     " \\
13.18 & +0.03 & 2.5 & Radio &     " \\
16.35 & -0.18 & 2.8 & Radio &     " \\
17.33 & -0.13 & 1.8 & Radio &     " \\
18.25 & -0.30 & 3.5 & Radio &     " \\
18.75 & -0.07 & 1.6 & Radio &     " \\
19.46 & +0.14 & 6.0 & Radio &     " \\
19.58 & -0.24 & 3.2 & Radio &     " \\
19.59 & +0.02 & 0.8 & Radio &     " \\
19.61 & -0.12 & 4.5 & Radio &     " \\
19.66 & -0.22 & 4.5 & Radio &     " \\
21.64 & +0.00 & 2.8 & Radio &     " \\
22.38 & +0.10 & 7.0 & Radio &     " \\
22.75 & -0.49 & 3.8 & Radio &     " \\
22.99 & -0.35 & 3.8 & Radio &     " \\
23.56 & -0.03 & 9.0 & Radio &     " \\
24.18 & +0.21 & 5.2 & Radio &     " \\
25.22 & +0.29 & 2.0 & Radio &     " \\
27.13 & +0.03 & 11.0 & Radio &     " \\
28.37 & +0.20 & 10.0 & Radio &     " \\
28.51 & +0.13 & 14.0 & Radio &     " \\
28.55 & -0.00 & 3.0 & Radio &     " \\
28.76 & -0.42 & 9.5 & Radio &     " \\
29.06 & -0.67 & 8.0 & Radio &     " \\
29.07 & +0.45 & 0.7 & Radio &     " \\
29.36 & +0.10 & 9.0 & Radio &     " \\
30.84 & +0.13 & 2.2 & Radio &     " \\
31.05 & +0.48 & 4.5 & Radio &     " \\
31.60 & +0.33 & 3.1 & Radio &     " \\
31.82 & -0.12 & 1.8 & Radio &     " \\
7.50 & -1.70 & 120.0 & Radio &  \cite{2008ApJ...681..320R} \\
51.00 & +0.10 & 13.0 & Radio &   \cite{2014A+A...565A...6S} \\
\hline
32.15 & +0.13 & 0.2 & Optical & \cite{1991PASP..103..487T}  \\
348.10 & -1.80 & 10.0 & Optical & \cite{2008MNRAS.390.1037S} \\
18.70 & -2.20 & 30.0 & Optical &    "           \\
\hline
189.60 & +3.30 & 90.0 & X/g-ray & \cite{1994A+A...284..573A} \\
18.00 & -0.69 & 5.0 & X/g-ray & \cite{1996ApJ...466..938F} \\
0.57 & -0.02 & 0.3 & X/g-ray & \cite{2002ApJ...565.1017S} \\
359.79 & -0.26 & 10.0 & X/g-ray & \cite{2003ANS...324..151S} \\
359.77 & -0.09 & 10.0 & X/g-ray &   "            \\
11.00 & +0.00 & 20.0 & X/g-ray & \cite{2003ApJ...589..253B} \\
25.50 & +0.00 & 12.0 & X/g-ray &    " \\
26.60 & -0.10 & 12.0 & X/g-ray &    " \\
22.00 & +0.00 & 5.0 & X/g-ray & \cite{2006IAUS..230..333U} \\
23.50 & +0.10 & 4.0 & X/g-ray &    " \\
25.00 & +0.00 & 5.0 & X/g-ray &    " \\
26.00 & +0.00 & 5.0 & X/g-ray &    " \\
35.50 & +0.00 & 5.0 & X/g-ray &    " \\
36.00 & +0.00 & 5.0 & X/g-ray &    " \\
37.00 & -0.10 & 5.0 & X/g-ray &    " \\
0.61 & +0.01 & 3.5 & X/g-ray & \cite{2007PASJ...59S.221K}  \\
0.42 & -0.04 & 2.5 & X/g-ray & \cite{2008PASJ...60S.191N}  \\
356.80 & -1.70 & 6.0 & X/g-ray & \cite{2009ApJ...701..811T} \\
359.41 & -0.12 & 4.0 & X/g-ray & \cite{2009PASJ...61S.219T} \\
1.20 & -0.00 & 9.0 & X/g-ray & \cite{2009PASJ...61S.209S} \\
0.13 & -0.12 & 3.0 & X/g-ray & \cite{2013MNRAS.434.1339H} \\
0.22 & -0.03 & 3.3 & X/g-ray & \cite{2015MNRAS.453..172P} \\
0.52 & -0.04 & 3.5 & X/g-ray &     " \\
0.57 & -0.00 & 2.1 & X/g-ray &     " \\
26.40 & -0.10 & 10.0 & X/g-ray & \cite{2015AdSpR..55.2493N} \\
\end{longtable}
\twocolumn

\section{Regions used to determine SNR flux densities}

These plots follow the format of \Fig~\ref{fig:snr_rgb_example}, indicating where the polygons were drawn in \polyflux~ to measure SNRs. The top two panels of each figure show the GLEAM 170--231\,MHz images; the lower two panels show the RGB cube formed from the 72--103\,MHz (R), 103--134\,MHz (G), and 139--170\,MHz (B) images. As described in \Sect~\ref{sec:fluxes}, the annotations on the right two panels consist of: white polygons to indicate the area to be integrated in order to measure the SNR flux density; blue dashed lines to indicate regions excluded from any background measurement; the light shaded area to show the region that is used to measure the background, which is then subtracted from the final flux density measurement. Figures proceed in order of Galactic longitude, first for the outer-Galactic ``oG'' region ($180^\circ < l < 240^\circ$) and then the inner-Galactic ``iG'' region ($345^\circ < l < 60^\circ$). SNRs for which no GLEAM spectra was extracted are excluded from this list.

\begin{figure}
    \centering
    \includegraphics[width=0.5\textwidth]{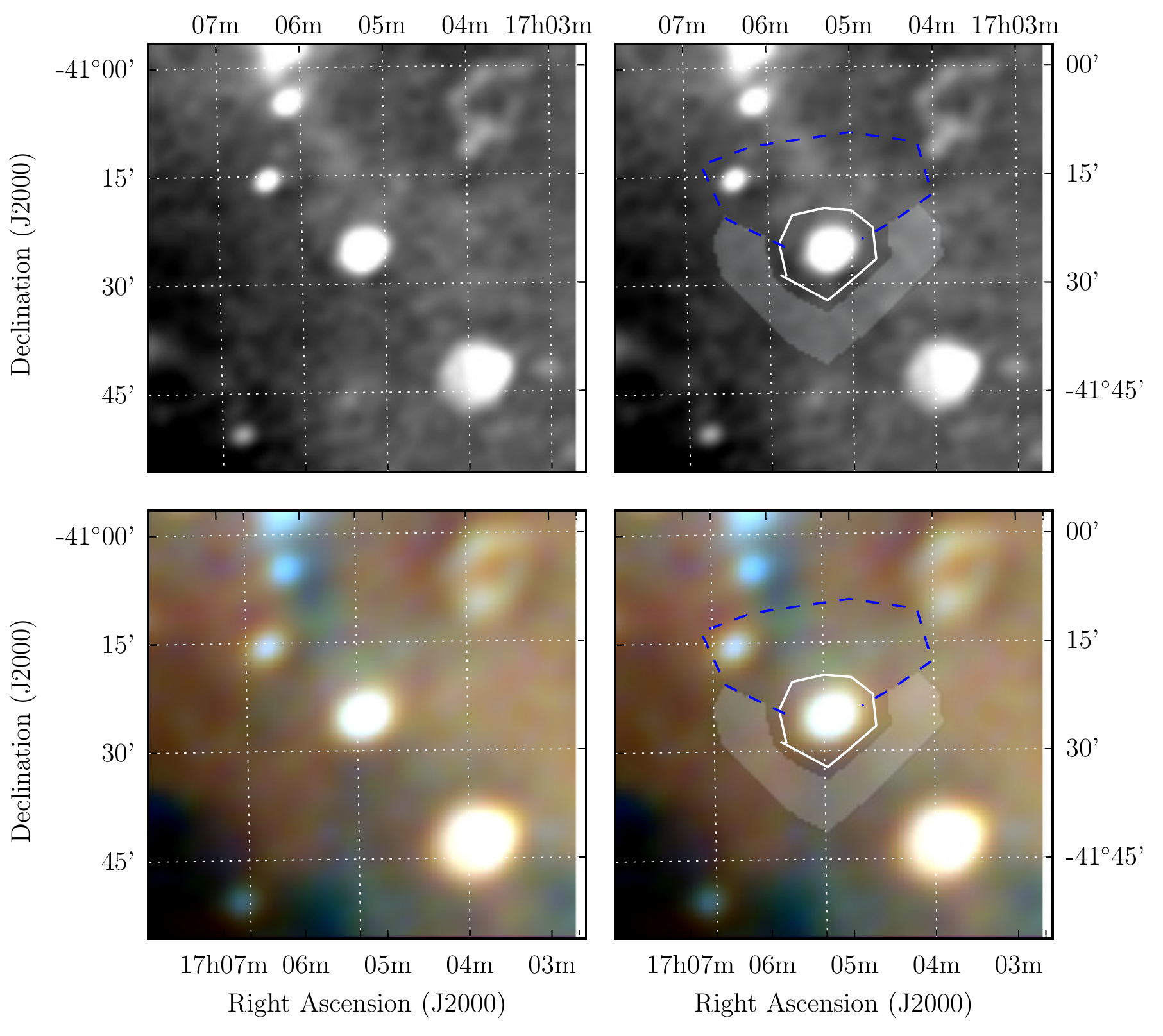}
    \caption{\polysummary G345.1-0.2. \polysuffix}
    \label{fig:SNR_G345.1-0.2_poly}
\end{figure}

\begin{figure}
    \centering
    \includegraphics[width=0.5\textwidth]{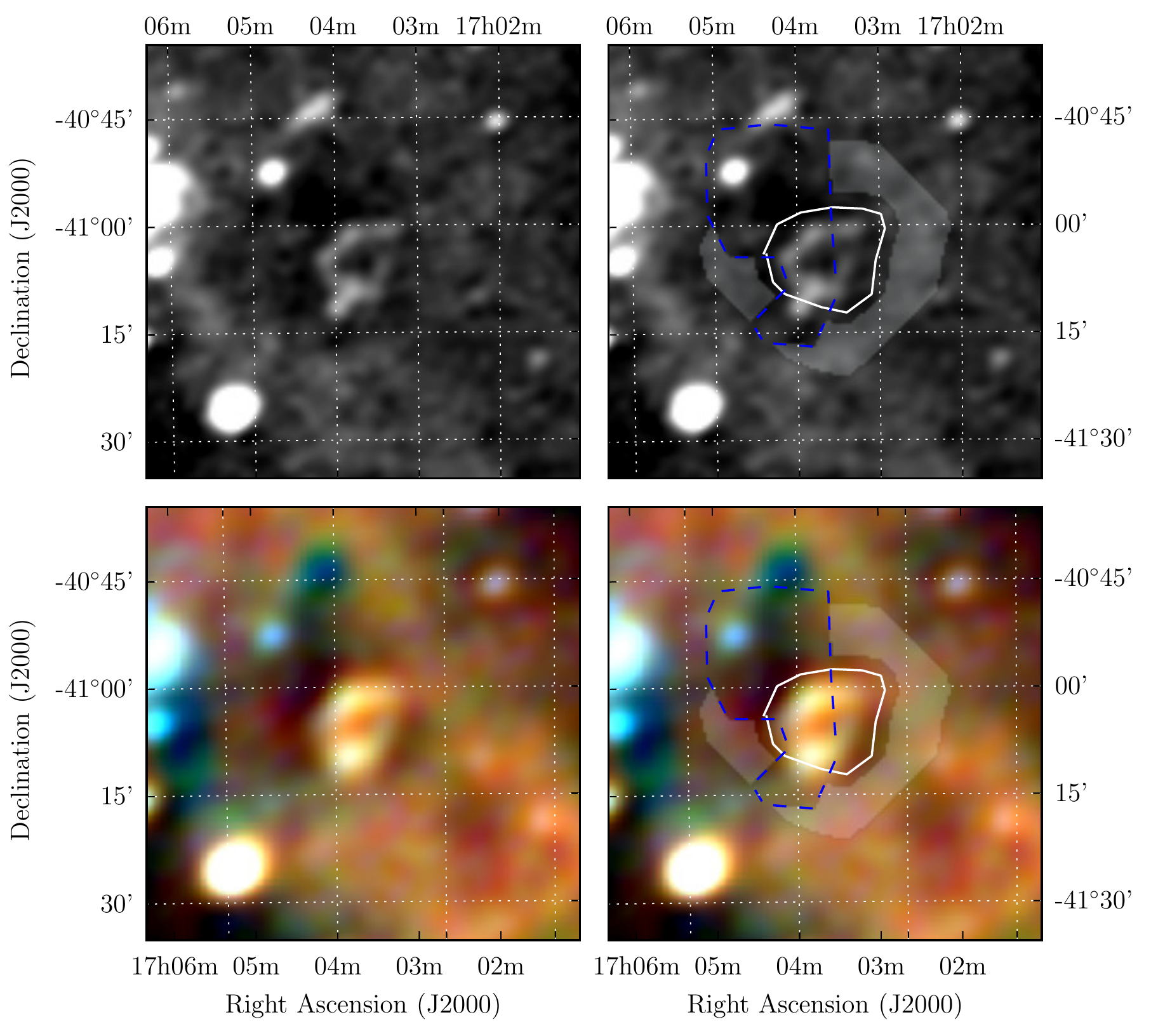}
    \caption{\polysummary G345.1+0.2. \polysuffix}
    \label{fig:SNR_G345.1+0.2_poly}
\end{figure}

\begin{figure}
    \centering
    \includegraphics[width=0.5\textwidth]{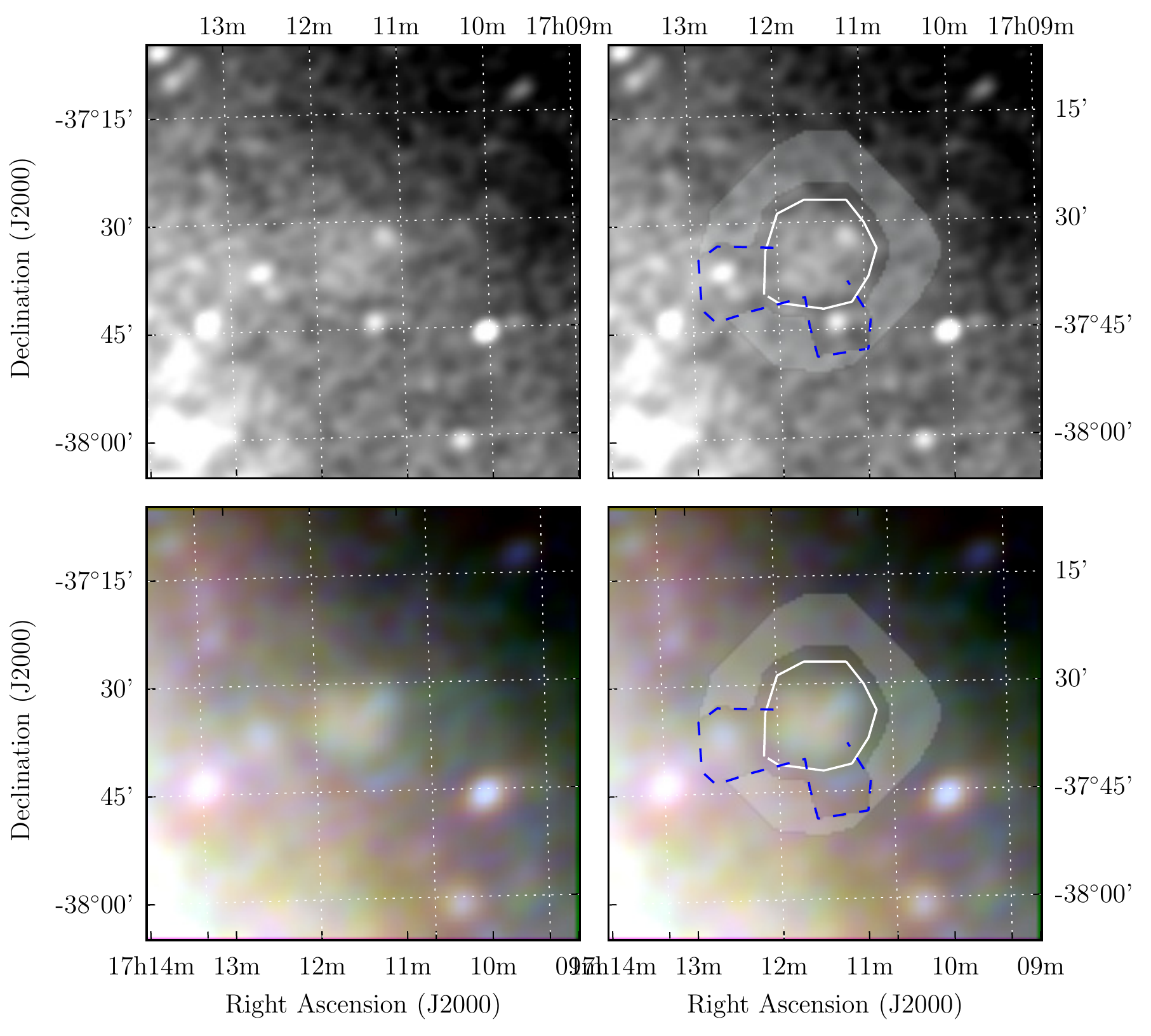}
    \caption{\polysummary G348.8+1.1. \polysuffix}
\end{figure}

\begin{figure}
    \centering
    \includegraphics[width=0.5\textwidth]{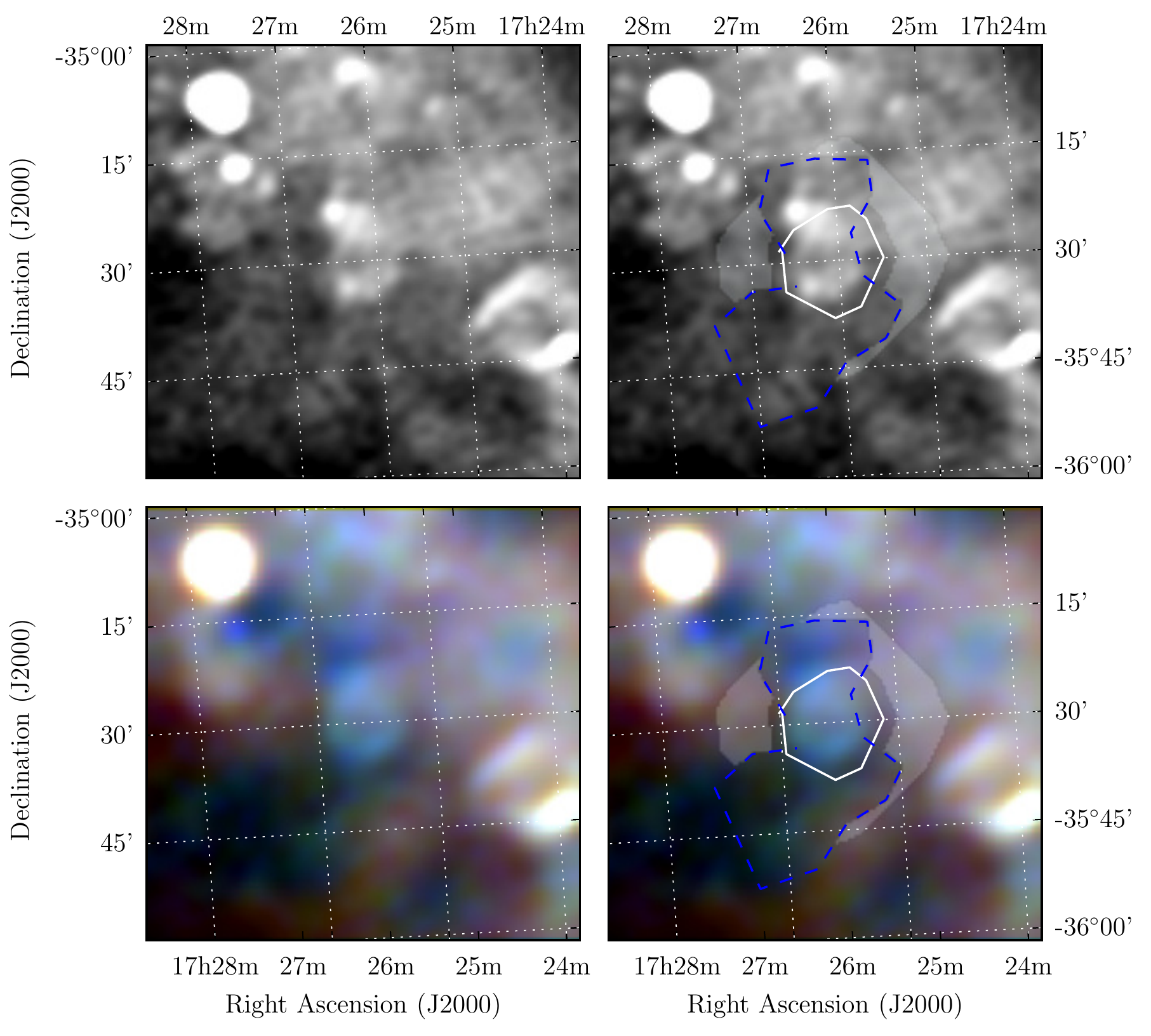}
    \caption{\polysummary G352.2-0.1. \polysuffix}
\end{figure}

\begin{figure}
    \centering
    \includegraphics[width=0.5\textwidth]{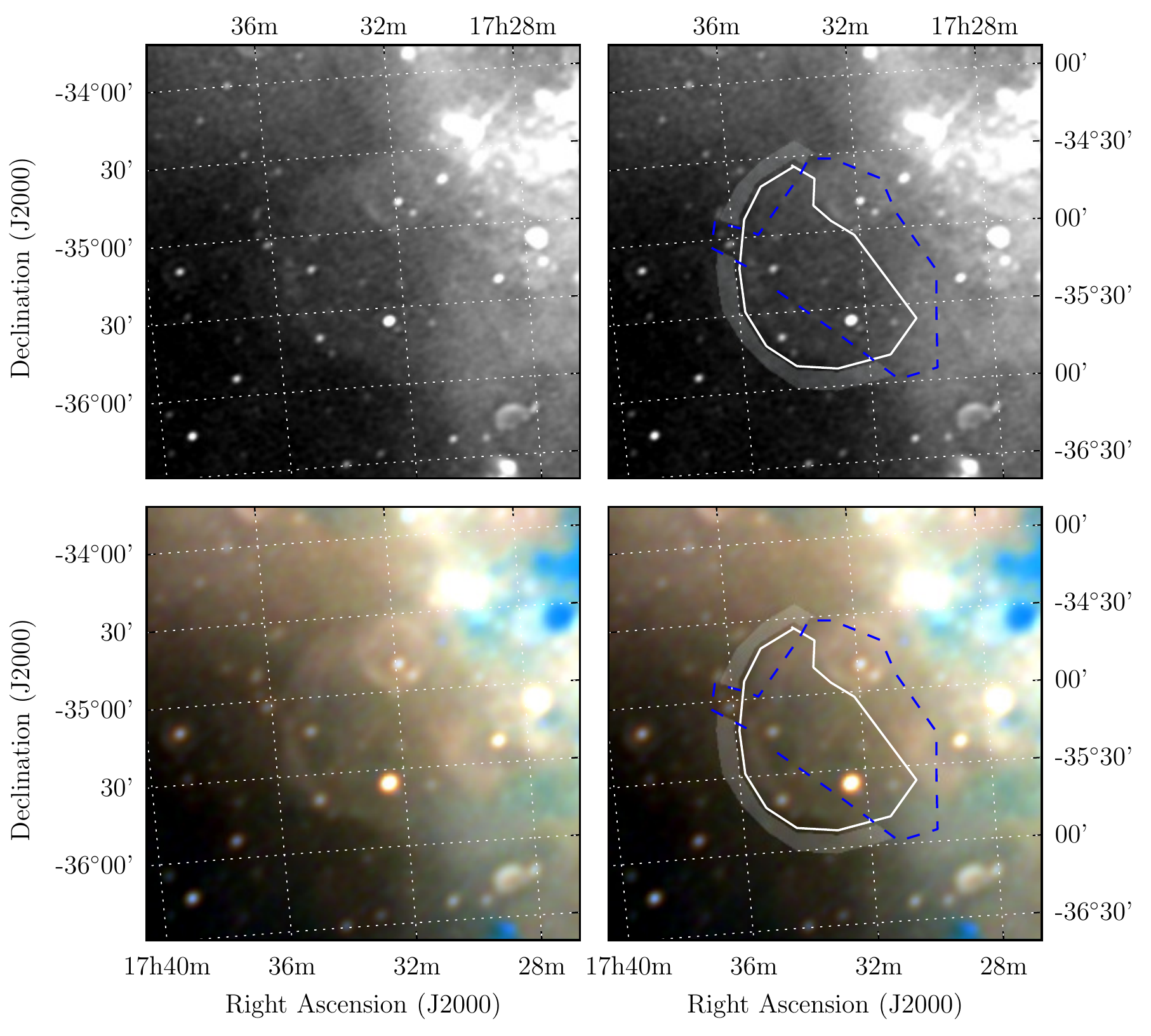}
    \caption{\polysummary G353.3-1.1. \polysuffix}
    \label{fig:SNR_G353.3-1.1_poly}
\end{figure}

\begin{figure}
    \centering
    \includegraphics[width=0.5\textwidth]{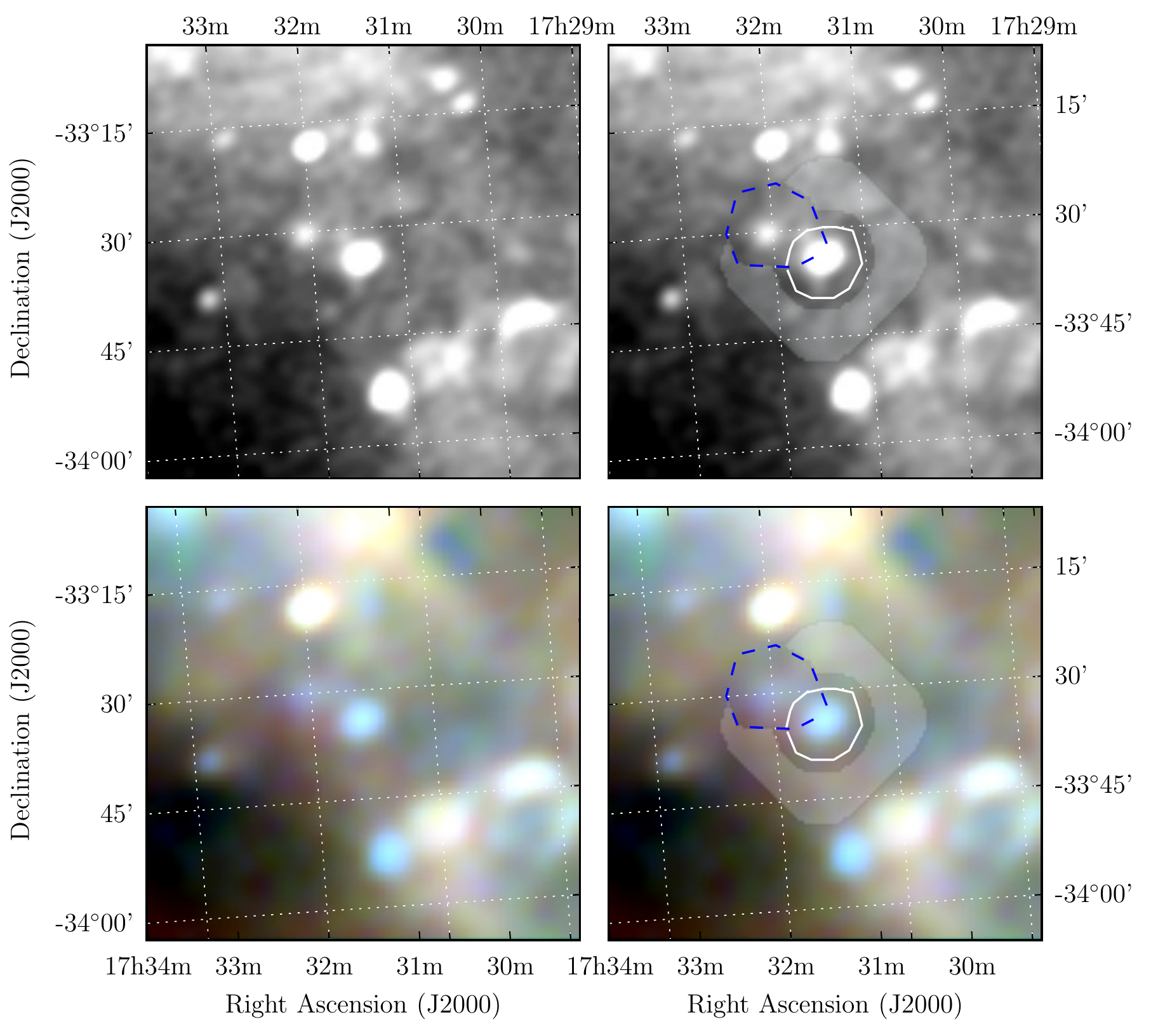}
    \caption{\polysummary G354.46+0.07. \polysuffix}
\end{figure}

\begin{figure}
    \centering
    \includegraphics[width=0.5\textwidth]{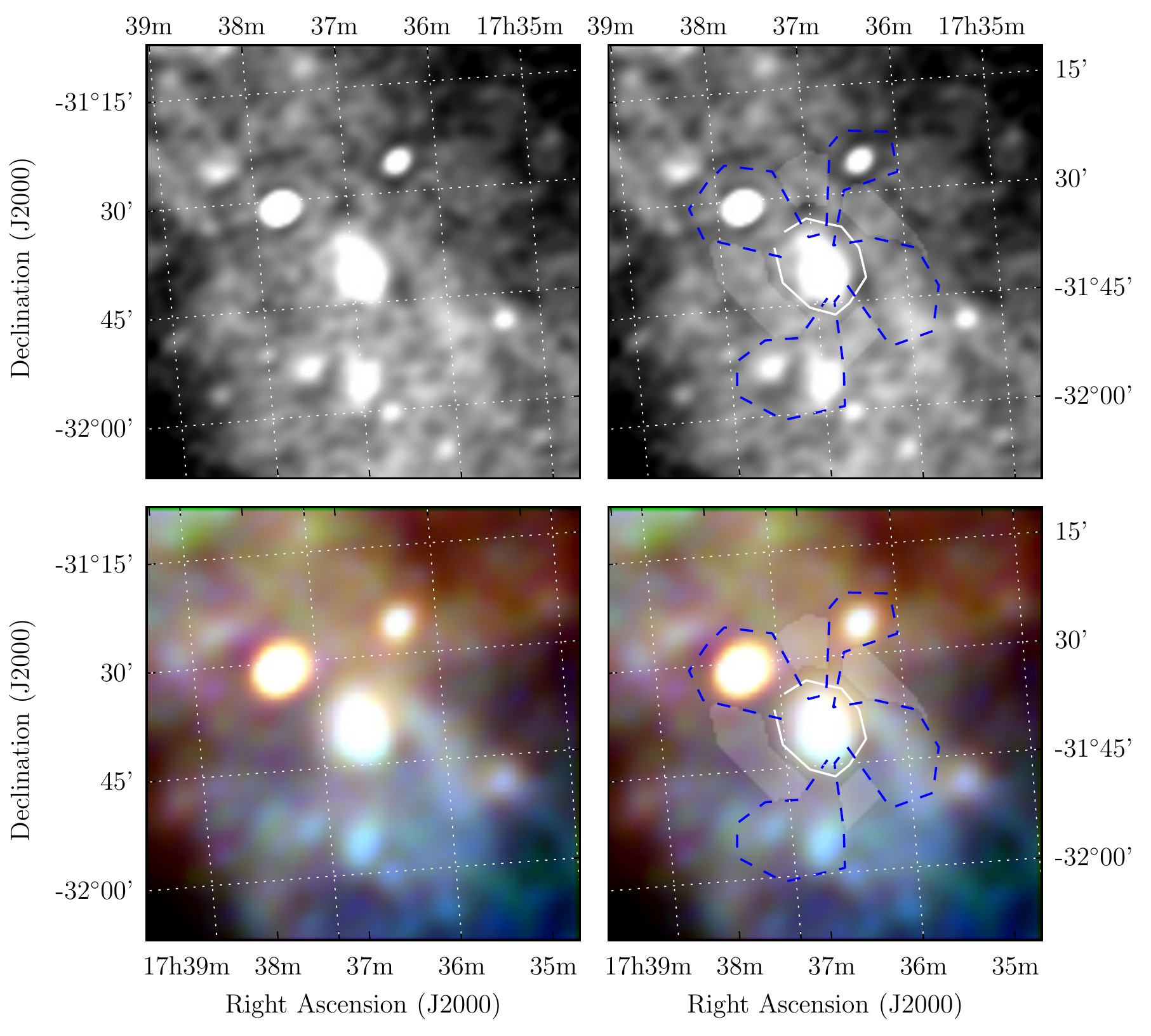}
    \caption{\polysummary G356.6+00.1. \polysuffix}
\end{figure}

\begin{figure}
    \centering
    \includegraphics[width=0.5\textwidth]{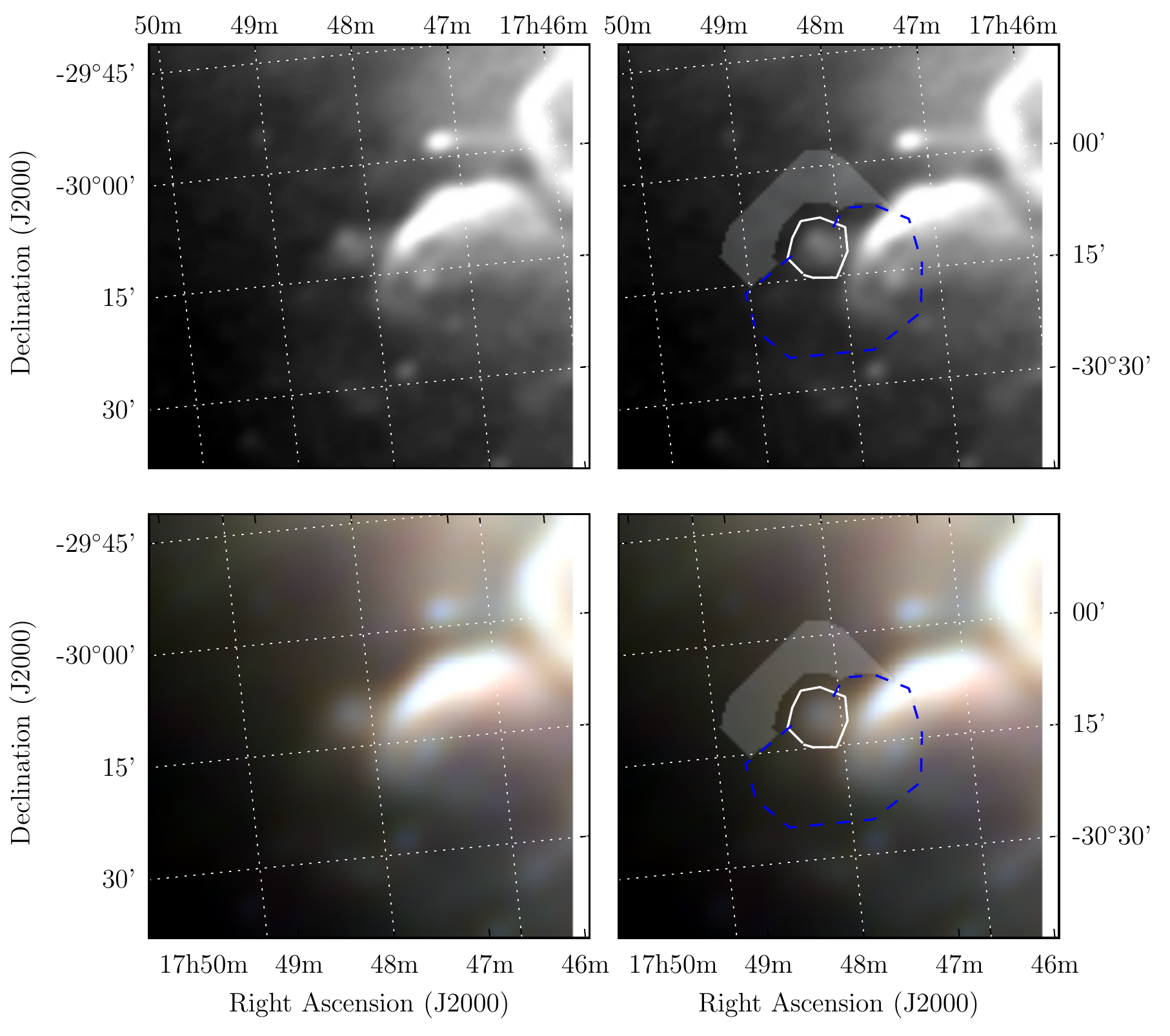}
    \caption{\polysummary G359.2-01.1. \polysuffix}
\end{figure}

\begin{figure}
    \centering
    \includegraphics[width=0.5\textwidth]{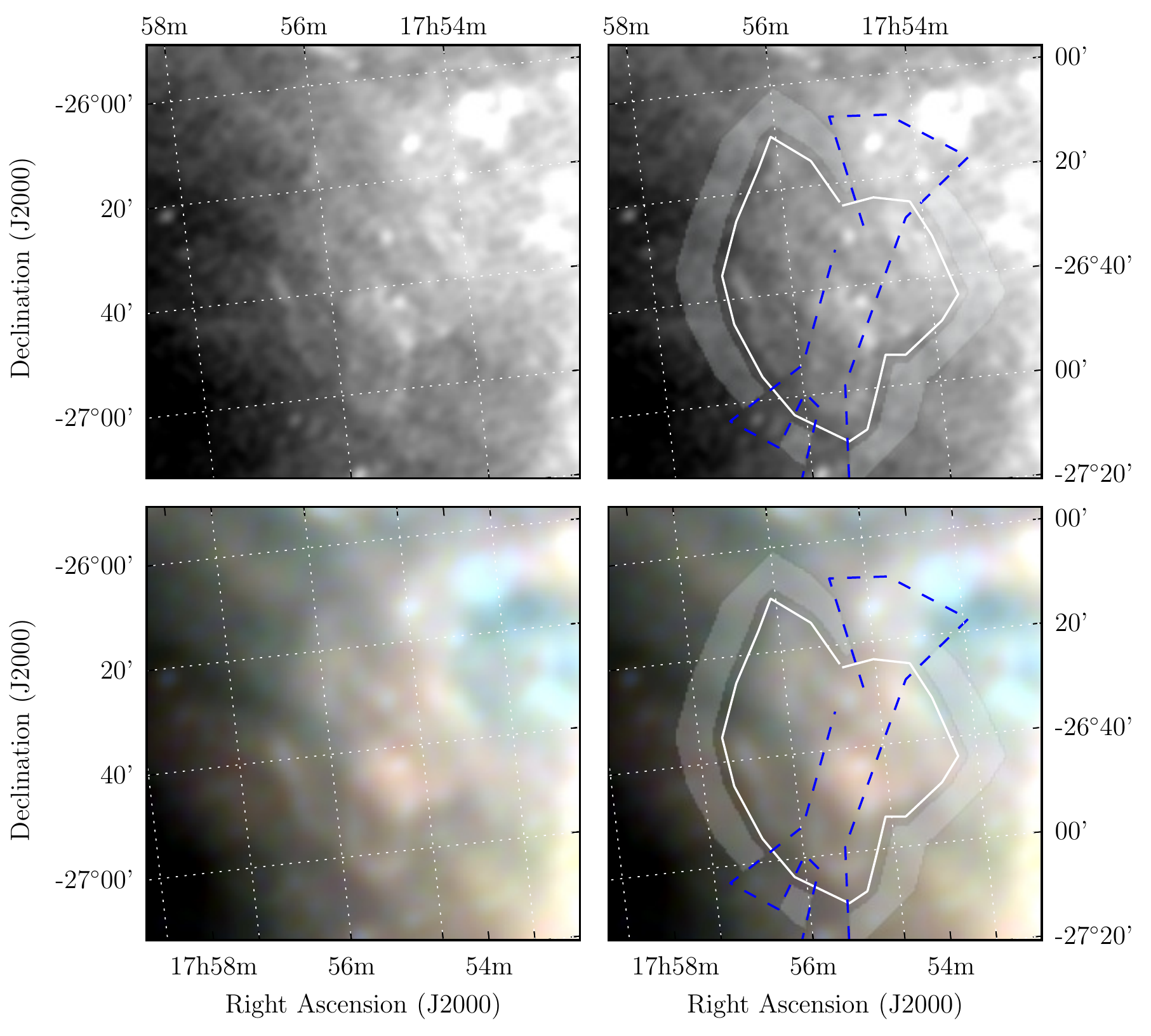}
    \caption{\polysummary G3.1-0.7. \polysuffix}
\end{figure}

\begin{figure}
    \centering
    \includegraphics[width=0.5\textwidth]{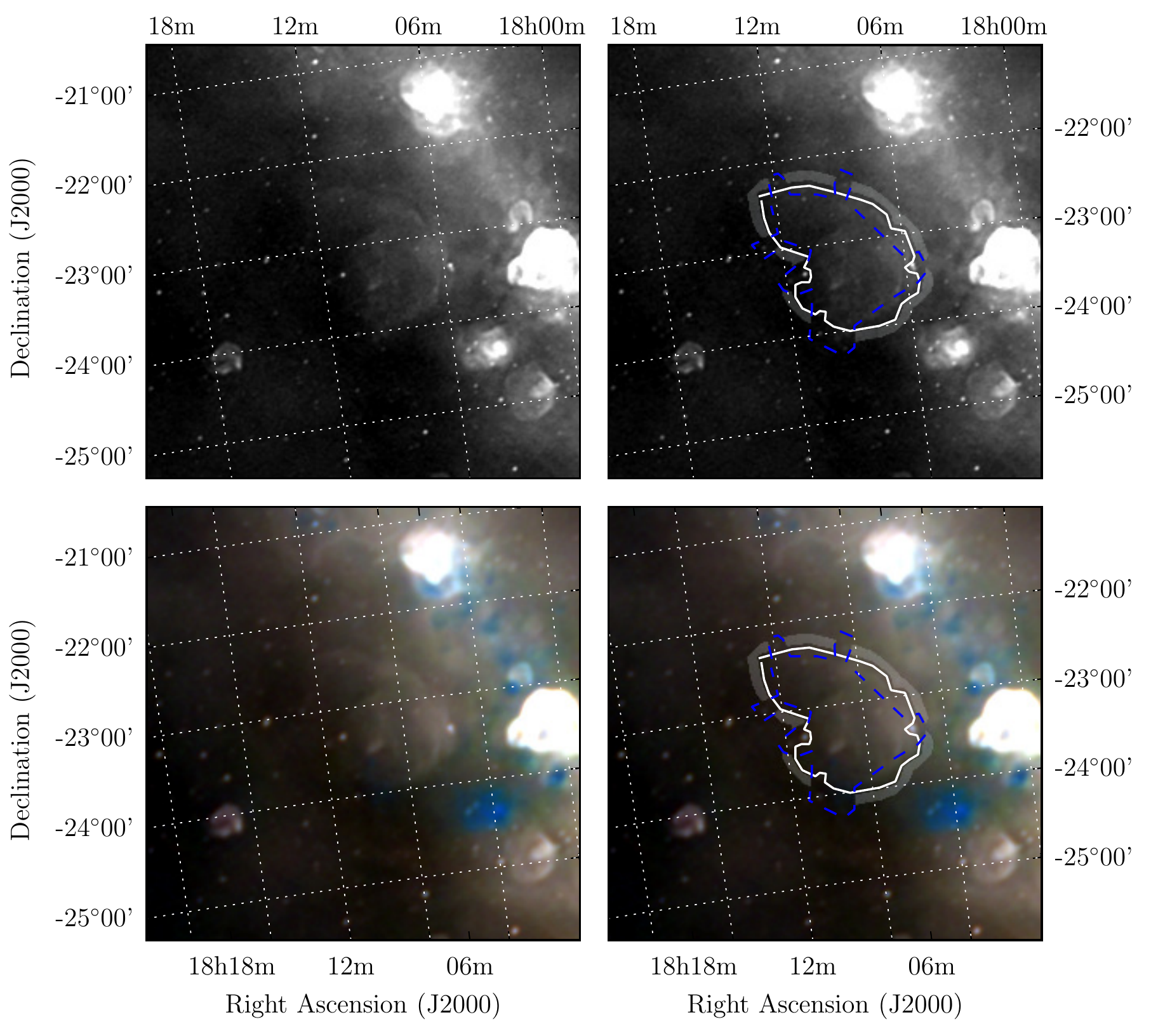}
    \caption{\polysummary G7.5-1.7. \polysuffix}
\end{figure}

\begin{figure}
    \centering
    \includegraphics[width=0.5\textwidth]{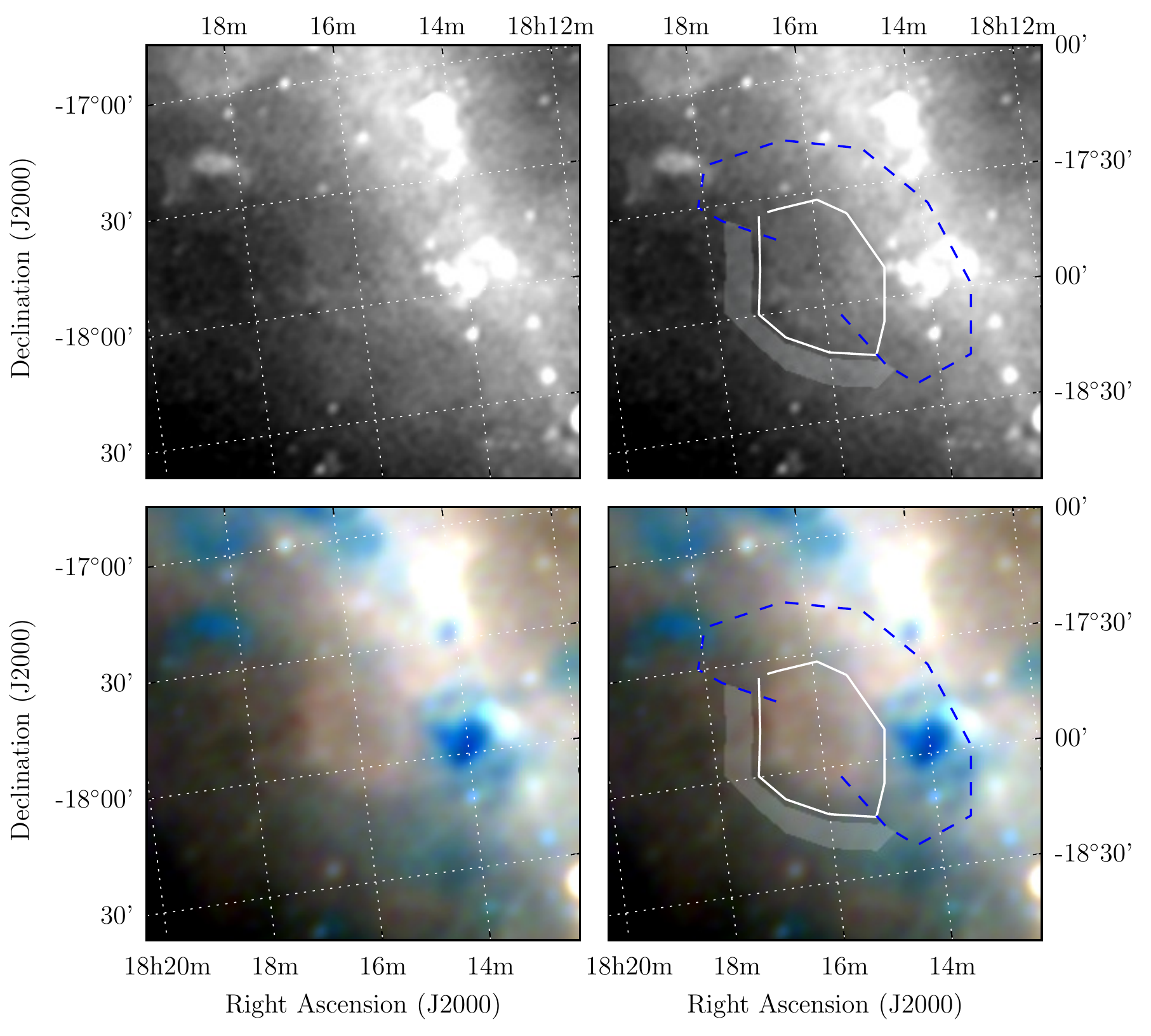}
    \caption{\polysummary G13.1-0.5. \polysuffix}
\end{figure}	
		
\begin{figure}
    \centering
    \includegraphics[width=0.5\textwidth]{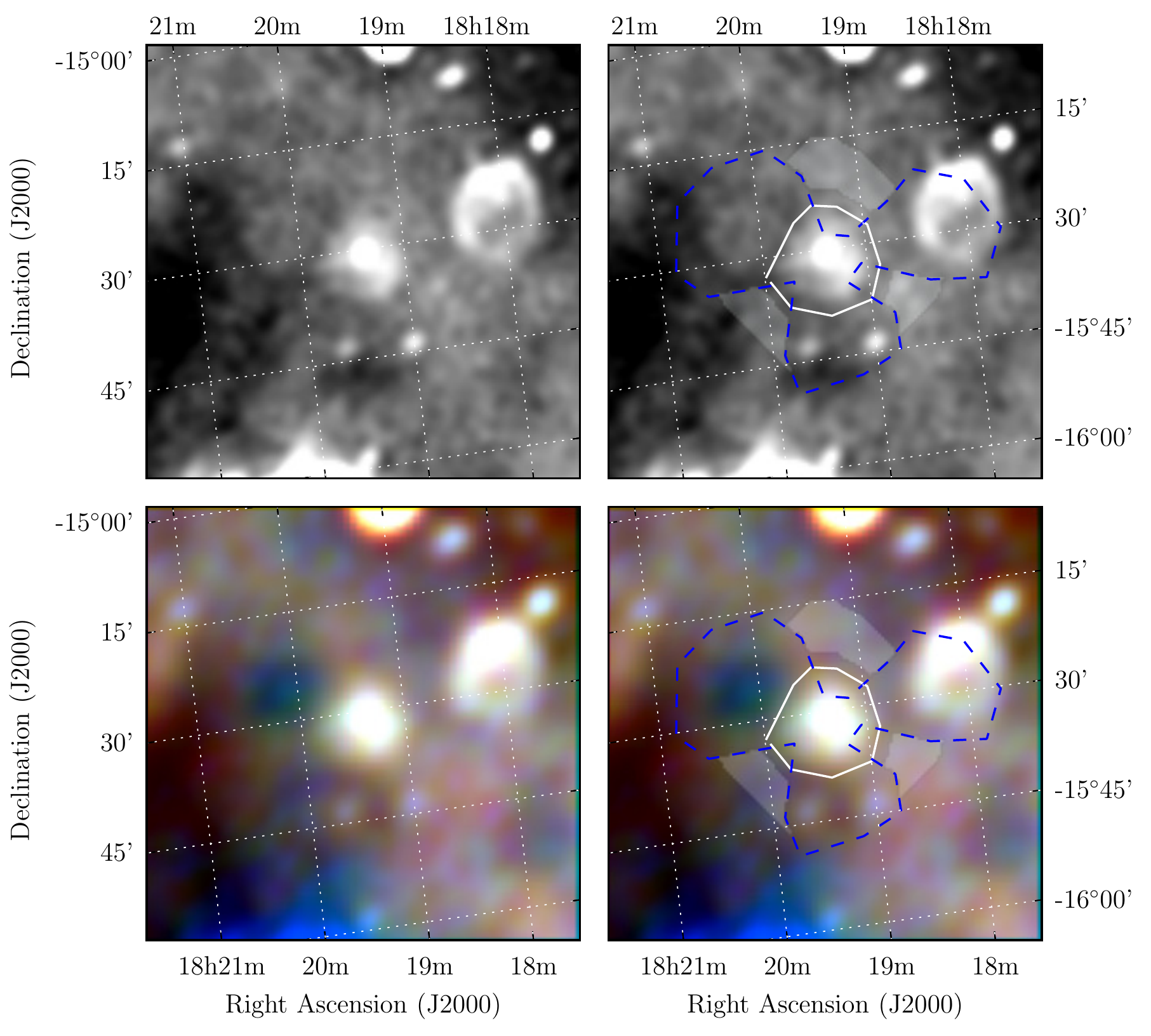}
    \caption{\polysummary G15.51-0.15. \polysuffix}
\end{figure}

\begin{figure}
    \centering
    \includegraphics[width=0.5\textwidth]{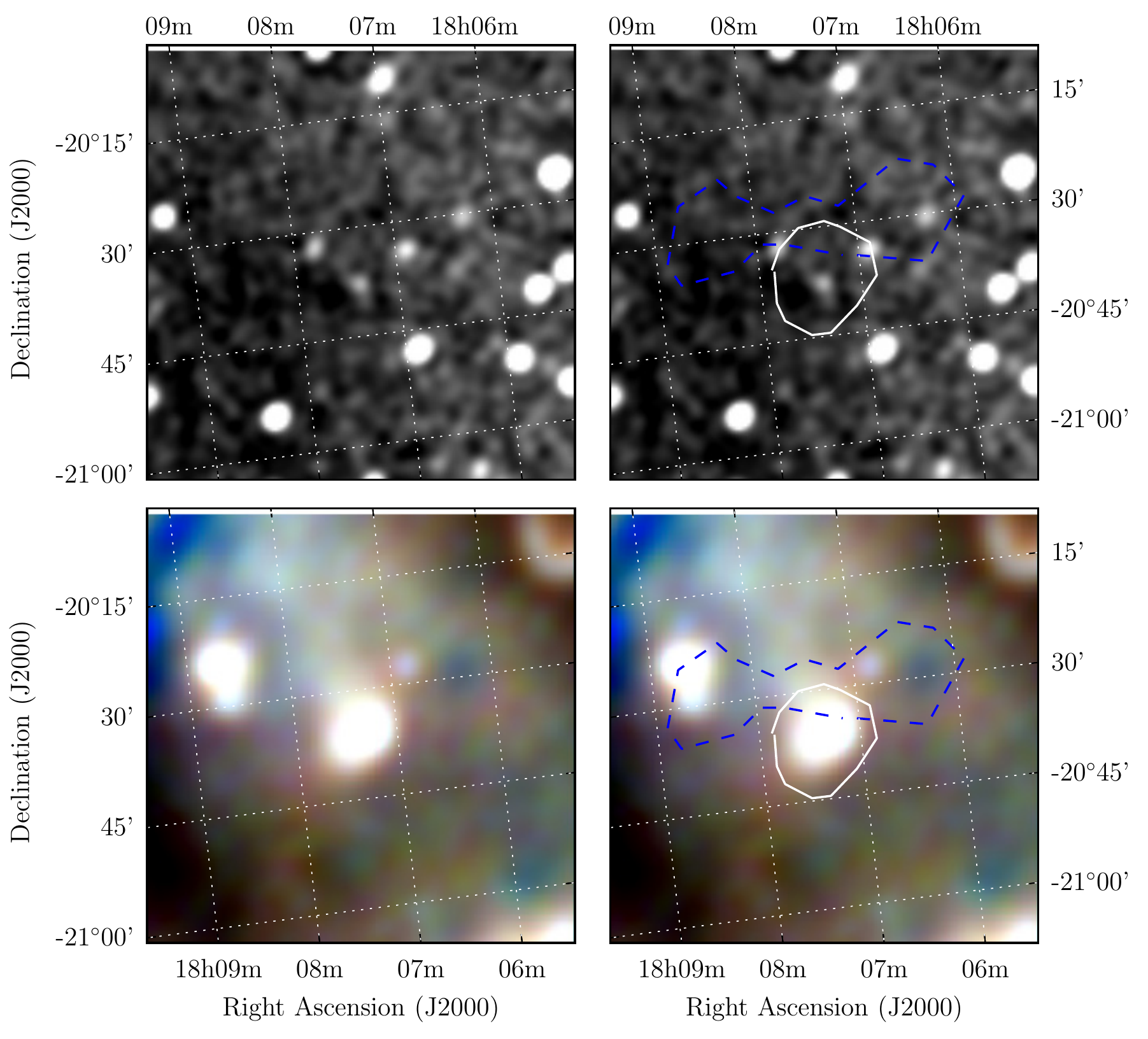}
    \caption{\polysummary MAGPIS\,$9.6833-0.0667$. \polysuffix}
\end{figure}

\begin{figure}
    \centering
    \includegraphics[width=0.5\textwidth]{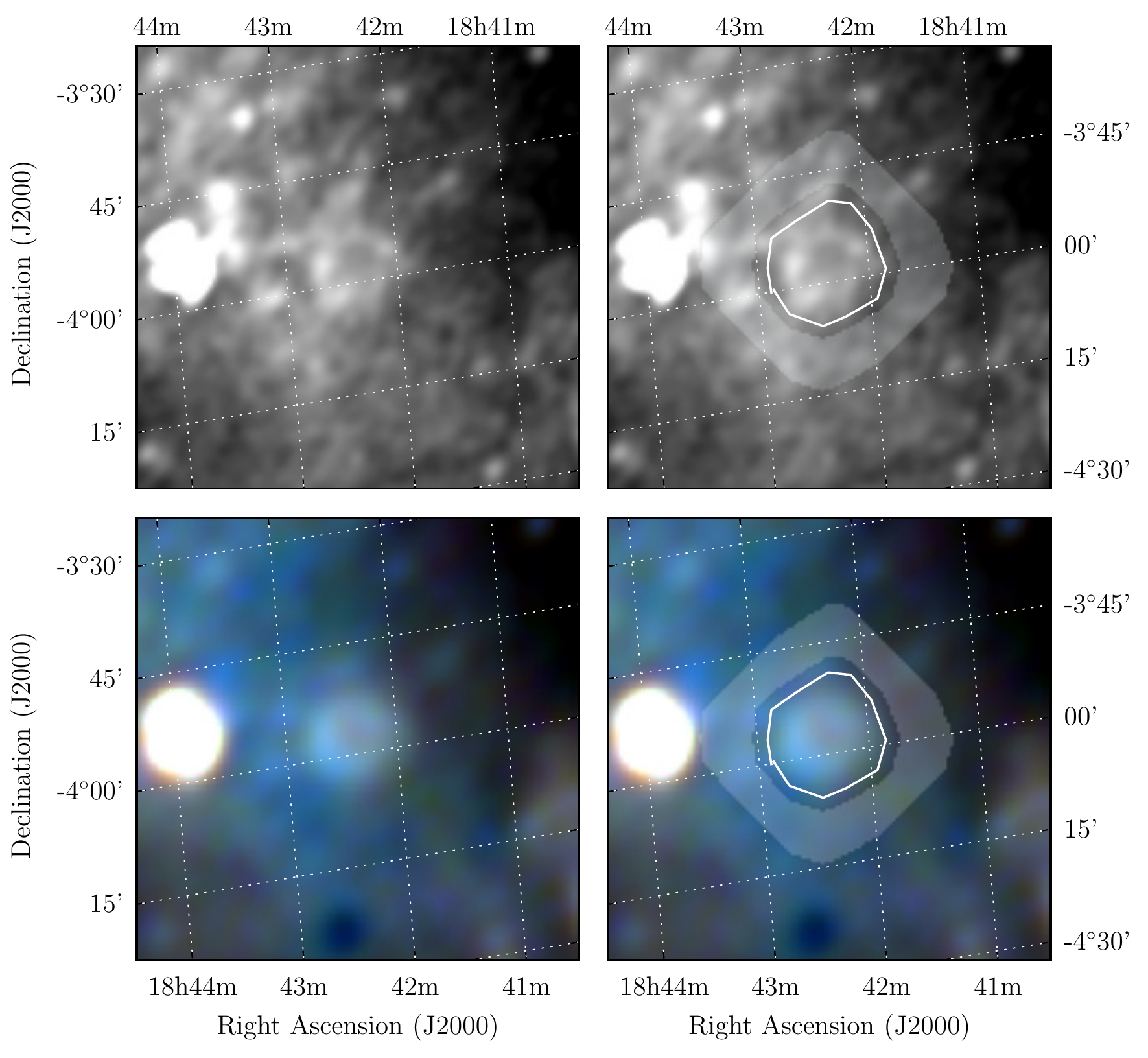}
    \caption{\polysummary MAGPIS\,$28.3750+0.2028$. \polysuffix}
\end{figure}

\begin{figure}
    \centering
    \includegraphics[width=0.5\textwidth]{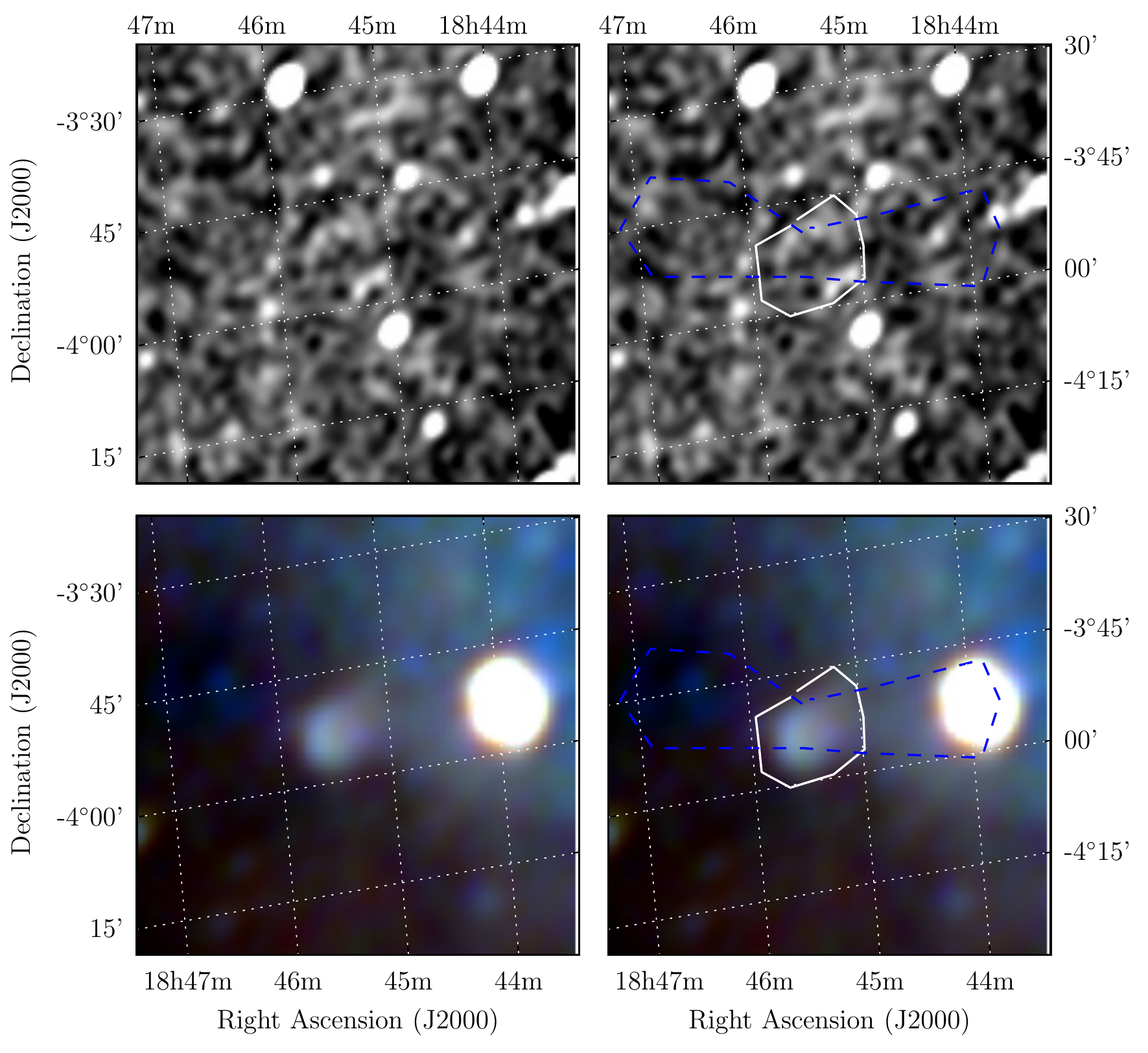}
    \caption{\polysummary MAGPIS\,$28.7667-0.4250$. \polysuffix}
\end{figure}

\begin{figure}
    \centering
    \includegraphics[width=0.5\textwidth]{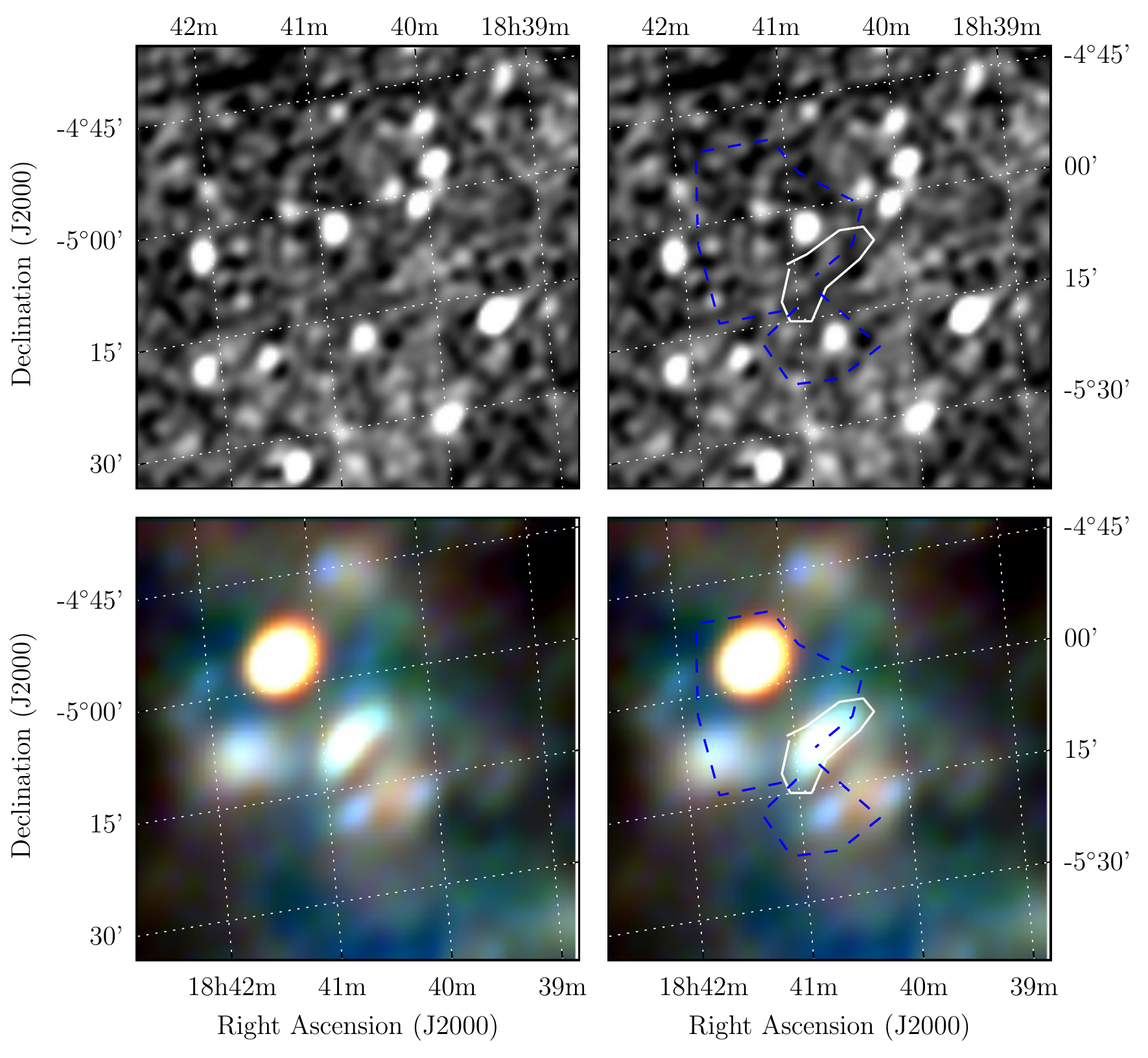}
    \caption{\polysummary MAGPIS\,$27.1333+0.0333$. \polysuffix}
\end{figure}

\begin{figure}
    \centering
    \includegraphics[width=0.5\textwidth]{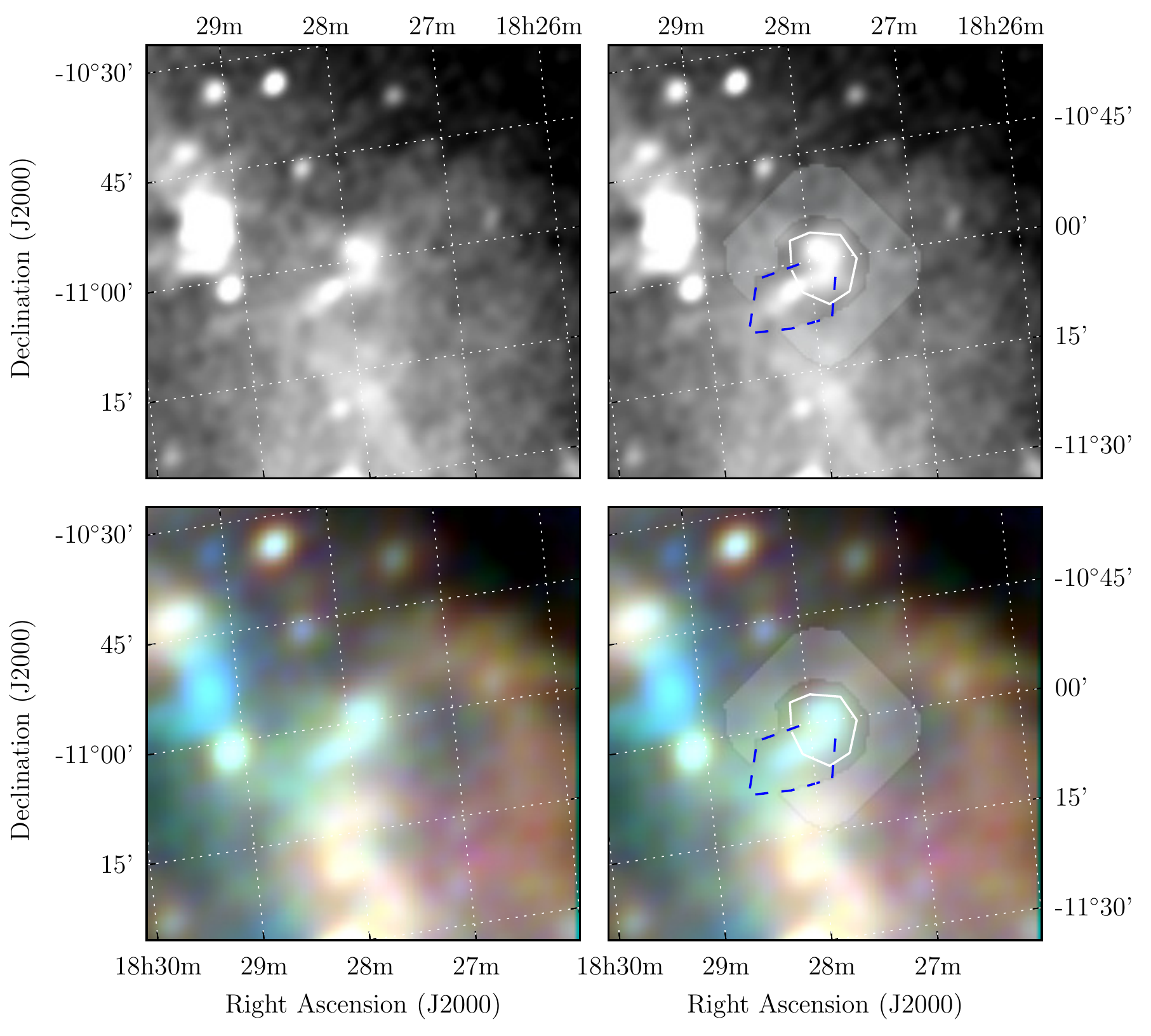}
    \caption{\polysummary MAGPIS\,$20.4667+0.1500$. \polysuffix}
\end{figure}

\section{Spectra}

The spectra of the measured SNR using the backgrounding and flux summing technique described in \Sect~\ref{sec:fluxes}. The left panels show flux density against frequency with linear axes while the right panels show the same data in log. (It is useful to include both when analysing the data as a log plot does not render negative data points, which occur for faint SNRs). The black points show the (background-subtracted) SNR flux density measurements, the red points show the measured background, and the blue curve shows a linear fit to the log-log data (i.e. $S_\nu \propto \nu^\alpha$). The fitted value of $\alpha$ is shown at the top right of each plot.

\begin{figure}
    \centering
    \includegraphics[width=0.5\textwidth]{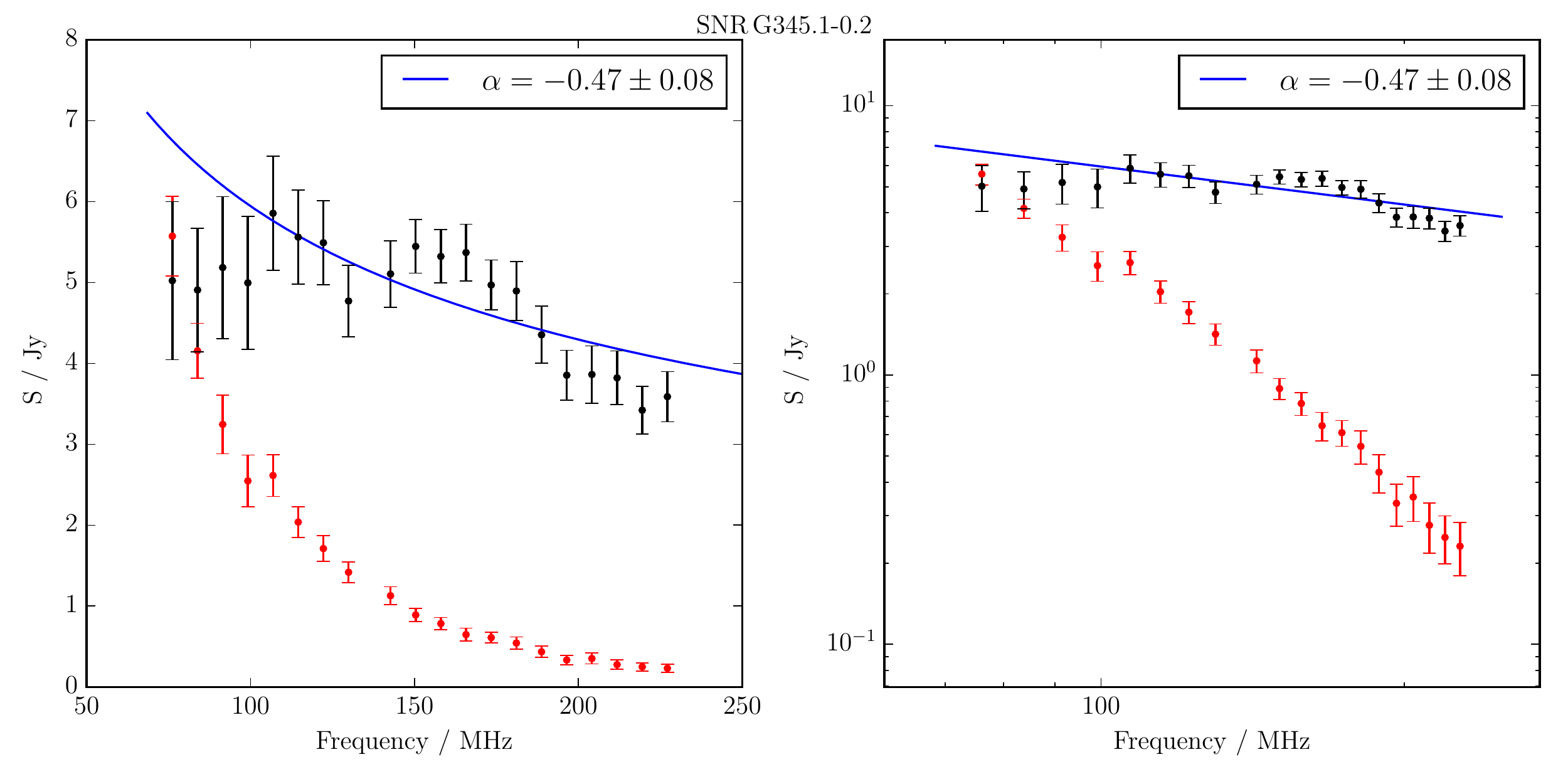}
    \caption{\spectrasummary G345.1-0.2. \spectrasuffix}
    \label{fig:SNR_G345.1-0.2_spectrum}
\end{figure}

\begin{figure}
    \centering
    \includegraphics[width=0.5\textwidth]{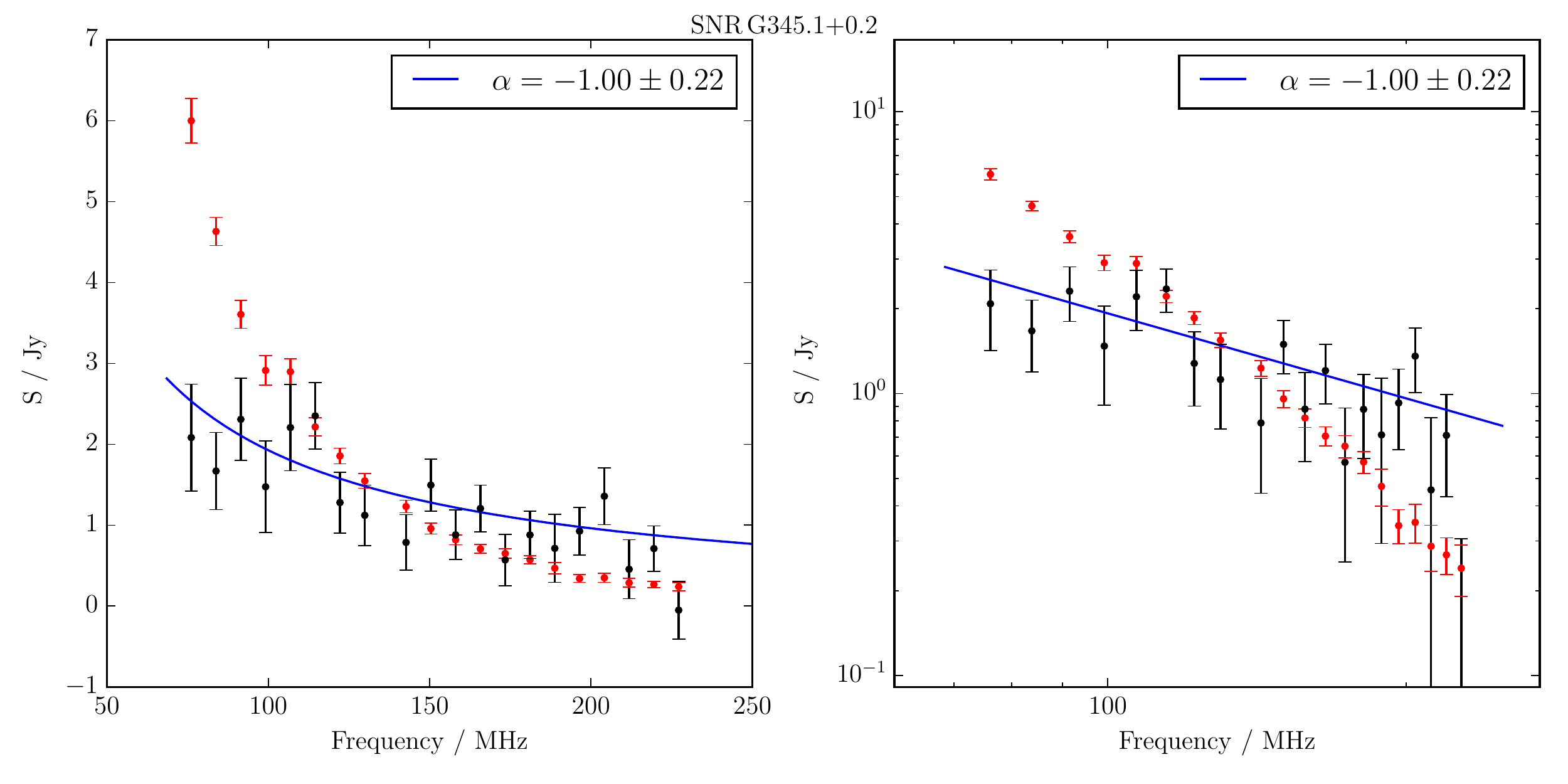}
    \caption{\spectrasummary G345.1+0.2. \spectrasuffix}
    \label{fig:SNR_G345.1+0.2_spectrum}
\end{figure}

\begin{figure}
    \centering
    \includegraphics[width=0.5\textwidth]{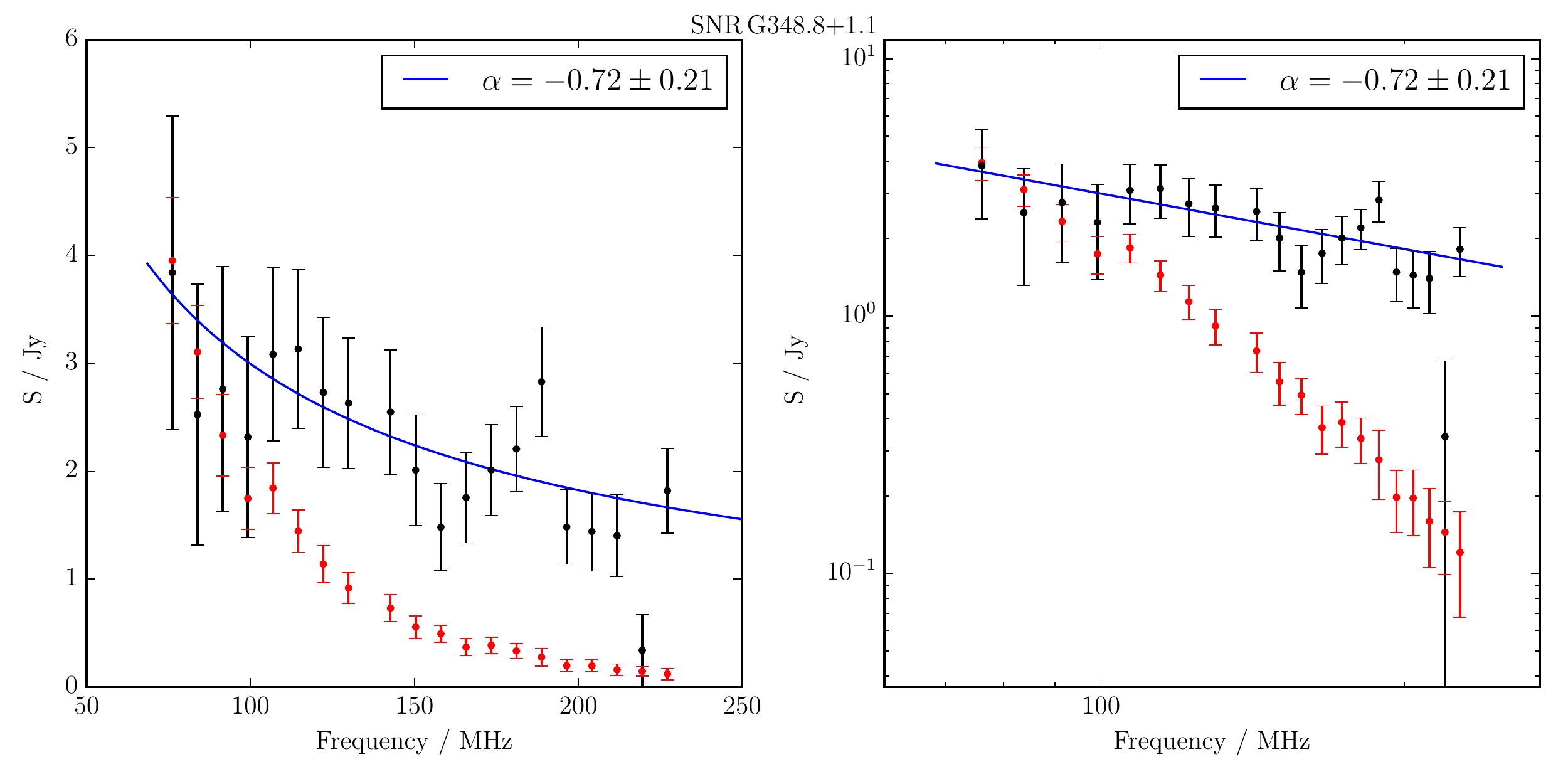}
    \caption{\spectrasummary G348.8+1.1. \spectrasuffix}
\end{figure}

\begin{figure}
    \centering
    \includegraphics[width=0.5\textwidth]{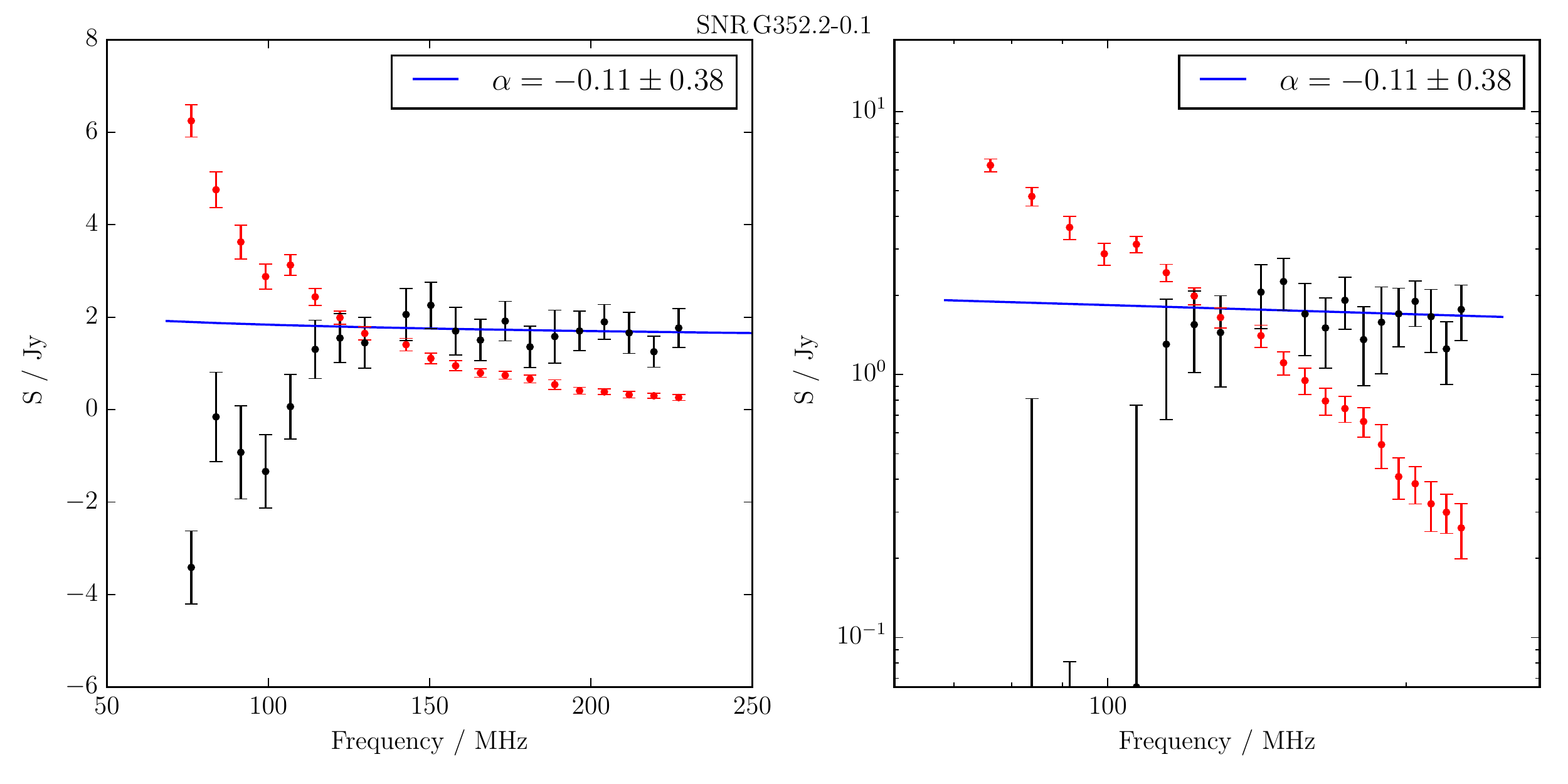}
    \caption{\spectrasummary G352.2-0.1. \spectrasuffix}
    \label{fig:SNR_G352.2-0.1_spectrum}
\end{figure}

\begin{figure}
    \centering
    \includegraphics[width=0.5\textwidth]{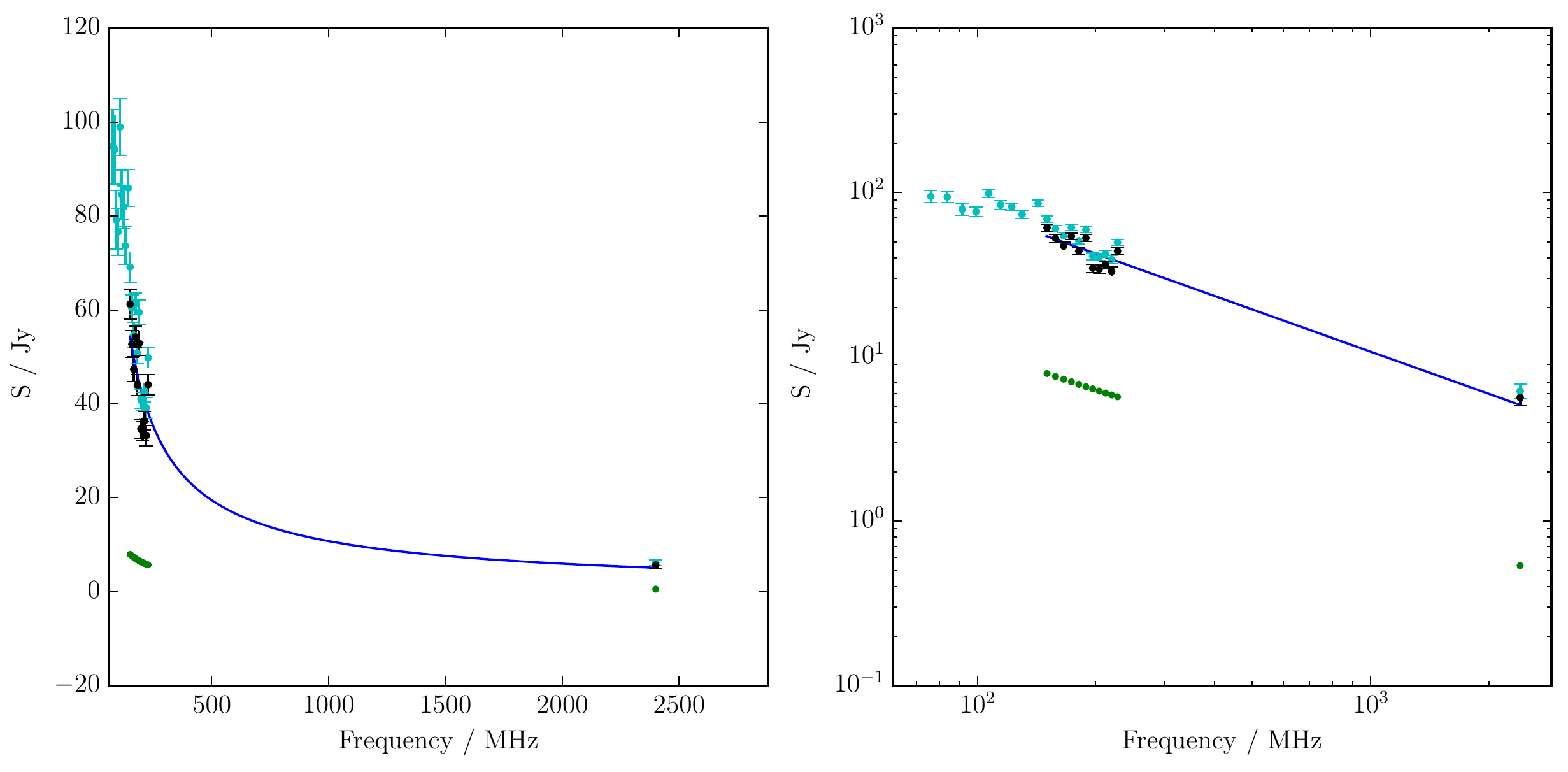}
    \caption{Spectral fit to the GLEAM and Parkes 2.4\,GHz data integrated flux densities for SNR\,G$353.3-1.1$. Raw flux densities are shown with grey points; measured and extrapolated compact source flux densities are shown by green points, and the source-subtracted data (for $\nu>150$\,MHz) used to fit the spectrum of the SNR are shown in black. As per the other plots, the fit is shown with a blue line. \spectrasuffix}
     \label{fig:SNRG353.3-1.1_spectrum}
\end{figure}

\begin{figure}
    \centering
    \includegraphics[width=0.5\textwidth]{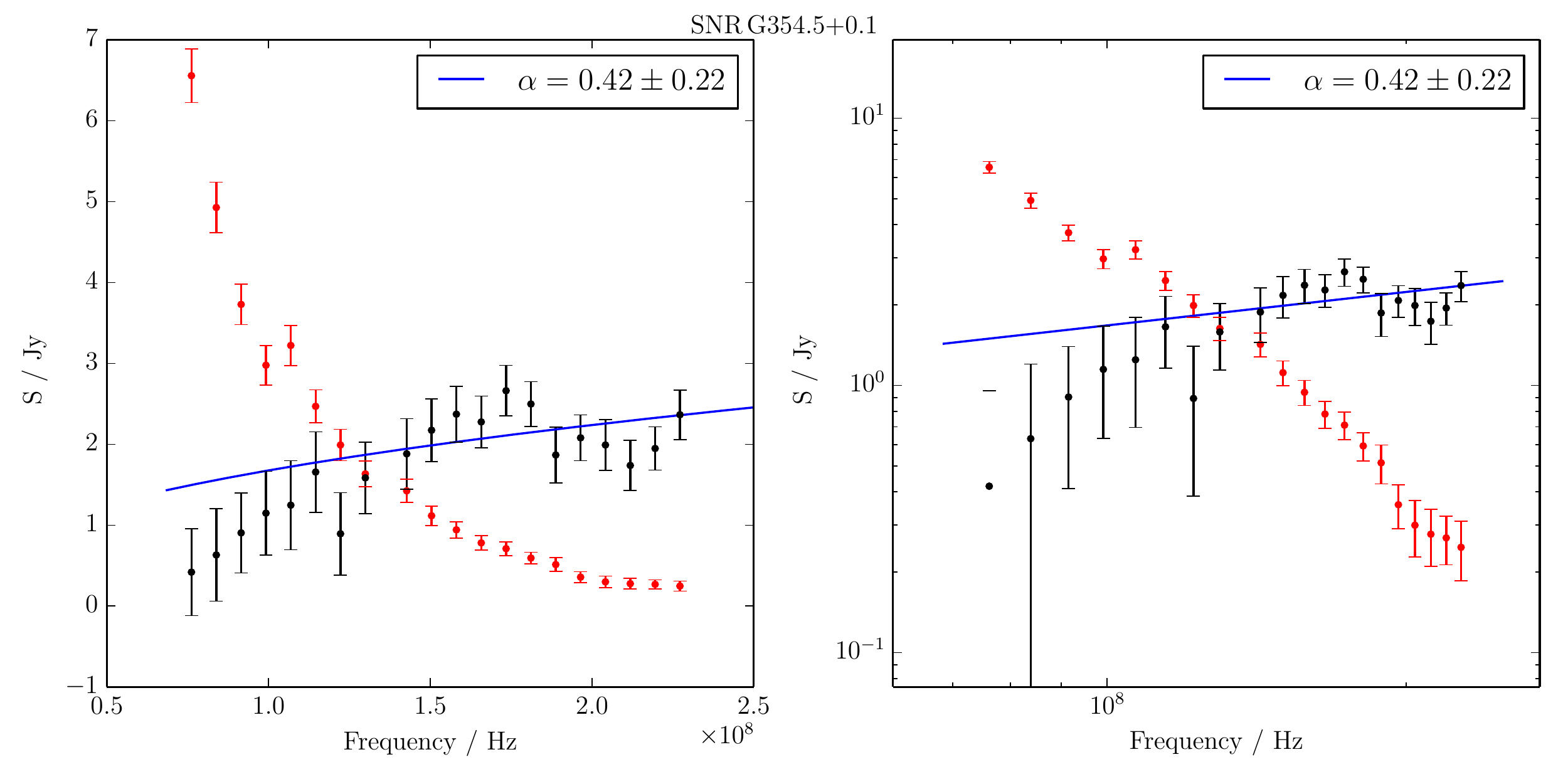}
    \caption{\spectrasummary G354.46+0.07. \spectrasuffix}
    \label{fig:SNR_G354.5+0.1_spectrum}
\end{figure}

\begin{figure}
    \centering
    \includegraphics[width=0.5\textwidth]{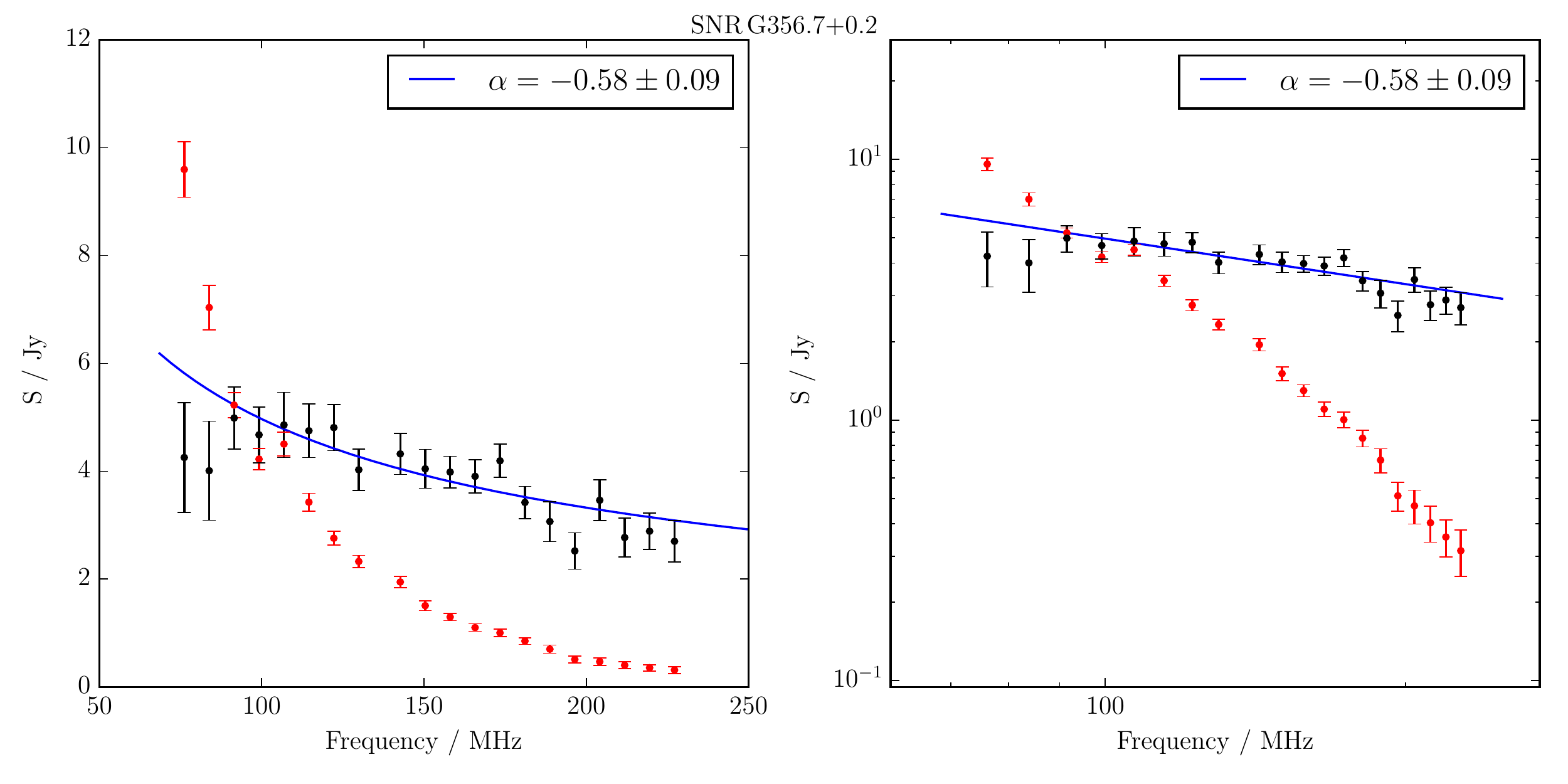}
    \caption{\spectrasummary G356.6+00.1. \spectrasuffix}
\end{figure}

\begin{figure}
    \centering
    \includegraphics[width=0.5\textwidth]{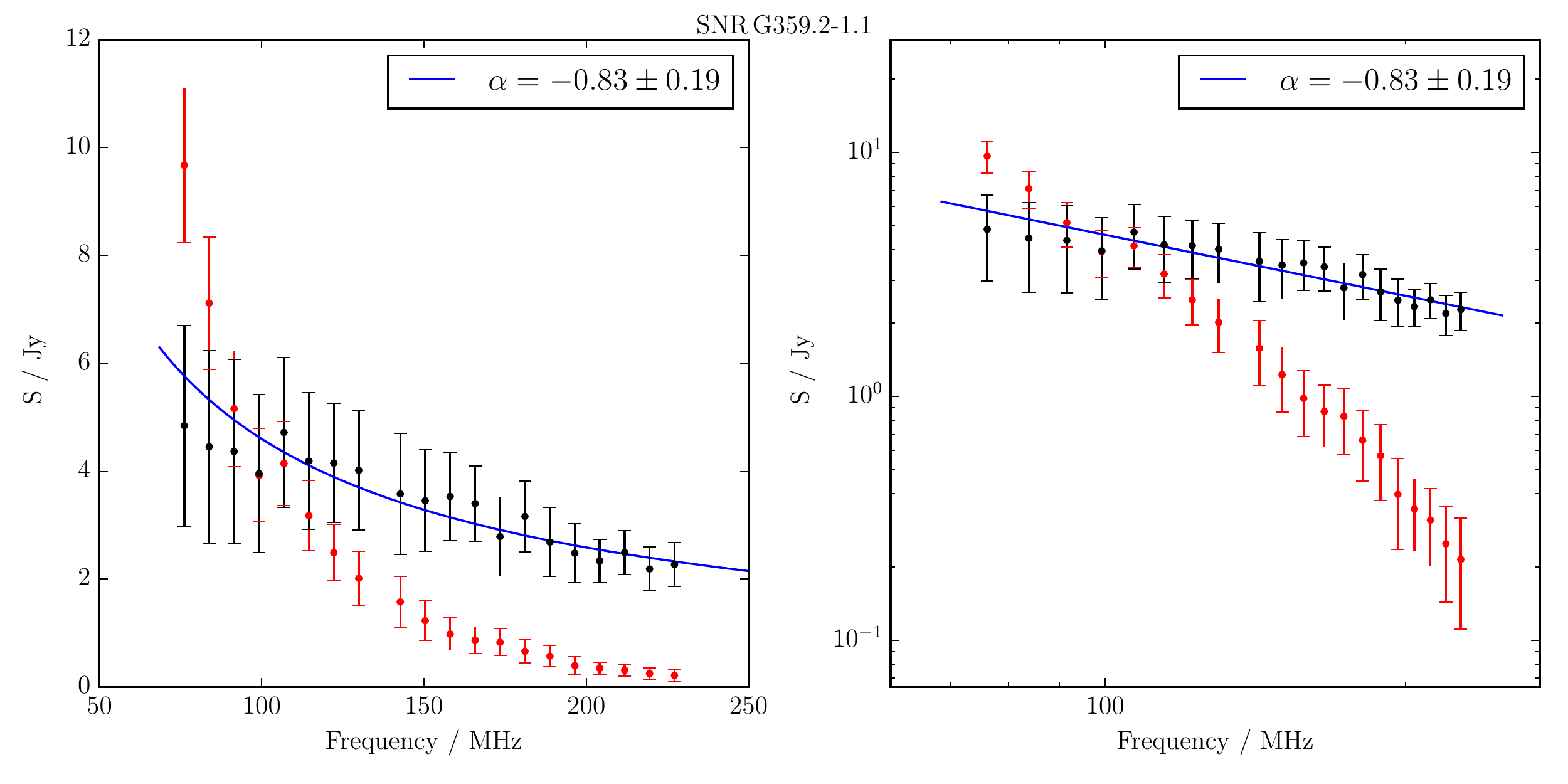}
    \caption{\spectrasummary G359.2-01.1. \spectrasuffix}
\end{figure}

\begin{figure}
    \centering
    \includegraphics[width=0.5\textwidth]{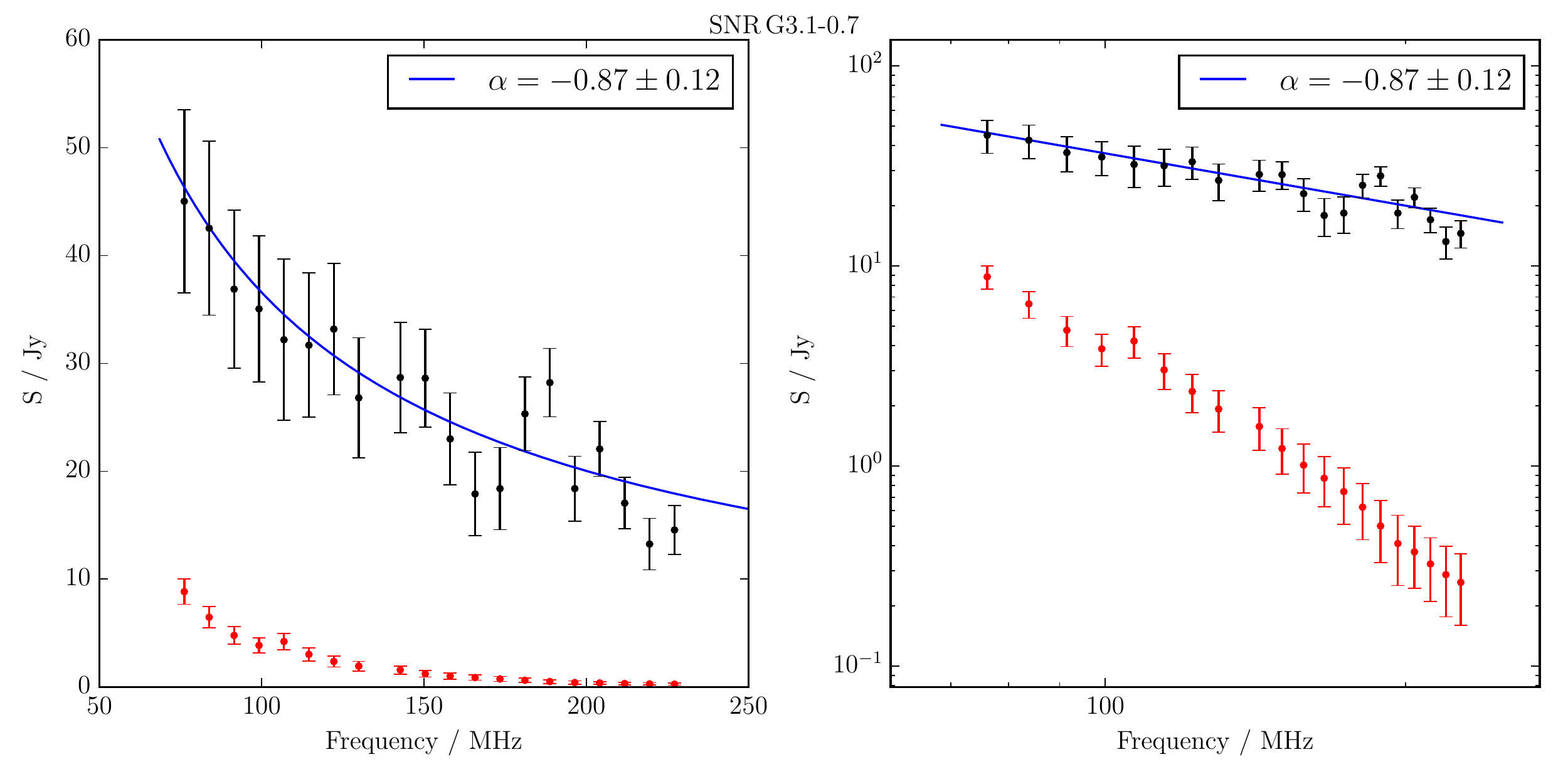}
    \caption{\spectrasummary G3.1-0.7. \spectrasuffix}
\end{figure}
		
\begin{figure}
    \centering
    \includegraphics[width=0.5\textwidth]{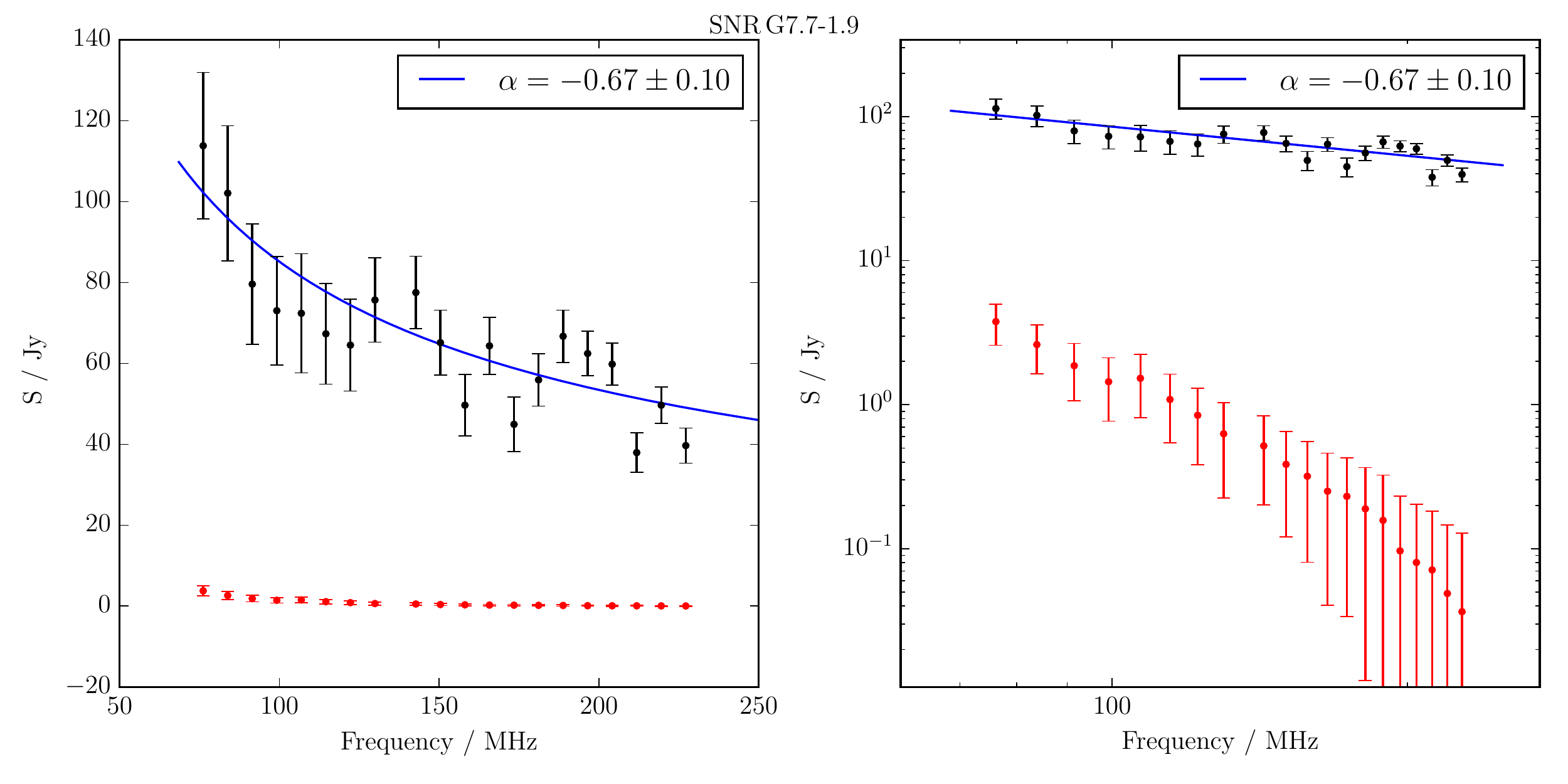}
    \caption{\spectrasummary G7.5-1.7. \spectrasuffix}
\end{figure}	

\begin{figure}
    \centering
    \includegraphics[width=0.5\textwidth]{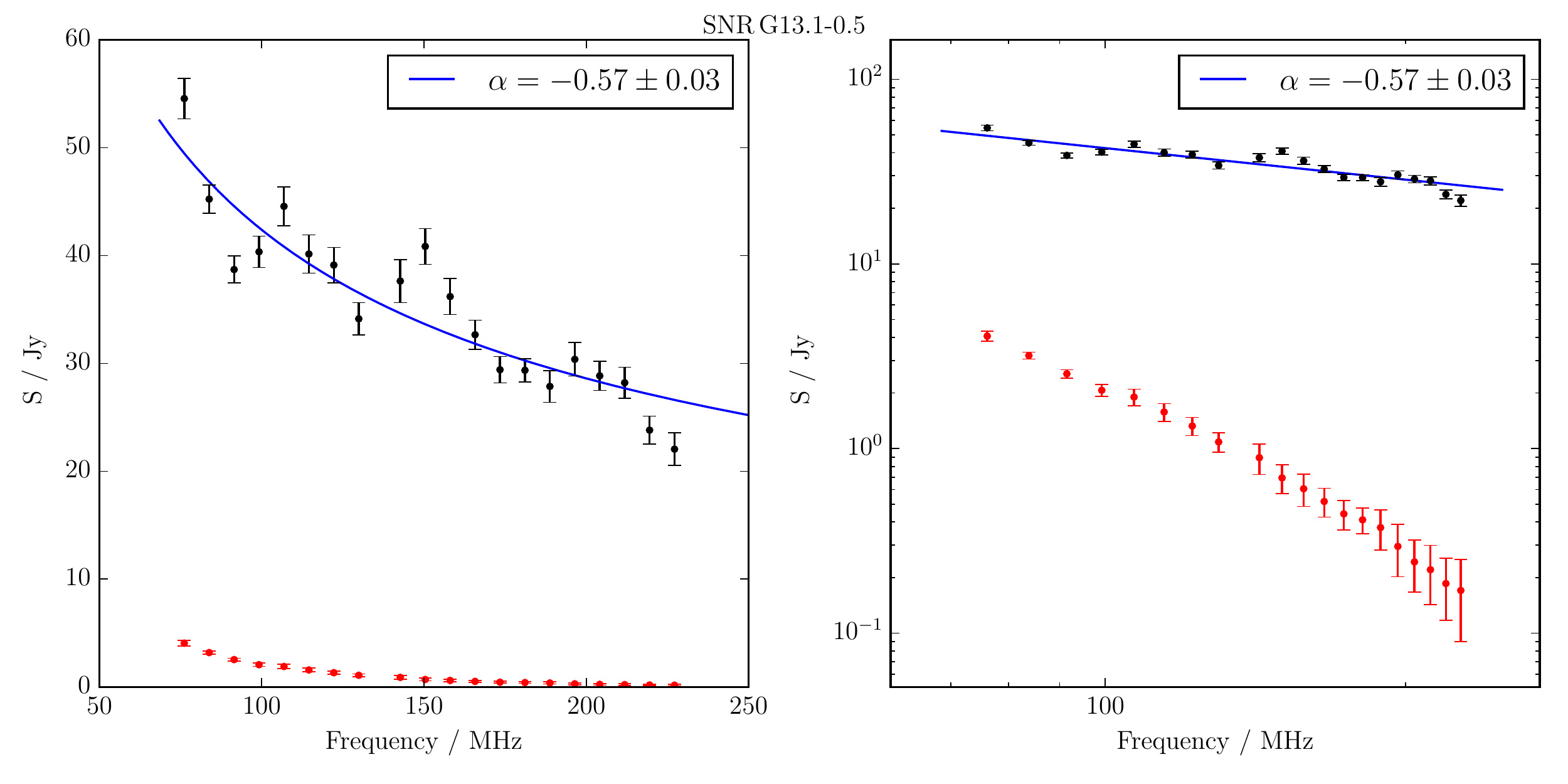}
    \caption{\spectrasummary G13.1-0.5. \spectrasuffix}
\end{figure}	
		
\begin{figure}
    \centering
    \includegraphics[width=0.5\textwidth]{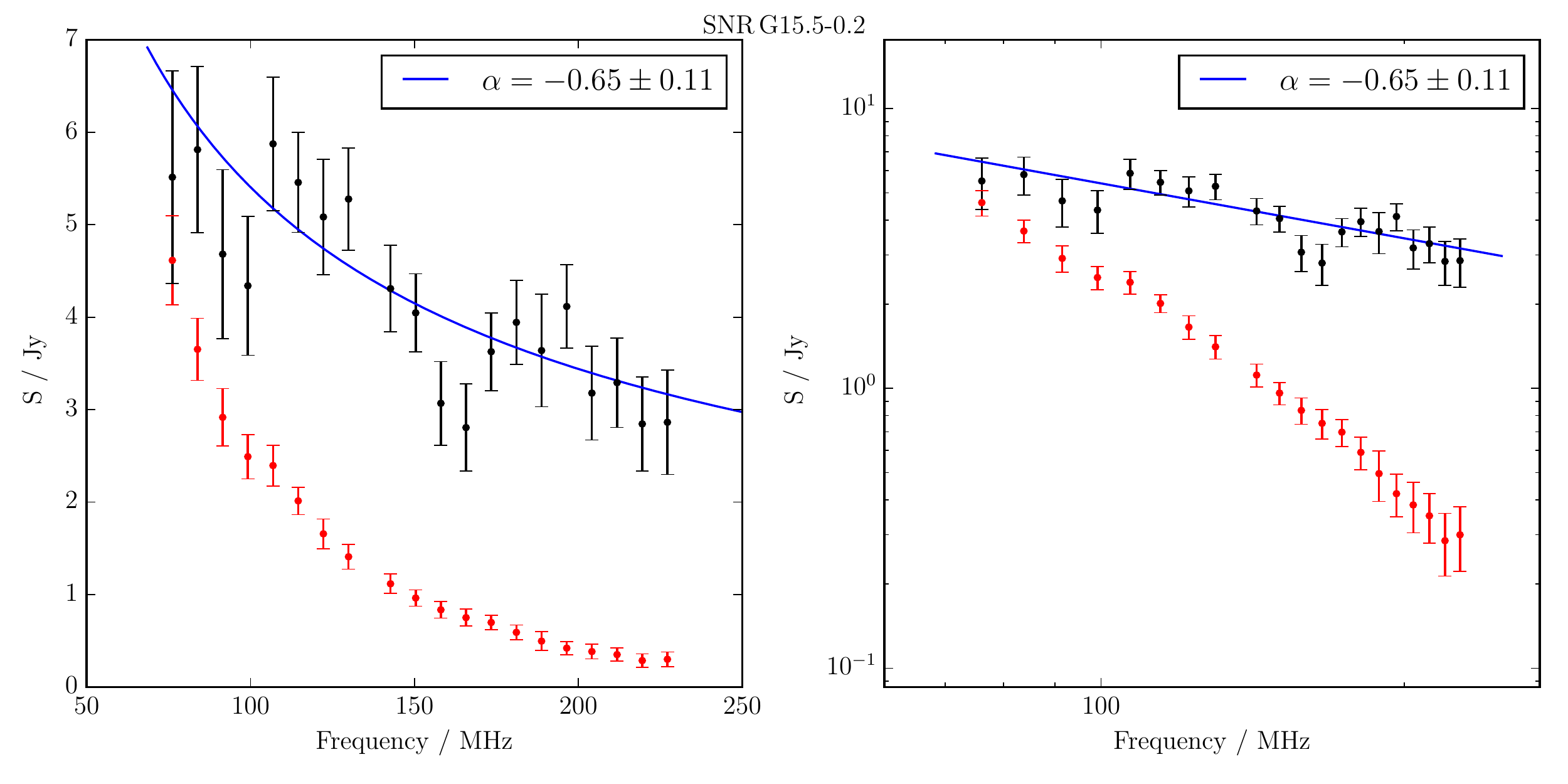}
    \caption{\spectrasummary G15.51-0.15. \spectrasuffix}
\end{figure}

\begin{figure}
    \centering
    \includegraphics[width=0.5\textwidth]{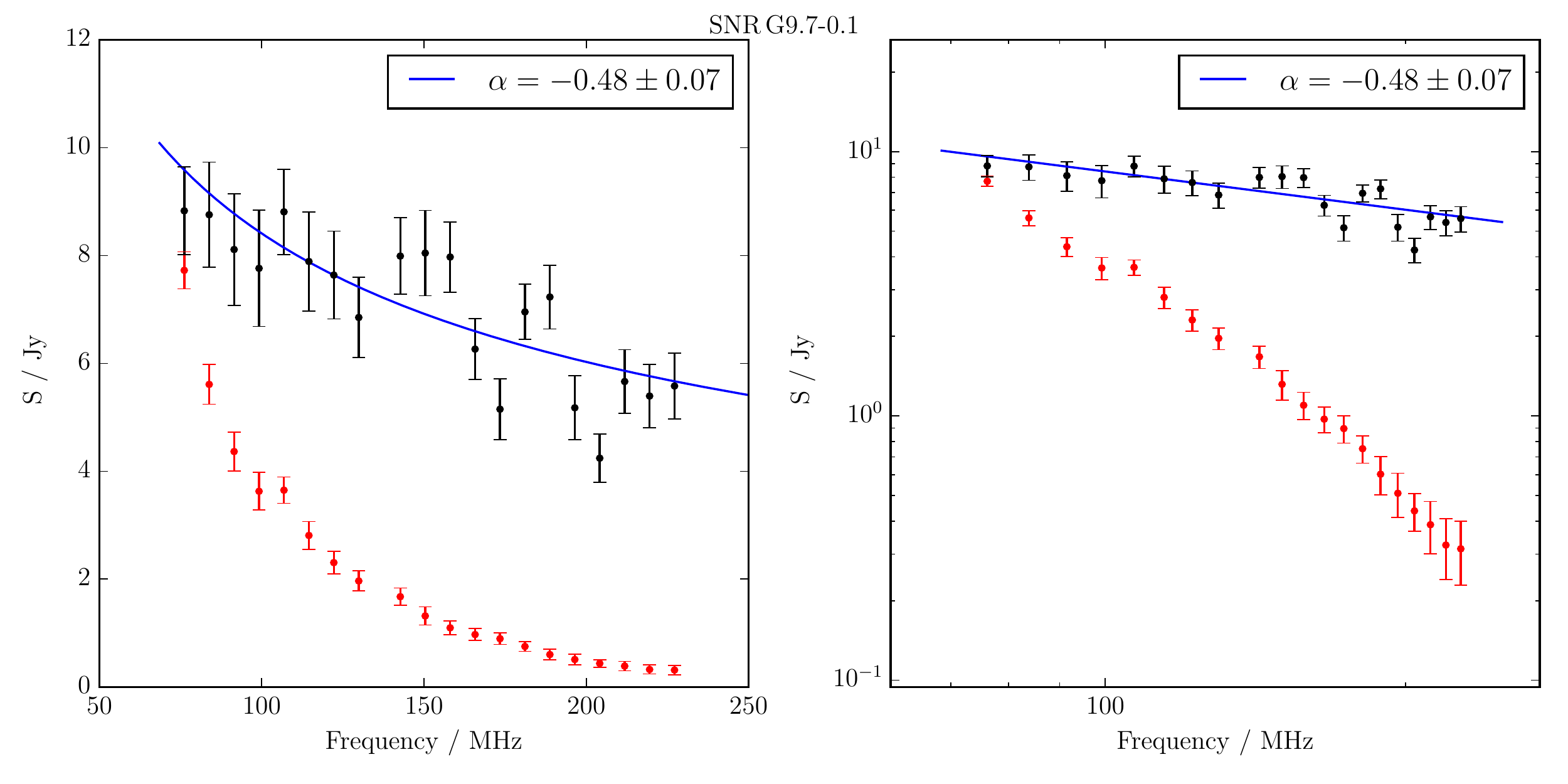}
    \caption{\spectrasummary MAGPIS\,$9.6833-0.0667$. \spectrasuffix}
\end{figure}

\begin{figure}
    \centering
    \includegraphics[width=0.5\textwidth]{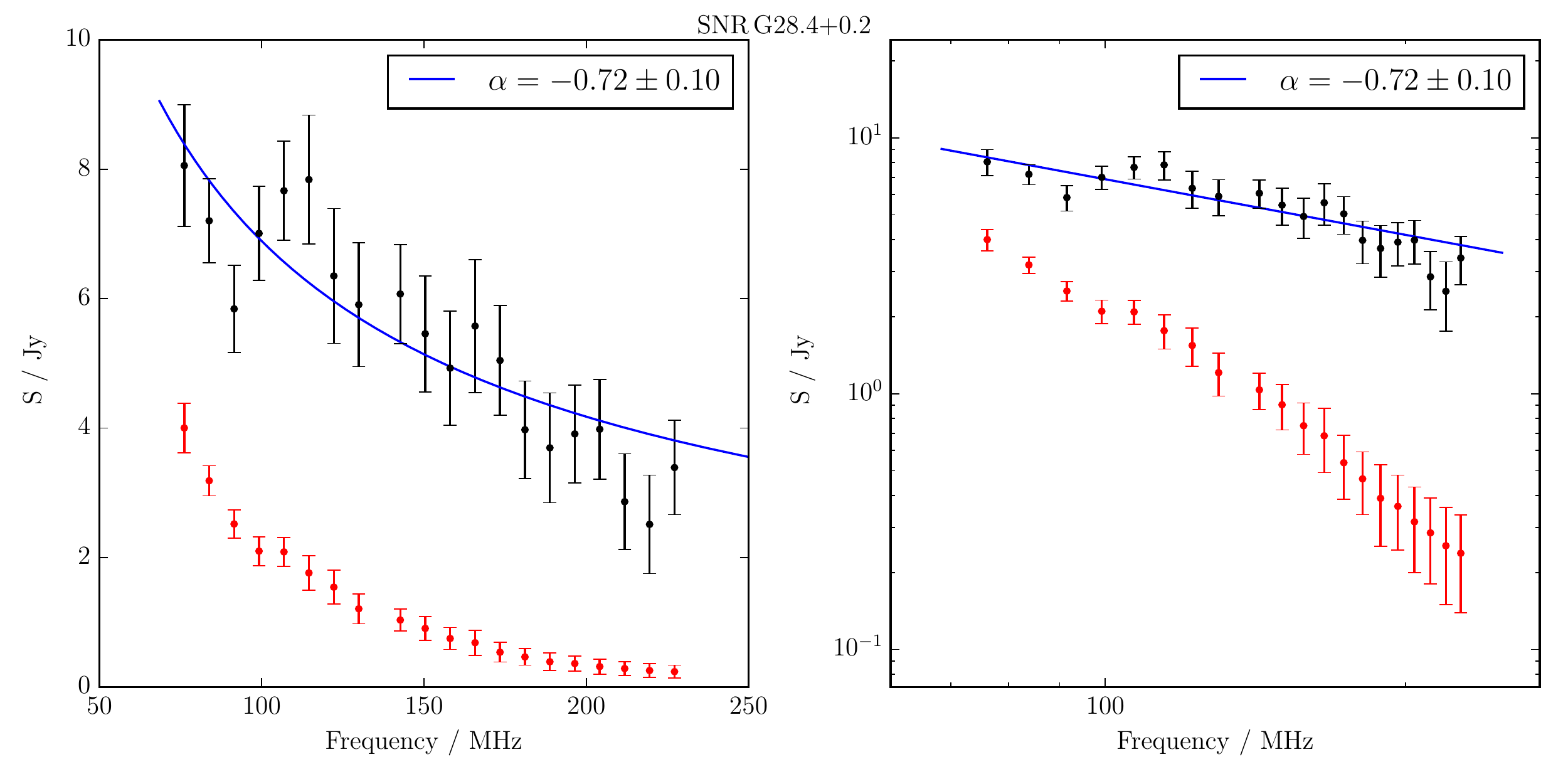}
    \caption{\spectrasummary MAGPIS\,$28.3750+0.2028$. \spectrasuffix}
\end{figure}

\begin{figure}
    \centering
    \includegraphics[width=0.5\textwidth]{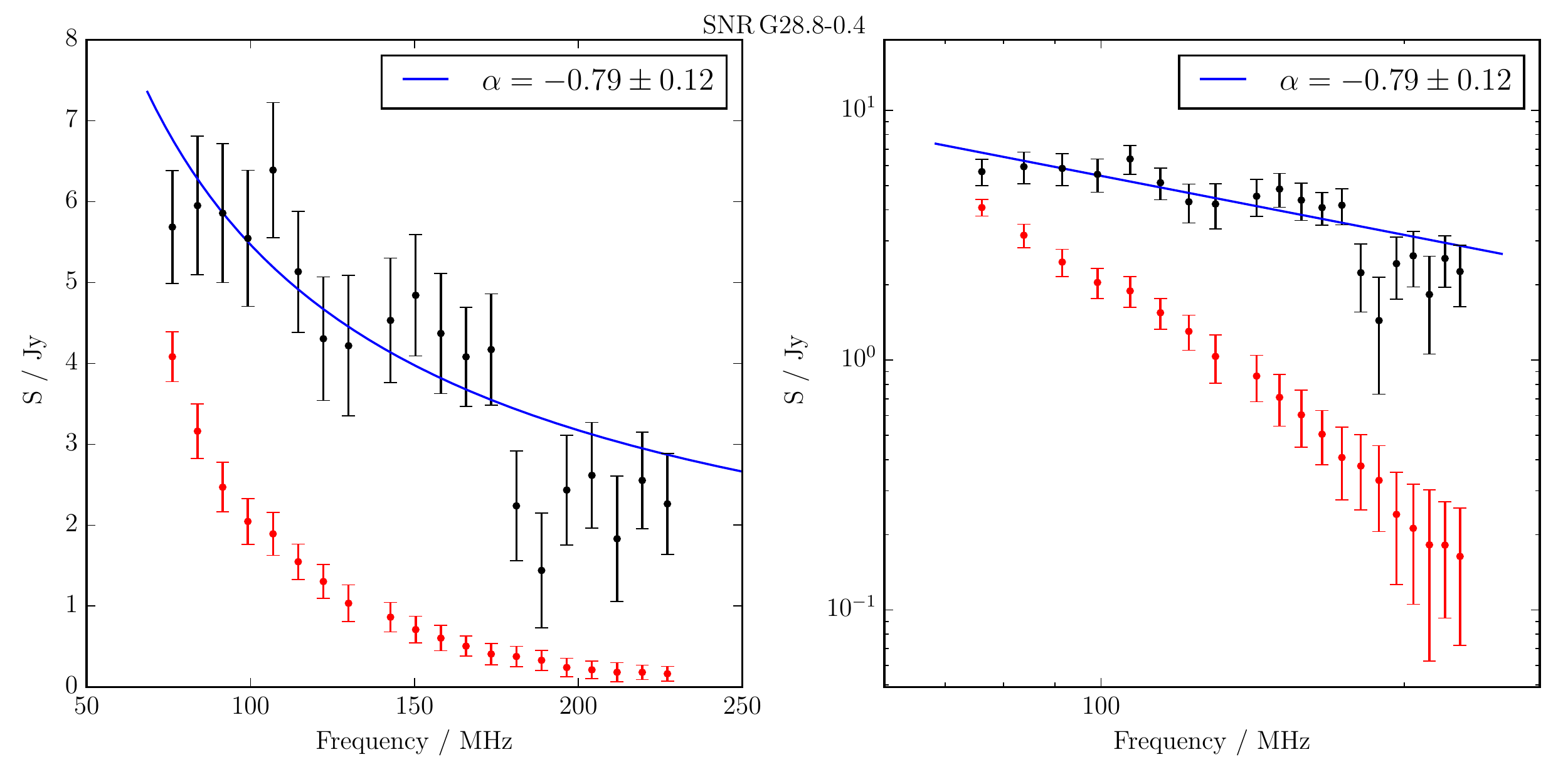}
    \caption{\spectrasummary MAGPIS\,$28.7667-0.4250$. \spectrasuffix}
\end{figure}

\begin{figure}
    \centering
    \includegraphics[width=0.5\textwidth]{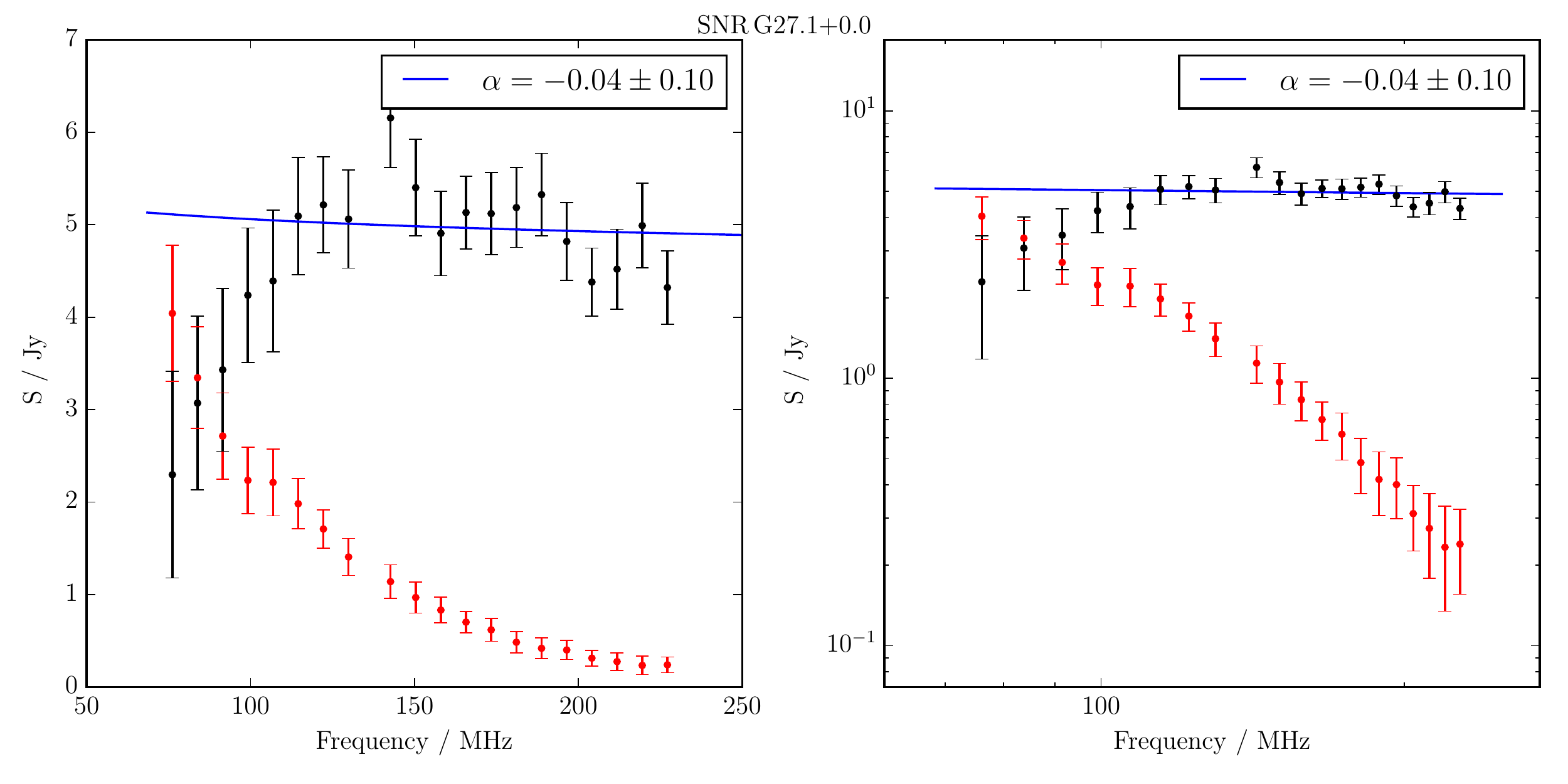}
    \caption{\spectrasummary MAGPIS\,$27.1333+0.0333$. \spectrasuffix}
\end{figure}

\begin{figure}
    \centering
    \includegraphics[width=0.5\textwidth]{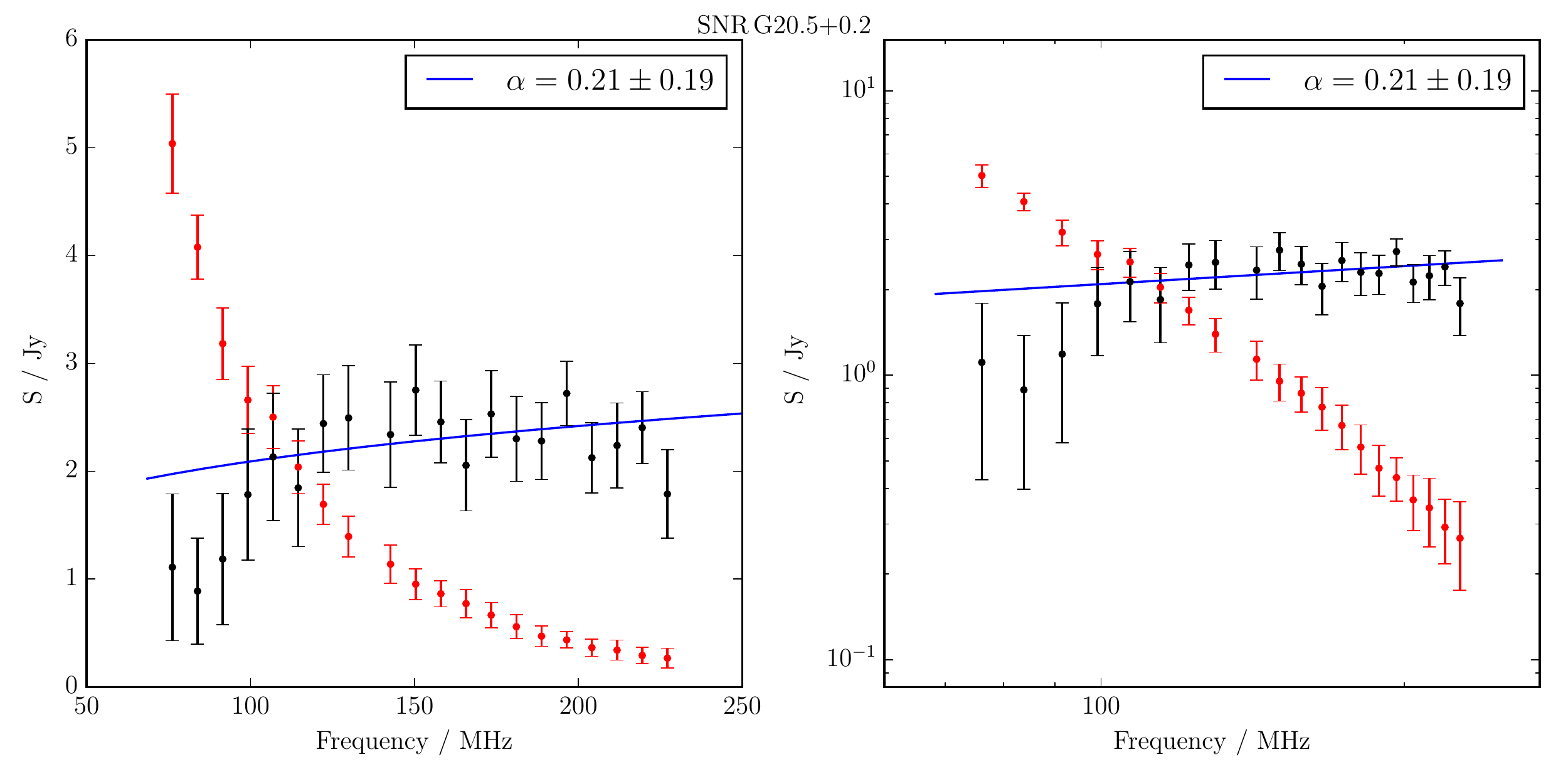}
    \caption{\spectrasummary MAGPIS\,$20.4667+0.1500$. \spectrasuffix}
    \label{fig:SNR_G20.5+0.2_spectrum}
\end{figure}

\end{appendix}

\bibliographystyle{pasa-mnras}
\bibliography{refs}

\end{document}